# Phonon-based partition of (ZnSe-like) semiconductor mixed crystals on approach to their pressure-induced structural transition

M. B. Shoker,[1] O. Pagès,[1,*] V. J. B. Torres,[2] A. Polian,[3,4] J.-P. Itié,[4] G. K. Pradhan,[5] C. Narayana,[6] M. N. Rao,[7] R. Rao,[7] C. Gardiennet,[8] G. Kervern,[8] K. Strzałkowski[9] and F. Firszt[9]

[1] Université de Lorraine, LCP-A2MC, ER 4632, F-57000 Metz, Fr
[2] Departamento de Fisica and I3N, Universidade de Aveiro, 3810 – 193 Aveiro, Portugal
[3] Institut de Minéralogie, de Physique des Matériaux et de Cosmochimie, Sorbonne Université — UMR CNRS 7590, F-75005 Paris, France
[4] Synchrotron SOLEIL, L'Orme des Merisiers Saint-Aubin, BP 48 F-91192 Gif-sur-Yvette Cedex, France
[5] Department of Physics, School of Applied Sciences, KIIT Deemed to be University, Bhubaneswar, Odisha, 751024, India
[6] Jawaharlal Nehru Centre for Advanced Scientific Research (JNCASR), Jakkur P.O. Bangalore 560064, India
[7] Solid State Physics Division, Bhabha Atomic Research Centre, Mumbai, 400085, India
[8] Université de Lorraine, Laboratoire de Cristallographie, Résonance Magnétique et Modélisations, UMR 7036, Vandoeuvre-lès-Nancy, F-54506, France
[9] Institute of Physics, N. Copernicus University, 87-100 Toruń, Poland

The generic 1-bond→2-mode "percolation-type" Raman signal inherent to the short bond of common $A_{1-x}B_xC$ semiconductor mixed crystals with zincblende (cubic) structure is exploited as a sensitive "mesoscope" to explore how various ZnSe-based systems engage their pressure-induced structural transition (to rock-salt) at the sub-macroscopic scale – with a focus on $Zn_{1-x}Cd_xSe$. The Raman doublet, that distinguishes between the AC- and BC-like environments of the short bond, is reactive to pressure: either it closes ($Zn_{1-x}Be_xSe$, $ZnSe_{1-x}S_x$) or it opens ($Zn_{1-x}Cd_xSe$), depending on the hardening rates of the two environments under pressure. A partition of II-VI and III-V mixed crystals is accordingly outlined. Of special interest is the "closure" case, in which the system resonantly stabilizes *ante* transition at its "exceptional point" corresponding to a virtual decoupling, by overdamping, of the two oscillators forming the Raman doublet. At this limit, the chain-connected bonds of the short species (taken as the minor one) freeze along the chain into a rigid backbone. This reveals a capacity behind alloying to reduce the thermal conductivity.

---

[*] Correspondence to olivier.pages@univ-lorraine.fr

$A_{1-x}B_xC$ semiconductor mixed crystals with zincblende structure[1] are benchmark systems for the experimental study of the chemical disorder due to alloying, which relates to the percolation site theory[2-4]. The C-invariant and (A,B)-substituting sublattices intercalate through a tetrahedral bonding, resulting in two bond species (A-C and B-C) arranged with maximal (cubic) symmetry. This forms the most "simple" three-dimensional (3D) disordered system of chemical bonds one can imagine, comparable to an ideal object, which can be experimentally tested and confronted with models ran at the ultimate atom scale. The simplicity gives grounds for hope to solve certain critical issues behind alloying. One refers to the nature of the A↔B atom substitution[5], as to whether this is ideally random, or not. Another one, tackled in this work, is to elucidate how lattice-supported complex media engage their pressure-induced structural transition at the local scale[6], this being already an issue for the pure compounds[7]. Yet in order to address experimentally the raised issues, one not only needs a suitable system as described above but also a local probe, such as the bond force constant, as conveniently measured at the laboratory scale by Raman scattering.

The historical models used for the discussion of the Raman spectra of $A_{1-x}B_xC$ zincblende mixed crystals, namely the modified-random-element-isodisplacement (MREI) and the cluster ones (both summarized in Ref. 8), are not well suited to tackle the raised issues, for different reasons detailed elsewhere[9]. In brief, the MREI model is blind to the local environment of a bond, by construction. As for the cluster model, to our view it suffers from a conceptual bias that was shown to generate a misleading insight into the nature of the atom substitution. Moreover, both models are based on the virtual crystal approximation in which the A and B substituents are replaced by virtual $A_{1-x}B_x$ atoms with physical properties averaged over the A and B ones depending on the composition x. By doing so the virtual crystal approximation virtually restores a perfect chemical/structural order, which comes to view a $A_{1-x}B_xC$ ternary mixed crystal at the macroscopic scale in terms of a pseudo-binary $(A_{1-x}B_x)C$ compound. Somehow, this denies the essence of substitution, leaving the impression that the semiconductor mixed crystals are saved from disorder, and are thus special among complex media.

Over the past two decades we introduced an alternative so-called percolation model that views an $A_{1-x}B_xC$ zincblende ternary at the mesoscopic scale in terms of an AC/BC-like composite[9]. A given bond vibrates at different frequencies in the AC- and BC-like regions giving rise to a generic bimodal Raman signal *per* bond (1-bond→2-mode). This is especially clear for the short bond (say, *e.g.*, B-C) that usually involves the substituent with a small covalent radius. As such, the short bond has more room to distort than the long one, and hence is the one that mostly accommodates the local strain due to the contrast in the A-C and B-C bond length/stiffness. This generates significant variation in the (B-C) bond force constant from site to site in the crystal depending on the local AC- or BC-like environment, with concomitant impact on the (B-C) Raman frequency, being accordingly diversified.

Just like its MREI and cluster predecessors, the percolation model is a phenomenological one in which the lattice dynamics is grasped within the one-dimension (1D) paradigm. The justification is that, as an optical method, Raman scattering can only detect long-wavelength ($\lambda \to \infty$, $q=2\pi\lambda^{-1} \to 0$) lattice vibrations[10]. On such length scale the information is averaged over several crystal unit cells so that there is no point in trying to spot an atom in the real (3D) lattice. For the sake of consistency, the AC- and BC-like environments of a bond are likewise described at 1D, which, for the minor species of the crystal at least (say, *e.g.*, B-C), comes to distinguish between isolated bonds (dispersed inside the AC-like region) and self-connected ones (forming the BC-like region). In fact, a singularity is expected in the Raman frequency ($\omega$) on crossing the bond percolation threshold – in echo to a universal bond length anomaly[11], *i.e.*, ~19 at.% in the zincblende structure, at which critical composition the self-connection becomes suddenly infinite, a pure statistical effect of the random A↔B substitution[1]. More generally, in the percolation scheme the bond environments are defined up to second-neighbors at most. This is consistent with a basic feature of the bond charge model used to describe the phonon dispersion of semiconductor compounds (like AC and BC) with diamond and zincblende structures that phonons are essentially a matter of short range interactions[12,13].

So far, the percolation scheme was applied to different mixed crystals with zincblende, wurtzite and diamond structures[8,14] suggesting its universal character. Apparently, only wavelength-resolved vibration spectroscopies such as optical methods ($\lambda \to \infty$) – covering Raman scattering and infrared absorption[15] – are sensitive to the local environment of a bond as formalized within the percolation scheme. Advanced temperature-dependent extended-X-ray-fine-structure-absorption measurements, while offering a useful overview of the entire lattice dynamics averaged over all possible wavelengths in one shot[16,17], fail to do so.

Seen from the angle of the percolation model, Raman scattering breaks new ground for studying mixed crystals. For instance, the first issue raised above has been solved recently: when exploited within the percolation scheme, the Raman spectra of zincblende[18] as well as diamond[19] mixed crystals can be used to shed light on the nature of the atom substitution (random vs. clustering/anticlustering) on a quantitative basis (using an *ad hoc* order parameter – see below). In this work, we address the second issue and test the Raman doublet of the short bond as a sensitive chemical probe to "see" how mixed crystals engage their pressure-induced structural transition at the mesoscopic scale.

Preliminary Raman studies have been conducted in this spirit on $Zn_{1-x}Be_xSe$[20] ($x \leq 0.52$) and $ZnSe_{1-x}S_x$[21] ($x=0.32$), though without being able, yet, to understand the observed phenomena, listed below:

(i) The Be-Se ($Zn_{1-x}Be_xSe$) and Zn-S ($ZnSe_{1-x}S_x$) Raman doublets close under pressure, due to a progressive convergence of the lower "mode 2" onto the upper "mode 1".

(ii) During the convergence process, mode 2 gradually collapses down to full extinction on crossing mode 1, apparently due to a dead loss of oscillator strength.

(iii) The actual crossing occurs around the same critical pressure $P_c$ at any $x$ value, falling close to the pressure-induced zincblende↔rock-salt structural transition of pure ZnSe.

(iv) Beyond $P_c$ only mode 1 survives; mode 2 "freezes", testified by *ab initio* calculations.

Features (i-iv), originally evidenced with $Zn_{1-x}Be_xSe$, were qualitatively attributed to the large contrast in the pressure-induced structural transitions of ZnSe (to the rock-salt structure[22], at $P_{ZnSe}$~13 GPa) and BeSe (to the NiAs structure[22], at $P_{BeSe}$~56 GPa). This seemed coherent with an *ab initio* trend that the Ga-P doublet of $GaAs_{1-x}P_x$ – characterized by nearly identical pressure transitions of its parent compounds[22] (~15±3 GPa, zincblende→*Cmcm*) – remains quasi stable under pressure[20]. However, this was a wrong track since features (i-iv) lately repeated with the Zn-S doublet of $ZnSe_{1-x}S_x$[21] even though the alleged contrast is suppressed in this case (ZnS transforms to rock-salt[22] at ~13 GPa, as ZnSe). Since then, we fall short of an adequate explanation for any of the features (i-iv)[21].

In this work a deeper insight into the origin of (i-iv) is searched for by extending our high-pressure Raman study of ZnSe-based mixed crystals to $Zn_{1-x}Cd_xSe$ taken in its zincblende structure (x<0.3, the structure is wurtzite otherwise) using a single crystal. The 3-mode {1×(Cd-Se),2×(Zn-Se)} Raman behavior of $Zn_{1-x}Cd_xSe$ at ambient pressure has been clarified recently[18], a prerequisite to its high-pressure study. As features (i-iv) are not x-dependent, for the current experimental case studies we focus on $Zn_{0.83}Cd_{0.17}Se$ whose Raman behavior at ambient pressure was studied in detail. $Zn_{0.83}Cd_{0.17}Se$ is further interesting in that its optical band gap is nearly resonant with the green laser line used to excite the Raman spectra[1]. This offers a chance to play with the gap-related singularity in the dispersion of the refractive index in view to access the phonon-polaritons besides the conventional phonons (a similar approach with $Zn_{1-x}Mg_xSe$[14] proved to be much rewarding in this respect), hence offering a Raman overview. Generally, the phonon-polaritons propagating in the volume of mixed crystals remain unexplored experimentally – not to mention under high pressure, apart from our own recent studies[14,18,21] (and Refs. therein).

The as-completed series of ZnSe-based systems encompasses a broad range of contrasts in bond physical properties:

(v) The Be-Se (0.420), Zn-S (0.764) and Cd-Se (0.841) bond ionicities are smaller, similar and larger, respectively, than that of Zn-Se (0.740)[23]. Therefore, the Be and Cd incorporations stiffen and soften the ZnSe lattice, respectively, whereas the S incorporation is neutral with this respect. The zincblende↔rock-salt transition pressures are correspondingly larger, smaller and comparable to $P_{ZnSe}$[22].

(vi) The Zn-Se bond stands out among the II-VI's in that its capacity to stiffen under pressure hits the lowest level, judged by the minimal volume derivative of its bond ionicity

(vii) ($df_i^*/dlnV$=0.127), that makes roughly half the value for Zn-S and Cd-Se, and one sixth as much as the Be-Se one[23].

(vii) The three systems cover various percolation-type Raman patterns, *i.e.*, with well-separated (~200 cm$^{-1}$, Zn$_{1-x}$Be$_x$Se[20]), close (~50 cm$^{-1}$, ZnSe$_{1-x}$S$_x$[21]) and degenerate (~0 cm$^{-1}$, Zn$_{1-x}$Cd$_x$Se[18]) AC-singlet and BC-doublet – to a point, in the latter case, that the spacing is larger for the doublet ($\delta$~20 cm$^{-1}$) than between the doublet and the singlet ($\Delta$~10 cm$^{-1}$) – (see, *e.g.*, Supplementary Fig. S6b).

(viii) Zn-Se is either the "passive" (AC-singlet, Zn$_{1-x}$Be$_x$Se and ZnSe$_{1-x}$S$_x$) or the "active" (BC-doublet, Zn$_x$Cd$_{1-x}$Se) bond; the active bond is sensitive to its local environment either up to first- (Zn$_{1-x}$Be$_x$Se) or second-neighbors (ZnSe$_{1-x}$S$_x$, Zn$_{1-x}$Cd$_x$Se), and the active bond is either the minor (Be-Se, Zn-S) or the dominant (Zn-Se of Zn$_{1-x}$Cd$_x$Se) species[18,20,21].

We hope that such variety of contrasts (v-viii) will help to get sufficient hindsight – that was sorely lacking so far – to elucidate features (i-iv). Such forward step is needed prior to validating the percolation-type Raman doublet of a bond as a useful "mesoscope" for study of semiconductor mixed crystals under pressure.

An overview of the pressure dependence of various Raman doublets for the current series of ZnSe-based systems, useful to fix ideas, is sketched out in Fig. 1. This reveals in advance the primary outcome of this work that the Zn-Se doublet (Zn$_{1-x}$Cd$_x$Se) opens under pressure, contrary to the Be-Se (Zn$_{1-x}$Be$_x$Se) and Zn-S (ZnSe$_{1-x}$S$_x$) ones, that close[20,21]. We detail below how the opening trend for Zn$_{1-x}$Cd$_x$Se was evidenced combining experimental and *ab initio* Raman insights into its polar and non-polar modes, respectively. Also, the variety of trends in Fig. 1 appears to be sufficient to trace the origin of the pressure-induced closure/opening of a percolation-type Raman doublet.

To complete the picture, the disconcerting features (i-iv) in the closure case are re-examined within a basic model of coupled/damped harmonic oscillators defined at 1D[24,25] – for the sake of consistency with the 1D-percolation scheme, focusing on Zn$_{~0.5}$Be$_{~0.5}$Se as a case study. A pivotal feature in this model is the so-called *exceptional point*[24,25], characterized by a perfect balance between gain (mechanical coupling in this case) and loss (due to overdamping). Such singular points are being actively sought in optics and photonics as the source of exotic behaviors[26], and also in phononics, notably in view to minimize the thermal conductivity in semiconductor-based devices[27,28] (achieved through nanoarchitecturing in the cited works).

## Results and discussion

By placing the study (at 17 at.% Cd) close to the Cd-Se bond percolation threshold (*i.e.*, $x_p$~19 at.% Cd, in case of a random substitution), we can be assured of a top-diversified mesostructure – thus a representative one, *i.e.*, with maximum variety in the topology of the minor (Cd-Se) species: isolated as well as self-connected Cd-Se bonds are expected to coexist in significant proportions, forming, in the latter case, small as well as large (nearly infinite, on the verge of percolation) clusters, also in significant proportions[2].

Now, Zn$_{1-x}$Cd$_x$Se is prone to clustering on approach to its composition-induced zincblende→wurtzite structural transition[18] ($x$~0.3), which might significantly impact its mesostructure, especially at compositions near the Cd-Se bond percolation threshold as in the present case. A Raman insight into the clustering rate $\kappa$ of Zn$_{1-x}$Cd$_x$Se with moderate-to-large Cd content – on a scale of 0 (random substitution) to 1 (full clustering, *i.e.*, phase separation) – using the terminology of the $\kappa$-based formalism developed within the cluster model[8] – has earlier been obtained within the percolation scheme based on a description of Zn$_{1-x}$Cd$_x$Se in terms of a CdSe/ZnSe-like "composite" (see above). Such Raman insight operated at the "mesoscopic" scale, leading to $\kappa$~0.5, was independently supported by *ab initio* calculations[18]. A more refined "microscopic" insight relying on a description of the current Zn$_{0.83}$Cd$_{0.17}$Se crystal in terms of five elementary Se-centered 3D-tetrahedron units with (Zn,Cd) at the vertices is currently gained by performing $^{77}$Se solid-state nuclear magnetic resonance (NMR) measurements on a powdered sample – along the approach earlier used with Zn$_{1-x}$Cd$_x$Te[29], leading to $\kappa$~0.12. The latter value is only used to fix ideas since the $\kappa$-based formalism eventually appears to be non-transferable to the NMR data (Supplementary Section I).

A subsequent statistical analysis of the (Cd,Zn)–arrangement in large (10×10×10) $A_xB_{1-x}C$ zincblende supercells reveals that the minor Cd-Se bonds percolate with the same probability in large ($x$=0.17, $\kappa$~0.5)-clustered and ($x$=0.19, $\kappa$~0)-random supercells, meaning that the bond percolation is hastened by clustering (Supplementary Section I). However, as $\kappa$=0.5 seems to be an upper estimate for $Zn_{0.83}Cd_{0.17}Se$ (compare the Raman vs. NMR insights), we can safely state that our crystal has not yet crossed the Cd-Se bond percolation threshold. Hence, its mesostructure presumably consists of a CdSe-like dispersion embedded in a Swiss cheese-like ZnSe-like matrix (and not of two finely interwined CdSe- and ZnSe-like treelike 3D-continua).

The limit for the planned high-pressure Raman study of $Zn_{0.83}Cd_{0.17}Se$ is the zincblende→rock-salt (fourfold→sixfold coordination) structural transition[22], identified at ~12 GPa by high-pressure X-ray diffraction. Representative diffractograms *ante* and *post* transition in the upstroke (pressure increase) and downstroke (pressure decrease) regimes are shown in the Supplementary Section I. Remarkably, the composition dependence of the $Zn_{1-x}Cd_xSe$ (x<0.4) bulk modulus $B_0(x)$ at ambient pressure derived from the X-ray data exhibits a percolation-type singularity at 17 at.% Cd. This offers an insight into the mechanical properties of the studied $Zn_{0.83}Cd_{0.17}Se$ mixed crystal at the macroscopic scale besides that achieved hereafter at the mesoscopic scale by Raman scattering.

A selection of pressure-reversible backward (reflection-like) and forward (transmission-like) Raman spectra taken at (nearly) the same spot at low (a, ~0 GPa), intermediate (b, ~5 GPa) and high (c, ~9 GPa < $P_{CdZnSe}$) pressures through/onto the (110)-faces of a tiny $Zn_{0.83}Cd_{0.17}Se$ single crystal inserted in a diamond anvil cell is shown in Fig. 2 (more spectra are shown in Supplementary Section I together with additional ones taken on a powder – Fig. S6). Such scattering geometries probe the transverse optic (TO) modes of a zincblende crystal in their purely-mechanical (PM-TO strictly speaking, but abbreviated TO hereafter) and phonon-polariton (PP) regimes[10,30], respectively. Though the longitudinal optic (LO) modes are theoretically forbidden[31], they also show up due to multi-reflection of the laser beam between parallel crystal faces[32], completing an overview into the $Zn_{0.83}Cd_{0.17}Se$ optic modes.

The optic modes are linked to each other: the TO and LO modes mark the asymptotic limits of the "$\omega$ vs. $q$" PP dispersion far off the center $\Gamma$ ($q$=0) of the Brillouin zone (being clear that as an optical technique Raman scattering operates close to $\Gamma$ anyway) and near $\Gamma$, respectively[10,30,31]. The PP dispersion (curves) and Raman intensities (thickness of curves) calculated for $Zn_{0.83}Cd_{0.17}Se$ by using the generic expression of the Raman cross section given in Ref. 18 (details are given in the Supplementary Section I) are shown in Fig. 2.

We are mostly interested in the pressure dependence of the $TO_{Zn-Se}^{Zn} - TO_{Zn-Se}^{Cd}$ doublet, the subscript specifying the bond vibration and the superscript the local environment. The focus on the TOs is justified because, generally, being non polar (purely-mechanical) vibrations they hardly couple and thus reflect the intrinsic phonon pattern at a given pressure. However, a direct experimental TO insight (as with $Zn_{1-x}Be_xSe$[20]) is difficult in the case of $Zn_{1-x}Cd_xSe$ because its three-mode TO pattern is so compact (see, *e.g.*, Fig. S6b) – to a point that $Zn_{1-x}Cd_xSe$ was long considered to exhibit the ultimate one-mode (2-bond→1-phonon) behavior in its Raman spectra[18]. Further the $TO_{Zn-Se}^{Cd}$ Raman mode is hardly discernible (Fig. 1) due to a non-favorable sharing of the available Zn-Se available oscillator strength between the two Zn-Se submodes at x~0.17[18]. Last, on its sensitive high frequency side the TO signal is screened by the LO modes that show up strongly.

In view of such drawbacks with the TO modes we are forced to proceed with the polar PP (as with $ZnSe_{1-x}S_x$[21]) and LO modes, with a difficulty that they couple via their macroscopic transverse $\vec{E}_T(q)$ and longitudinal $\vec{E}_L$ electric fields, respectively. Due to the $\vec{E}_{T,L}$-coupling, neither a specific bond nor a specific environment can be assigned to the PPs and LOs. Therefore, these modes are simply labeled as $\{PP^-, PP^{int}, PP^+\}$ and $\{LO^-, LO^{int}, LO^+\}$ in ascending frequency (lower, intermediate, upper). Generally, the $\vec{E}_T(q)$- and $\vec{E}_L$-couplings channel the available (Cd-Se and Zn-Se) oscillator strength towards low and high frequencies, respectively[21]. Hence the PPs and the LOs are naturally well suited to investigate the compact Zn$_{1-x}$Cd$_x$Se TO pattern on its low- (CdSe-like) and high-frequency (ZnSe-like) sides, respectively.

At (nearly) ambient pressure, the situation is as follows (Fig. 2): the $\{TO_{Cd-Se}, TO_{Zn-Se}^{Zn}, TO_{Zn-Se}^{Cd}\}$ modes of Zn$_{0.83}$Cd$_{0.17}$Se show up at {~190, ~200 and ~220} cm$^{-1}$, with their LO replicas at {~190, ~220 and ~245} cm$^{-1}$. The PPs are not visible, because, as soon as they depart from their native TOs, they interfere destructively with a two-phonon continuum involving transverse acoustic modes from the Brillouin zone edge (2×TA) that emerges nearby[18] – as independently observed with Zn$_{1-x}$Be$_x$Se[20] and ZnSe$_{1-x}$S$_x$[21]. Under pressure, the 2×TA zone-edge band softens (shifts to low frequency), in contrast with all zone-center (Γ-like) optic modes that harden (shift upward)[33]. This opens a path for PP detection at minimal scattering angle – see Methods. Only $PP^-$ is observed at ~5 GPa (close to $TO_{Zn-Se}^{Zn}$), whereas both $PP^-$ (emerging well beneath $TO_{Zn-Se}^{Zn}$ then) and $PP^{int}$ come about at ~9 GPa. The increased PP diversity can be explained only if the native $TO_{Cd-Se}$ behind $PP^-$ breaks away from the native $TO_{Zn-Se}^{Zn} - TO_{Zn-Se}^{Cd}$ doublet of $PP^{int}$ under pressure (see clear arrows in Fig. 2).

An indirect insight into the remaining upper/minor $TO_{Zn-Se}^{Cd}$ end of the Zn-Se doublet is achieved via its (quasi) degenerate $LO^{int}$ mode. At ambient pressure, $LO^{int}$ emerges as a residual feature slightly remote from the mid-[$TO_{Zn-Se}^{Zn} - LO^+$] band on the TO-side[18]. By increasing pressure, $LO^{int}$ is shifted away from $TO_{Zn-Se}^{Cd}$ towards $LO^+$ and strengthens until arriving at quasi intensity matching with $LO^+$ at ~9 GPa (judging by the areas of the Raman peaks). The pressure-induced upward-shift/strengthening of $LO^{int}$ are intrinsic to cubic Zn$_{1-x}$Cd$_x$Se since they are also visible in the LO-like Raman spectra of disordered and semi-ordered (cubic) thin films (~45 at.% Cd)[34]. Their combination cannot be merely fortuitous, suggesting a common origin. This is searched for by trying a blind test on the pressure dependence of the underlying $TO_{Zn-Se}^{Cd}$ frequency behind $LO^{int}$, with several options: the ($TO_{Zn-Se}^{Zn} - TO_{Zn-Se}^{Cd}$) doublet closes (scenario 1), remains stable (scenario 2), or widens (scenario 3) under pressure. There is no need to speculate on the $TO_{Zn-Se}^{Zn}$ frequency that is readily accessible at any pressure. Both the upward-shift/strengthening of $LO^{int}$ under pressure nicely fit into scenario 3 (Fig. 2), whereas scenarios 1 and 2 fail to generate any of those trends (Fig. S8).

In brief, the experimental PP and LO Raman insights on each side of the compact TO pattern of Zn$_{0.83}$Cd$_{0.17}$Se converge to reveal that the $\{TO_{Cd-Se}, TO_{Zn-Se}^{Zn}, TO_{Zn-Se}^{Cd}\}$ triplet splits off under pressure. This applies in particular to the Zn-Se doublet (Zn$_{1-x}$Cd$_x$Se) – of central interest, hence contrasting with the Zn-S (ZnSe$_{1-x}$S$_x$) and Be-Se (Zn$_{1-x}$Be$_x$Se) doublets, which are closing under pressure. Independent support to the overall splitting arises from *ab initio* calculation of the high-pressure pure-TO Raman spectra related to a large (216-atom) disordered ($\kappa$~0) Zn$_{0.5}$Cd$_{0.5}$Se cubic-supercell (see Methods) corresponding to well-resolved Cd-Se and Zn-Se Raman signals (Fig. 2, bottom).

Out of the listed contrasts (v-viii), only that related to (vi) $df_i^*/dlnV$ can explain the pressure-induced closure/opening of the Raman doublet of the short bond depending on the system. Consider, *e.g.*, the Zn-Se doublet of Zn$_{1-x}$Cd$_x$Se that distinguishes between Zn-Se TO vibrations in homo (ZnSe-like, lower mode) and hetero (CdSe-like, upper mode) environments. Under pressure, the Cd-Se bonds stiffen up faster ($df_i^*/dlnV$ is large) than the Zn-Se ones ($df_i^*/dlnV$ is small), with concomitant impact on the Zn-Se Raman shifts, being large for the upper mode and small for the lower one, meaning that the Zn-Se gap widens. In this line the Ga-P doublet of GaAs$_{1-x}$P$_x$ is also expected to widen – though to a less extent since the $df_i^*/dlnV$-contrast is less in GaAs$_{1-x}$P$_x$ than in Zn$_{1-x}$Cd$_x$Se[23]. The trend line is actually there, as revealed by a careful examination of existing *ab initio* data[20]. The Be-Se (Zn$_{1-x}$Be$_x$Se) and Zn-S (ZnSe$_{1-x}$S$_x$) closures can be explained in the same way, which solves issue (i).

Generally, the frequency gap between the two Be-Se TO sub-modes of Zn$_{1-x}$Be$_x$Se is stable throughout the x-domain (within ~10%), that is the reason why the closure occurs around the same critical pressure $P_c$ at any x value. The gap within the Zn-S gap doublet of ZnSe$_{0.68}$S$_{0.32}$[21] is roughly one-third the Be-Se one in Zn$_{1-x}$Be$_x$Se[20]; at the same time Zn-S rigidifies under pressure at a faster rate than Be-Se does, also by about one-third (Fig. 1)[23]. Hence, the Zn-S (small gap, low rate) and Be-Se (large gap, high rate) doublets close around the same $P_c$, by chance falling close to $P_{ZnSe}$[22]. This resolves issue (iii).

Remaining issues in the closure case relate to the (ii) Raman extinction of the lower mode and to (iv) its phonon freezing (i) on crossing the upper one at ($\omega_c$,$P_c$). A common origin is discussed below by using the well documented Be-Se doublet of Zn$_{1-x}$Be$_x$Se as a case study (Supplementary Section II).

Feature (ii) reveals that the lower TO mode actually "feels" the upper one when forced into its proximity by pressure. This suggests some coupling, *i.e.*, a mechanical one then, owing to the purely-mechanical nature of the TOs. In this case one would a priori expect the repulsion of the two TOs in their tight-coupling at the resonance. However, this contradicts experimental findings demonstrating that the crossing, within experimental resolution, actually occurs (Fig. S7). In principle, the crossing makes sense only if the two TOs do not "see" each other, and hence are not coupled. This opposes to our basic premise.

The contradiction is removed by considering a model of two coupled but damped harmonic (mass + spring) 1D-oscillators with identical masses but different force constants standing for the Be-Se bonds vibrating in homo (BeSe-like, lower mode) and hetero (ZnSe-like, upper mode) environments. A recent TO-like version of such model[25] is generalized to LOs in this work (Supplementary Section II). A pivotal feature in this model is the so-called exceptional point characterized by exact screening of the mechanical coupling by overdamping right at the resonance ($\omega_c$, $P_c$) – corresponding to a perfect tuning of the bare-uncoupled TO oscillators. Such screening leads, in fact, to a virtual decoupling of the TOs, so that an actual crossing is allowed – in reference to (i).

The picture which emerges is that the mechanical coupling between the two TOs is progressively undermined by the increasing damping of the lower mode on approach to the upper one – testified by (ii), until the as-overdamped system of coupled TOs resonantly locks into its exceptional point from $P_c$ onwards. In fact, no crossing can occur except in this point: the prevalence of either the mechanical coupling or the overdamping at the resonance leads to TO-anticrossing (Supplementary Section II)[25].

The unique coupled-overdamped normal mode at the exceptional point, abbreviated exceptional mode below, manifests a compromise between the distinct symmetric and antisymmetric normal modes of the coupled system formed at the resonance in absence of damping. As the latter refer to in-phase and out-of-phase motions of oscillators – with equal magnitude, respectively, only one oscillator can be active in the exceptional mode, the other one has to be passive, being merely involved as a frozen/inert body – as sketched out in Fig. 1 (Supplementary Section II). This nicely resonates with the *ab initio* insight (iv). As such, the exceptional mode vibrates at nearly the same frequency as the active oscillator, hence assigned as the upper Be-Se mode. As for the passive oscillator, by freezing it becomes Raman inactive: the Raman effect results from modulation of the electronic susceptibility by a bond vibration, absent in this case. This is consistent with the lower mode suffering a Raman extinction from $P_c$ onwards – referring to (ii).

In brief, the (Raman extinction, phonon-freezing, exceptional mode) triptych is consistent and outlines an appropriate framework to explain all pending issues (i-iv).

Retrospectively the above situation can be viewed as a damped variant of the non-linear 1D-lattice dynamics studied in a pioneering numerical experiment done on an extended chain of similar (in masses and spring constants) undamped oscillators[35]. In our case, the pointedly introduced anharmonicity – in reference to the mechanical coupling between the two TOs – fails to generate the chaotic dynamics observed by the cited authors, because on reaching the resonance at ($\omega_c$,$P_c$) the coupled system becomes overdamped and locks into its exceptional mode.

## Conclusion

Summarizing, this work reveals a partition between A$_{1-x}$B$_x$C zincblende mixed crystals depending on whether they engage their pressure-induced structural transition with/without a rigid in-chain-backbone of self-connected bonds of the short species – when this is minor in the crystal (at least up to ~50 at.% in the case of Zn$_{1-x}$Be$_x$Se – Supplementary Section II). A decisive test is the closure/opening of the related percolation-type Raman doublet under pressure. The closure/opening is explained around the notion of a phonon exceptional point being achieved/avoided depending on the pressure dependence of the mechanical properties of the AC- and BC-like host media behind the Raman doublet. Such partitions for the common II-VI and III-V mixed crystals, compiled on the basis of the theoretical estimates of bond length and volume dependence of bond ionicity provided in Ref. 23 (along the same line as discussed in this work), are shown in Fig. 3.

    We foresee potential applications in the closure case. The pressure can be tuned to release or inhibit the vibration along the backbone, offering an effective on/off phononic switch at the unusual mesoscopic scale. Additional flexibility arises in that the backbone can be continuous or segmented depending on whether the short bonds percolate or not. Generally, this is interesting in view to reduce the thermal conductivity. Besides, the pressure-induced freezing of part of the short/stiff bonds (the self-connected ones) vibrating at high frequency dramatically impacts the oscillator strength awarded to the related/upper Γ-like LO mode (Supplementary Section II). As the latter dominates the heat dissipation process of photoelectrons in zincblende semiconductors[36], the pressure emerges as a possible means to play with the thermalization rates of electrons in mixed-semiconductor-based photovoltaic devices.

## Methods

This Section provides detail for replication and interpretation of the reported data. Additional experimental insights gained by nuclear magnetic resonance, high-pressure X-ray diffraction and high-pressure Raman scattering, together with statistical issues related to bond percolation phenomena as well as theoretical ones concerning coupled systems of damped harmonic 1D-oscillators – applied to $Zn_{0.48}Be_{0.52}Se$ in this case – are reported as Supplementary Information (Sec. II).

**Sample growth and preparation.** The used $Zn_{1-x}Cd_xSe$ and $Zn_{1-x}Be_xSe$ samples were grown by using the Bridgman method[37] as large high-quality single crystals – as the $ZnSe_{1-x}S_x$ systems completing the current ZnSe-based series – and prepared for investigations as cylinders 3 mm in height (the as-grown crystals being 5-6 cm in length) and 8 mm in diameter. The purity of the ZnSe and CdSe compounds and of the Be material used to prepare the mixtures were 5N (99.9995) and 2N (99.5), respectively. The composition was determined by energy dispersive X-ray spectroscopy analysis. The composition gradient along the growth axis is negligible (less than 0.5 at. % Cd) for the considered length of the investigated samples. The particular $Zn_{0.83}Cd_{0.17}Se$ and $Zn_{0.48}Be_{0.52}Se$ crystals studied in the body of the manuscript and in the Supplementary Section II exhibit a pure zincblende structure at ambient pressure testified by X-ray diffraction (see Ref. 18 for $Cd_{0.17}Zn_{0.83}Se$) and are characterized by a trend towards clustering (based on the current nuclear resonance magnetic measurements) and by a quasi random Zn↔Be substitution, respectively. Corresponding references for $Zn_{0.48}Be_{0.52}Se$ are given in the course of the discussion.

**High-pressure Raman measurements.** High-pressure unpolarized near-forward and backward Raman spectra are taken on $Zn_{0.83}Cd_{0.17}Se$ by inserting, together with ruby chips used for pressure calibration (via the fluorescence linear scale[38]), a ~35 $\mu m$-thick single crystal with parallel (110)-oriented faces obtained by cleavage inside a stainless-steel gasket preindented to 60 $\mu m$ and drilled by spark-erosion to ~250 $\mu m$, placed between the large diamonds (with a 400 $\mu m$ culet) of a membrane Chervin type diamond anvil cell[39]. Methanol/ethanol/distilled-water (16:3:1), that remains hydrostatic up to ~10.5 GPa[40], i.e., slightly beneath the zincblende→ rock-salt pressure transition of the studied crystal (~12 GPa), is used as the pressure transmitting medium. Similar high-pressure Raman measurements performed with $Zn_{0.48}Be_{0.52}Se$ are detailed as Supplementary Information (Sec. II).

The phonon-polaritons are detected by adopting the (nearly) perfect forward scattering geometry in which the incident laser beam (with wavevector $\vec{k}_i$) enters the rear of the crystal at nearly normal incidence and the scattered light (the wavevector is $\vec{k}_s$) is detected in front along the same direction. However, the zero value of the scattering angle $\theta=(\vec{k}_i,\vec{k}_s)$ inside the crystal cannot be achieved experimentally. The limiting factor is the numerical aperture of the lens used to collect the scattered light. Outside the crystal, the detected light fits into a pencil-like cone with half top angle $\alpha$ smaller than 4°. When brought back to a unidirectional beam, this corresponds to an average deviation by $\bar{\alpha}$~0.4° from the normal to the crystal face (by averaging over $\sin\alpha \times d\alpha$). The angle for the corresponding scattered beam inside the crystal is scaled down to ~2.7 by the refractive index of the crystal, as measured by spectroscopic ellipsometry for the used green and blue laser lines at ambient pressure. With this, the minimal achievable $\bar{\theta}$ value for the (average) scattering angle inside the crystal falls down to $\bar{\theta}_{min}$~0.15°.

The dispersion and Raman intensities of the PP modes, including their TO and LO asymptotes, are obtained by using the same generic formula of the multi-mode Raman cross-section as in Ref. 18. Detail concerning the pressure dependence of various input parameters coming into this formula is given as Supplementary Information (Sec. I – $Zn_{1-x}Cd_xSe$ and Sec. II – $Zn_{1-x}Be_xSe$). The relevant $\bar{\theta}$ value *per* PP Raman spectrum is estimated theoretically, *i.e.*, via a fine tuning until the experimental "$\omega$ vs. $q$" Raman scan line derived from the wavevector conservation law that governs the Raman scattering process (*i.e.*, $\vec{k}_i - \vec{k}_s = \vec{q}$, once expressed in its $\bar{\theta}$-dependence using the relevant values of the refractive index for the incident and scattered lights) intercepts the "$\omega$ vs. $q$" PP dispersion right at the experimentally observed PP Raman frequencies. The as-obtained $\bar{\theta}$ values at intermediate (~5 GPa) and maximum (~9 GPa) pressures (see Fig. 2) fall close to $\bar{\theta}_{min}$, indicating that a nearly perfect forward scattering geometry is achieved experimentally.

A crucial ingredient in the calculation of the high pressure PP Raman cross-section is the dispersion of the refractive index of the crystal around the used laser lines at a given pressure. An experimental insight is a difficult task. A rough estimate, sufficient for our use, is obtained by translating as a whole the dispersion of the $Zn_{0.83}Cd_{0.17}Se$ refractive index measured by ellipsometry at ambient pressure using a large crystal piece. The translation is guided by the pressure-induced step increase in the optical band gap of pure ZnSe, constituting a natural reference given the moderate Cd content. In doing so we rely on a theoretical prediction[41] and proceed as earlier done with $ZnSe_{1-x}S_x$[21].

The sample geometry is crucial. If the scattering setup does not strictly conform to normal incidence/detection onto/from (110)-oriented crystal faces – as if, *e.g.*, the laser beam impinges on a non-oriented edge of the tiny piece of single crystal inserted in the diamond anvil cell – the high-pressure Raman signal may be spoiled by the 2×TA continuum, and may look extremely confusing. A decisive proof that the spurious 2×TA continuum has been "killed", meaning that all features in Fig. 2 can be safely discussed in terms of nominal one-phonon Raman-active modes, is the PP detection. Such modes can be readily identified experimentally based on their extreme sensitivity to change in the scattering angle and/or the laser line (Supplementary Section I, Fig. S6).

**High-pressure *ab initio* Raman calculations.** *Ab initio* calculation of the high-pressure TO (purely-mechanical) Raman spectrum of $Zn_{0.5}Cd_{0.5}Se$ is done by implementing the formula given in Ref. 42 within the *Ab initio* Modeling PROgram (AIMPRO) code[43,44] operated in the density functional theory along the local density approximation for the exchange-correlation potential, using a 216-atom zincblende-type supercell optimized to a random Cd↔Zn substitution ($\kappa$~0, 50 at.% Cd) by simulated annealing. The retained criterion for randomness is that the distribution of Se-centered tetrahedrons with Cd/Zn atoms at the vertices forming the zincblende crystal (five in total) matches the corresponding $\kappa$-dependent Binomial Bernouilli distribution at the considered composition (extensive detail is given in Ref. 18). The Raman calculations are done after full relaxation of the $Zn_{0.5}Cd_{0.5}Se$ supercell (lattice constant and atom positions) using the basis functions and pseudopotentials detailed, together with accuracy issues, in Ref. 18. In the high pressure study the third-order Birch-Murnaghan[45] equation of state was used to determine the supercell volume. To test this approach, 10 GPa applied to a reference pure-ZnSe supercell produces an increase in the TO Raman frequency of 44.4 cm$^{-1}$ in reasonable agreement with the experimental value[46] of ~50.0 cm$^{-1}$.

# References

1. Adachi, S. *Properties of Semiconductor Alloys: Group-IV, III-V and II-VI Semiconductors*. Chap. 6, 201 (John Wiley & Sons, Inc., Chichester, 2009).
2. Stauffer, D. & Aharony, A. *Introduction to Percolation Theory*. Chaps. 2 & 3 (Taylor & Francis, 1994).
3. Sahimi, M. *Applications of Percolation Theory* (Taylor & Francis, 1994).
4. D'Souza, R. M. & Nagler, J., Anomalous critical and supercritical phenomena in explosive percolation. *Nature Physics* **11**, 531–538 (2015).
5. Mascarenhas, A. *Spontaneous Ordering in Semiconductor Alloys*. (Kluwer Academic Press, Plenum Publishers, New York, 2002).
6. Loa, I., Bos, J.-W. G., Downie, R.A. & Syassen, K. Atomic ordering in cubic bismuth telluride alloy phases at high pressure. *Phys. Rev. B* **93**, 224109-1–224109-8 (2016).
7. Breidi, A., Postnikov, A. V. & Hajj Hassan, F. Cinnabar and SC16 high pressure phases of ZnSe: an *ab initio* study. *Phys. Rev. B* **81**, 205213-1–205213-9 (2010).
8. Chang, I. F. & Mitra, S. S., Long wavelength optical phonons in mixed crystals. *Advances in Physics* **20**, 359–404 (1971).
9. Pagès, O., Souhabi, J. Postnikov, A. V. & Chafi, A. Percolation versus cluster model for multimode vibration spectra of mixed crystals : GaAsP as a case study. *Phys. Rev. B* **80**, 035204-1–035204-12 (2009).
10. Born, M. & Huang, K. *Dynamical Theory of Crystal Lattices*. Chap. 2, p. 100. (Clarendon Press, 1954).
11. Bellaiche, L., Wei, S.-H. & Zunger, A., Localization and percolation in semiconductor alloys : GaAsN vs GaAsP. *Phys. Rev. B* **54**, 17568–17576 (1996).
12. Weber, W. New bond-charge model for the lattice dynamics of diamond-type semiconductors. *Phys. Rev. Lett*. **33**, 371–374 (1974).
13. Rustagi, K. C. & Weber, W. Adiabatic bond charge model for the phonons in $A^3B^5$ semiconductors. *Solid State Commun*. **18**, 673–675 (1976).
14. Dicko, H. *et al*. Defect-induced ultimately fast volume phonon-polaritons in the wurtzite $Zn_{0.74}Mg_{0.26}Se$ mixed crystal. *Sci. Rep*. **9**, 7817-1–7817-8 (2019), and refs. therein.
15. Wronkowska, A. A., Wronkowski, A., Firszt, F. & Łęgowski, S. Investigation of II-VI alloy lattice dynamics by IR spectroscopic ellipsometry. *Cryst. Res. Technol*. **41**, 580–587 (2006).
16. Eckner, S. *et al*. C. S. Bond-strength inversion in (In,Ga)As semiconductor alloys. *Phys. Rev. B* **97**, 195202-1–195202-6 (2018).
17. Eckner, S. *et al*. Bond-stretching force constants and vibrational frequencies in ternary zinc-blende alloys: A systematic comparison of (In,Ga)P, (In,Ga)As and Zn(Se,Te). *Europhys. Lett*. **126**, 36002-1–36002-7 (2019).
18. Shoker, M. B. *et al*. Multi-phonon (percolation) behavior and local clustering of $Cd_xZn_{1-x}Se$-cubic mixed crystals (x≤0.3): A Raman-*ab initio* study. *J. Appl. Phys.* **126**, 105707-1–105707-16 (2019), and refs. therein.
19. Torres, V. J. B., Hajj Hussein, R., Pagès, O. & Rayson, M. J. Clustering/anticlustering effects on the GeSi Raman spectra at moderate (Ge,Si) contents : Percolation scheme vs. *ab initio* calculations. *J. Appl. Phys.* **121**, 085704-1–085704-12 (2017).
20. Pradhan, G. K. *et al*. Pressure-induced phonon freezing in the $Zn_{1-x}Be_xSe$ alloy: A study via the percolation model. *Phys. Rev. B* **81**, 115207-1–115207-6 (2010).
21. Hajj Hussein, R. *et al*. Pressure-induced phonon freezing in the ZnSeS II-VI mixed crystal: phonon-polaritons and *ab initio* calculations. *J. Phys.: Condens. Matter* **28**, 205401-1–205401-13 (2016).
22. Mujica, A., Rubio, A., Muñoz, A. & Needs, R. High-pressure phases of group-IV, III-V and II-VI compounds. *Rev. Mod. Phys.* **75**, 863–912 (2003).
23. Christensen, N. E., Satpathy, S. & Pawlowska, Z. Bonding and iconicity in semiconductors. *Phys. Rev. B* **36**, 1032–1050 (1987).

24. Rodriguez, S. R.-K. Classical and quantum distinctions between weak and strong coupling. *Eur. J. Phys.* **37**, 025802-1– 025802-15 (2016).
25. Dolfo, G. & Vigué, J. Damping of coupled harmonic oscillators. *Eur. J. Phys.* **39**, 025005-1–025005-18 (2018).
26. Yin, X. & Zhang, X. Unidirectional light propagation at exceptional points. *Nature Mater*. **12**, 175–177 (2013).
27. Davis, B. L. & Hussein, M. I. Nanophononic metamaterial : thermal conductivity reduction by local resonance. *Phys. Rev. Lett.* **112**, 055505-1–055505-5 (2014).
28. Xiong, S. *et al*. Blocking phonon transport by structural resonances in alloy-based nanophononic metamaterials leads to ultralow thermal conductivity. *Phys. Rev. Lett.* **117**, 025503-1–025503-1 (2016).
29. Zamir, D., Beshah, K., Becla, P., Wolff, P. A. & Griffin, R. G. Nuclear magnetic resonance studies of II-VI semiconductor alloys. *J. Vac. Sci. Technol. A* **6**, 2612–2613 (1988).
30. Yu, P. Y. & Cardona, M. *Fundamentals of Semiconductors – Physics and Materials Properties*. Chap. 7, 381 (4$^{th}$ edition, Springer, 2010).
31. Henry, C. H. & Hopfield, J. J. Raman scattering by polaritons. *Phys. Rev. Lett*. **15**, 964 –966 (1965).
32. Claus, R., Merten & L., Brandmüller J. *Light Scattering by Phonon-Polaritons*. Chap. 3, 49 (Springer-Verlag, 1975).
33. Weinstein, B. A. Phonon dispersion of zinc chalcogenides under extreme pressure and the metallic transformation. *Solid State Commun*. **24**, 595–598 (1977).
34. Camacho, J. *et al*. Pressure dependence of optical phonons in ZnCdSe alloys. *Phys. Stat. Sol. B* **235**, 432–436 (2003).
35. Fermi, E., Pasta, J. & Ulam, S. Studies of non linear problems. Document Los Alamos 1940 (May 1955).
36. Uchiyama, H. *et al*. Phonon lifetime observation in epitaxial ScN film with inelastic X-ray scattering. *Phys. Rev. Lett.* **120**, 235901-1– 235901-7 (2018).
37. Firszt, F. *et al*. Growth and optical characterization of $Cd_{1-x}Be_xSe$ and $Cd_{1-x}Mg_xSe$ crystals. *Cryst. Res. Technol*. **40**, 386–394 (2005).
38. Chervin, J. C., Canny, B. & Mancinelli, M. Ruby-spheres as pressure gauge for optically transparent high pressure cells. *High. Press. Res*. **21**, 305–314 (2001).
39. Chervin, J. C., Canny, B., Besson, J. M. & Pruzan, P. A diamond anvil cell for IR microspectroscopy. *Rev. Sci. Instrum*. **66**, 2595–22598 (1995).
40. Klotz, S., Chervin, J. C., Munsch, P. & Le Marchand, G. Hydrostatic limits of 11 pressure transmitting media. *J. Phys. D Appl. Phys.* **42**, 075413-1–075413-7 (2009).
41. Khenata, R. *et al*. Elastic, electronic and optical properties of ZnS, ZnSe and ZnTe under pressure. *Comput. Mater. Sci*. **38**, 29–38 (2006).
42. De Gironcoli, S. Phonons in Si-Ge systems : An *ab initio* interatomic-force-constant approach. *Phys. Rev. B* **46**, 2412–2419 (1992).
43. Briddon, P. R. & Jones, R. LDA calculations using a basis of Gaussian orbitals. *Phys. Stat. Sol. B* **217**, 131–171 (2000).
44. Rayson, M. J. & Briddon, P. R. Rapid iterative method for electronic-structure eigenproblems using localized basis functions. *Comput. Phys. Commun*. **178**, 128–134 (2008).
45. Birch, F. Finite elastic strain of cubic crystals. *Phys. Rev.* **71**, 809–824 (1947).
46. Greene, R. G., Luo, H. & Ruoff, A. L. High pressure X-ray and Raman study of ZnSe. *J. Phys. Chem. Solids* **56**, 521–524 (1995).
47. Pagès, O. *et al*. Raman study of the random ZnTe-BeTe mixed crystal : Percolation model plus multi-mode decomposition. *J. Appl. Phys.* **99**, 063507-1–063507-8 (2006).
48. Pagès, O. *et al*. Unification of the phonon mode behavior in semiconductor alloys : theory and *ab initio* calculations. *Phys. Rev. B* **77**, 125208-1–125208-9 (2008).
49. Kozyrev, S. P. Features of the percolation scheme of transformation of the vibrational spectrum with varying alloy composition for Cd(TeSe) and (CdZn)Te alloys with soft bonds. *Semiconductors* **49**, 885–891 (2015).

## Figure captions

**Figure 1 | Pressure dependence of the percolation-type (purely-mechanical) TO Raman doublets of various ZnSe-based mixed crystals. a** Bond length and volume dependence of bond ionicity for the constituting species of the studied mixed crystals (indicated *via* dashed lines) – taken from Ref. 23, helping visualize which is the short bond (X-axis) and how its percolation-type Raman doublet changes (closure *vs.* opening) with pressure (Y-axis, see text). **b** Theoretical composition dependence of the Raman intensities of the $Zn_{1-x}Be_xSe$, $ZnSe_{1-x}S_x$ and $Zn_{1-x}Cd_xSe$ TO (purely-mechanical) triplets, as apparent at 0 GPa in a backscattering Raman experiment (as sketched out). **c** Pressure dependence of the corresponding Be-Se (this work), Zn-S (Ref. 21) and Zn-Se (this work) Raman doublets (dashed ovals) – reflecting sensitivity of bond vibrations up to first- or second neighbors (as specified) – at selected compositions. In the closure case ($Zn_{1-x}Be_xSe$, $ZnSe_{1-x}S_x$), the freezing of the lower oscillator – due to vibrations of self-connected bonds along the chain (central panel) – at the resonance (Res.) leads to a Raman extinction (collapse) transposing to inertia at 1D (left panel). In the opening case ($Zn_{1-x}Cd_xSe$), the oscillators remain independent at any pressure (right panel). The pressure dependencies of the bond force constants ($k_i$) and vibration dampings ($\gamma_i$) – governing the closure/opening and collapse processes of each system, respectively (see text) – are schematically indicated (using single or double arrows), together with the critical zincblende→rock-salt pressure transition ($P_T$).

**Figure 2 | $Zn_{1-x}Cd_xSe$ high-pressure Raman spectra. a** $Zn_{0.83}Cd_{0.17}Se$ near-forward Raman spectra at selected pressures – the star marks a Fano-type antiresonance. **b** Corresponding Raman scan lines (oblique lines) superimposed onto the relevant phonon-polariton dispersions (curves) including the Raman intensities (thickness of curves). The crossing points are emphasized (circles). **c** *Ab initio* TO (purely-mechanical) Raman spectra of a 216-atom $Zn_{0.5}Cd_{0.5}Se$ disordered cubic-supercell at ambient and high pressures. Paired arrows indicate changes in frequency gaps between the Cd-Se singlet and the Zn-Se doublet (hollow) and within the Zn-Se doublet (filled) with pressure.

**Figure 3 | Partitions of II-VI and III-V semiconductor mixed crystals based on the pressure dependence of the Raman "percolation" doublet. a** II-VI partition. **b** III-V partition. For each system the percolation-type Raman doublet refers to the short species that vibrates at high frequency with a distinction between homo (bottom) and hetero (top) environments, as sketched out. The pressure-induced closure/opening of a Raman doublet is inferred from the volume dependence of the bond ionicities given in Ref. 23. Left bars refer to potentially problematic systems in which the short bond involves the large/heavy substituent. Alternative bars equipped with relevant references identify systems in which the percolation doublet has been evidenced (right bars) and further studied under pressure (upward or downward bars).

## Acknowledgements

We acknowledge assistance from the PSICHE beamline staff of synchrotron SOLEIL for the high-pressure X-ray diffraction measurements, from the Nuclear magnetic resonance platform of Institut Jean Barriol (CNRS FR3843, Université de Lorraine – http://www.ijb.univ-lorraine.fr) and from A. EnNaciri and L. Broch from the Ellipsometry core facility of LCP-A2MC (Université de Lorraine – http://lcp-a2mc.univ-lorraine.fr). We thank José Lopez for constructing a large-size (~1m$^3$, 10×10×10 substituting sites) supercell replica used as visual background prior to MATLAB programming of site percolation on the zincblende and wurtzite lattices. We also thank Jean-Louis Bretonnet for useful discussions around the Fermi-Pasta-Ulam experiment, Andrei V. Postnikov for a critical reading of the manuscript, and Pascal Franchetti for technical assistance in the Raman measurements. This work is in part supported by the IFCPAR Project n°. 3204-1 led/co-led by Sudip K. Deb/O.P, and is in part developed within the scope of the project i3N, UIDB/50025/2020 & UIDP/50025/2020, financed by national funds through the FCT/MEC.

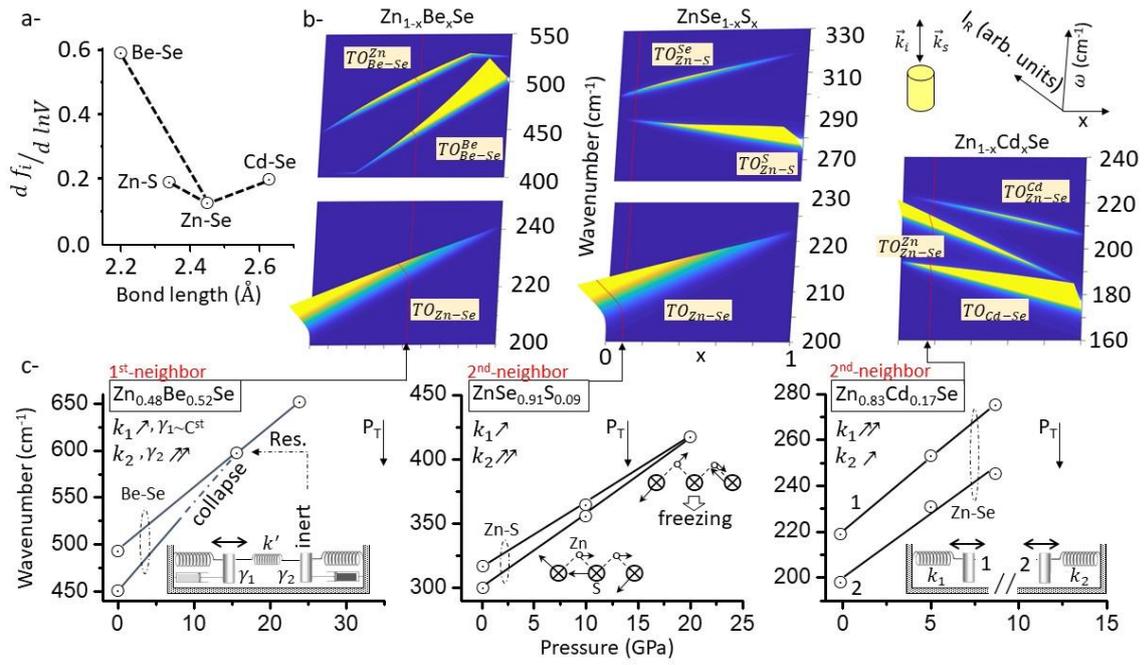

**Figure 1**

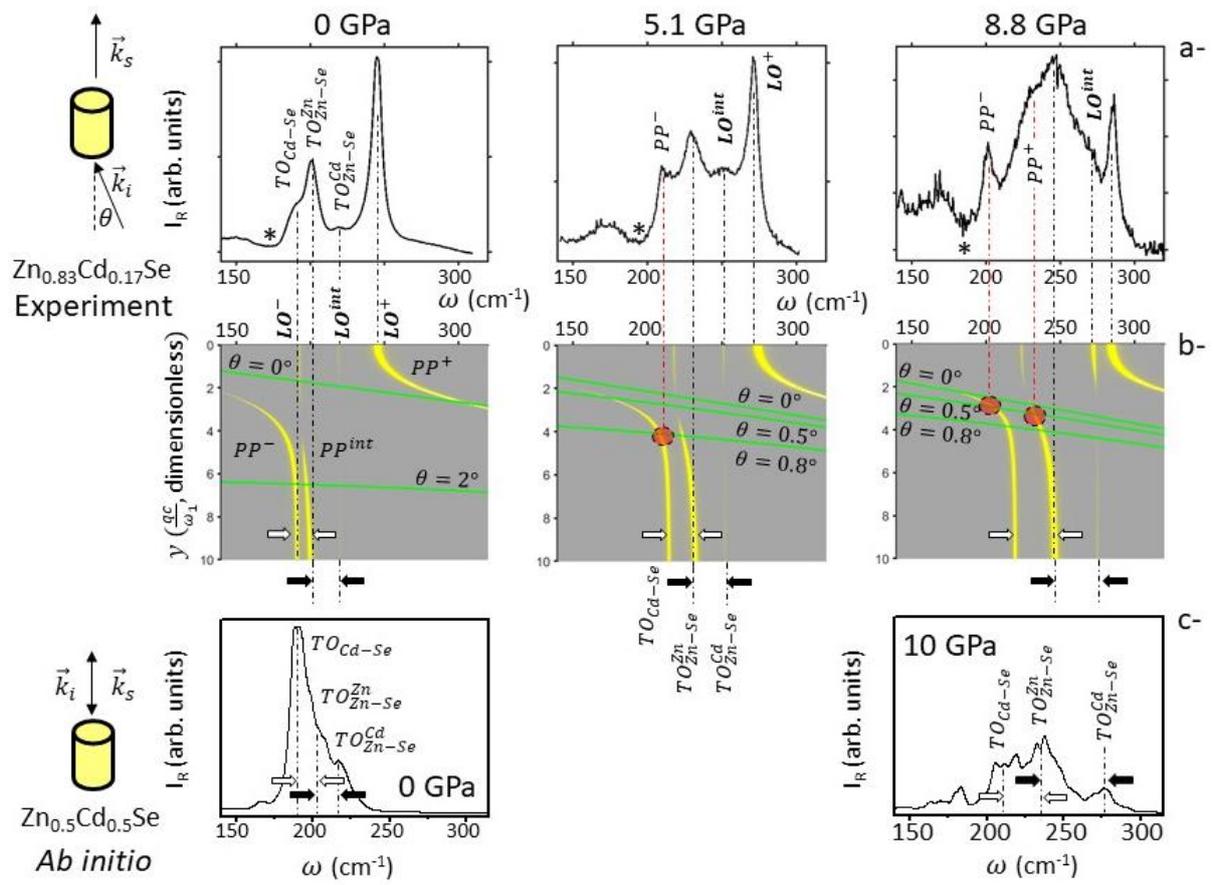

**Figure 2**

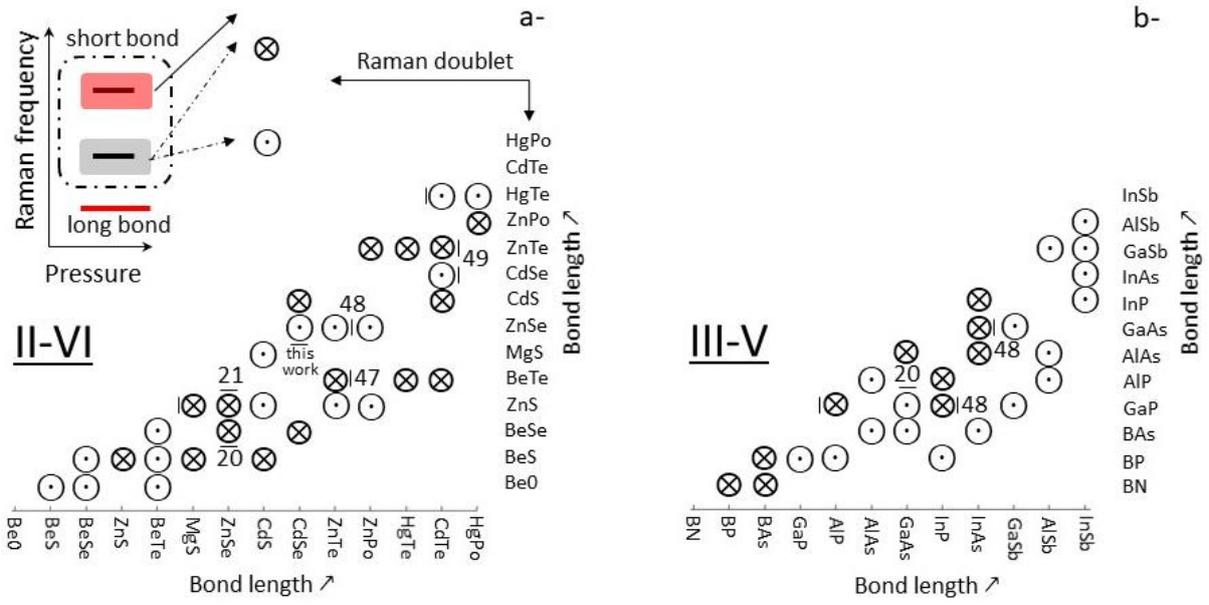

**Figure 3**

# Supplementary Information

**Phonon-based partition of (ZnSe-like) semiconductor mixed crystals on approach to their pressure-induced structural transition**

M. B. Shoker, O. Pagès, V. J. B. Torres, A. Polian, J.-P. Itié, G. K. Pradhan, C. Narayana, M. N. Rao, R. Rao, C. Gardiennet, G. Kervern, K. Strzałkowski and F. Firszt

In this annex we cover side aspects supporting the picture developed in the body of the manuscript. Sec. I is dedicated to $Zn_{1-x}Cd_xSe$. Sec. IA reports on the nature of the Cd↔Zn substitution in the studied $Zn_{0.83}Cd_{0.17}Se$ crystal, as to whether this is ideally random or not. A direct insight is achieved by performing solid-state nuclear magnetic resonance measurements concerned with the [77]Se-invariant. A comparative statistical study of the Cd-Se bond percolation near the nominal value of ~19 at.% Cd considering either a random Cd↔Zn substitution or a trend towards clustering (corresponding to $\kappa$~0.5 on a scale of 0 – random substitution to 1 – phase separation) further helps to gain insight into the $Zn_{0.83}Cd_{0.17}Se$ mesostructure. In Sec. IB a high pressure X-ray diffraction study of $Zn_{1-x}Cd_xSe$ at well-spanned compositions serves as a structural background for the vibrational insight achieved by high pressure Raman scattering in the main text. In particular, the pressure-induced structural transitions to rock-salt fixing the limit for the current Raman study are identified. Besides, a macroscopic insight into the mechanical properties of $Zn_{1-x}Cd_xSe$ is derived from the X-ray diffraction data, complementing that achieved at the mesoscopic scale by Raman scattering. In the latter respect, the representative series of high-pressure near-forward Raman spectra recorded with the (110)-oriented $Zn_{0.83}Cd_{0.17}Se$ single crystal, from which are taken the selected data displayed in Fig. 2, is shown in full in Sec. IC. A corresponding series of conventional backscattering Raman spectra taken with $Zn_{0.83}Cd_{0.17}Se$ crushed into powder is also shown, for reference purpose. On this occasion, detail is given concerning the pressure dependence of various input parameters coming into the generic (TO, PP, LO) multi-mode Raman cross section used for contour modeling of the $Zn_{0.83}Cd_{0.17}Se$ spectra. Sec. II focuses on $Zn_{0.5}Be_{0.5}Se$ as a reference system of the "closure" type. A generalization to the LO symmetry of the coupled model of two TO-like damped harmonic 1D-oscillators recently worked out in Ref. 25 is outlined for this system in Sec. IIA. The pressure-induced closure of the Be-Se Raman doublet of $Zn_{0.5}Be_{0.5}Se$ is then discussed based on this model in Sec. IIB, to clarify all disconcerting attributes of the "closure" process referenced as (i – iv) in the main text.

## I. $Zn_{1-x}Cd_xSe$

### A. Nature of the Cd↔Zn atom substitution in $Zn_{0.83}Cd_{0.17}Se$

In a recent Raman study of $Zn_{1-x}Cd_xSe$[18] we argued that the magnitude of the inverted $LO^{int} - TO_{Zn-Se}^{Cd}$ splitting due to the minor mode situated within the main $TO_{Zn-Se}^{Zn} - LO^+$ band can be used as a convenient marker with two respects. It primarily helps to decide whether the ($TO_{Zn-Se}^{Zn}$, $TO_{Zn-Se}^{Cd}$) doublet reflects a sensitivity of the Zn-Se vibrations at the first- (large splitting) or second-neighbor (small splitting) length scale. Should neither of the two options provide reasonable agreement between experiment and theory, then additional flexibility arises by considering a deviation with respect to the ideal Cd↔Zn substitution towards either clustering (reduction of splitting) or anticlustering (increase of splitting).

After close examination, in the above spirit, of the extended set of $Zn_{1-x}Cd_xSe$ $LO^{int}$ and $TO_{Zn-Se}^{Cd}$ frequencies carefully measured by applying far-infrared reflectivity to large size single crystals with small/moderate Cd contents (0≤x≤0.4) slightly exceeding the pure-zincblende domain (x<0.3)[S1] - an overview is given in Ref. 18 (see Fig. 1 therein) – we concluded to a sensitivity of the Zn-Se vibrations extending up to second-neighbors. At least this suffices to explain the $LO^{int} - TO_{Zn-Se}^{Cd}$ quasi degeneracy apparent in the reported data at small Cd content (x≤0.2). A sensitivity limited to first-neighbors would otherwise generate a finite splitting (as large as ~5 cm$^{-1}$ at x=0.2) exceeding by far the experimental resolution as (~1 cm$^{-1}$) as soon as departing from the Cd-dilute limit. Our Raman spectra lately recorded with similar single crystals at small Cd content (x=0.075, 0.170) consistently replicate the trend.

At moderate Cd-contents (0.2≤x≤0.4) the Vodopyanov's data reveal a finite $LO^{int} - TO_{Zn-Se}^{Cd}$ splitting, as expected in case of a Zn-Se sensitivity at the second-neighbor scale. However, the experimental splitting remains significantly smaller than that predicted in case of an ideally random Cd↔Zn substitution, *i.e.*, by as much as ~50% at x~0.4. Such discrepancy was explained by considering a pronounced trend towards local clustering corresponding to a value of ~0.5 on a scale of 0 (random substitution) to 1 (local phase separation) for the order parameter $\kappa$ introduced in Ref. 18. While such $\kappa$ value derived at moderate Cd content is generally compatible with the observed $LO^{int} - TO_{Zn-Se}^{Cd}$ quasi degeneracy throughout the composition domain, the quasi degeneracy at small Cd content is also there when considering a random substitution ($\kappa$=0) – see above. It follows that the magnitude of the inverted $LO^{int} - TO_{Zn-Se}^{Cd}$ splitting is not really informative regarding the nature of the Cd↔Zn substitution at small Cd content (x<0.2), of present interest At this limit, additional insight is needed to decide.

1. Zn$_{0.83}$Cd$_{0.17}$Se: Nuclear magnetic resonance measurements

A direct insight into the nature of the Cd↔Zn substitution of the studied Zn$_{0.83}$Cd$_{0.17}$Se crystal ground as a fine powder is achieved by $^{77}$Se solid-state nuclear magnetic resonance (NMR) measurements. The $^{77}$Se chemical shift being sensitive to the electronic environment, it changes with the first-neighbor tetrahedral environment of the Se nucleus, out of five possible ones (from now on labeled depending on the number of Cd at the vertices, *i.e.*, from 0 to 4, completed by Zn) in a Zn$_{1-x}$Cd$_x$Se zincblende-type mixed crystal. The relative abundance of the various (0 to 4-Cd) Se-cluster units forming Zn$_{0.83}$Cd$_{0.17}$Se can be derived from the integrals of the related $^{77}$Se signals, presently modeled as Gaussian functions, using the dmfit program[S2].

All NMR spectra are acquired on a Bruker Avance III 600 MHz spectrometer (14T) equipped with a Bruker 1.3 mm triple resonance MAS probe. The Y channel is not used (no insert) and the X channel is tuned to $^{77}$Se frequency (114.42 MHz). The $^1$H channel is not used, as no proton decoupling is necessary. The NMR spectra are processed by using a Gaussian apodization function with a 1000 Hz line broadening. The $^{77}$Se chemical shift is externally calibrated using a ZnSe powder sample (containing Se-clusters of the 0-Cd type only) resonating at -350 ppm (Fig. S1a, bottom spectrum), consistently with existing measurements in the literature[S3]. $^{77}$Se 90° pulse is 2.7 µs long, as determined using a L-selenomethionine powder sample (Acros Organics).

One-dimensional $^{77}$Se spectra taken with the Zn$_{0.83}$Cd$_{0.17}$Se powder exhibit three distinct NMR signals at -330 ppm, -380 ppm and -430 ppm (Fig. S1a, central and top spectra). Considering the $^{77}$Se chemical shifts of -350 ppm and -480 ppm found for ZnSe (Ref. S3) and CdSe (Ref. S4), respectively, the observed signals can be traced back to the Se-clusters of 0-, 1- and 2-Cd types, respectively – in order of increasing magnitude of the $^{77}$Se chemical shift. A similar ordering/deshielding effect has independently been observed with Zn$_{1-x}$Cd$_x$Te[29]. A further common trend between both systems is that the parent signal(s) (presently limited to ZnSe) emerge(s) within (and not outside) the spectral domain covered by the mixed crystal(s) (Zn$_{0.83}$Cd$_{0.17}$Se in this case).

A prerequisite in view to achieve a reliable quantitative insight into the individual Se-clusters fractions *via* the areas of the corresponding NMR signals is the knowledge of the longitudinal relaxation time T$_1$. Standard T$_1$ measurement with a saturation recovery experiment is hampered by the low sensitivity and long $T_1$. A precise $T_1$ measurement could not be achieved as the first non-zero spectrum in the series of saturation-recovery experiments showed at 1000 seconds. Longer recovery time is impractical, as it might have taken a week to get the full recovery. However, this result gives at least an order of magnitude for $T_1$, further confirmed by a series of three direct acquisition experiments with variable tilt-angle (15°, 30°, 60°) and constant recycle delay (300 s) converging towards a T1 estimate of around 10000 s. A direct acquisition experiment using short pulses (15° tilt) is recorded in 90 hours (Fig. S1a – central spectrum). The 300 s recycle delay chosen for this experiment is between 0.1 and 0.15 $T_1$ for each signal. This ensures that the relative intensity of each signal is within 10% of the maximum possible intensity under these conditions.

Despite efforts the signal-to-noise ratio remains rather poor – even though not negligible – in our direct acquisition experiment. This motivated an alternative approach to exploit the long homogeneous transverse relaxation time ($T_2$*) by implementing direct-acquisition Carr-Purcell-Meiboom-Gil (CPMG) experiment[S5]. In this method, several spin echoes are acquired over one scan, folded and stacked together to get an increase in the signal-to-noise ratio. The spin echoes are rotor-synchronized over 128 rotor periods for a half-echo (2.56 ms) and 32 full echoes and a half were acquired for a total acquisition time of 166.4 ms. The recycle delay chosen for this experiment (10 h) was between $3T_1$ and $4T_1$ for each signal. In this way, the relative intensity of each signal is within 10% of the signal intensity after full recovery. The quantitativity of the study is not challenged by any potential difference in $T_2$* between signals as the stacking of each echo exhibits a linear increase in the signal-to-noise ratio. Hence, each echo is of similar intensity with respect to the first one, and $T_2$* for each signal is much longer than the total free inductance decay acquisition time. The 90 h long direct CPMG acquisition experiment gives a much better signal-to-noise ratio than the standard direct acquisition one recorded over the same total experimental time, as apparent in Fig. S1a (compare the top and central spectra).

Hence, our approach in the following analysis is to focus on the more sensitive CPMG spectrum, which is deconvoluted by using the Dmfit program[S2] (red curve superimposed onto the top raw NMR spectrum in Fig. S1a). The NMR-signal integrals derived thereof lead to relative proportions of 38%, 51% and 11% for the three detected Se-clusters of 0-, 1- and 2-Cd types, respectively. A Montecarlo analysis leads to a maximal error of 1% due to the deconvolution process, which is negligible in view of the 10% error on the integrals related to the $T_1$ estimate. The corresponding cluster fractions derived for the same system by using the $\kappa$-dependent Binomial Bernouilli distribution[8] are shown in Fig. S1b, for comparison. The operative $\kappa$-domain for $Zn_{0.83}Cd_{0.17}Se$ is bound to 0 (random substitution) – fortuitously corresponding to full disappearance of the minor 4-Cd type Se-cluster at 17 at.% Cd (fixing the validity limit for the current $\kappa$-dependent approach) – and 1 (full clustering, *i.e.*, local phase separation) on its low and high estimate cases, respectively. In other words any deviation from the ideal Cd↔Zn random substitution ($\kappa$=0) in $Zn_{0.83}Cd_{0.17}Se$ should reflect a trend towards local clustering ($\kappa$>0); anticlustering ($\kappa$<0) is forbidden in principle. However, this is only theoretical, and valid provided the Cd *vs.* Zn arrangement in $Zn_{0.83}Cd_{0.17}Se$ actually fits into the general $\kappa$-dependent substitution process formalized in Ref. 8.

The experimental Se-cluster (of 0-, 1- and 2-type) distribution measured from the CPMG spectrum is somehow disconcerting in that it doesn't match any theoretical $\kappa$-dependent Bernouilli distribution in Fig. S1b, whichever $\kappa$ value is considered – even negative ones (in Fig. S1b, the $\kappa$-domain is artificially expanded to the slightly negative $\kappa$-value corresponding to disappearance of the next minor Se-cluster species, *i.e.*, of 3-Cd type, to get at least a minor insight into the effect of anticlustering on the Se-cluster fractions of $Zn_{0.83}Cd_{0.17}Se$). In fact, the significantly larger representation of the Se-cluster of 1-Cd type with respect to the 0-Cd one points towards anticlustering ($\kappa$<0). However, in this case, the development of the Se-cluster of 2-Cd type should be favored as well. This contradicts the CPMG experimental findings revealing, in fact, a sub-representation of the 2-Cd Se-cluster type with respect to the random case ($\kappa$=0), as expected in case of local clustering ($\kappa$>0). We conclude that the Cd↔Zn substitution process in the current Zn$_{0.83}$Cd$_{0.17}$Se crystal cannot be grasped within the $\kappa$-dependent scheme[8].

On this basis we re-examine hereafter the CPMG-NMR data by taking the Se-cluster distribution in the ideal/random case (refer to $\kappa$=0 in Fig. S1b) as the only reliable reference. While the Se-cluster fraction of the 0-Cd type species nearly matches the nominal ($\kappa$=0) value (within a few percent), the Se-cluster species of 1-Cd and 2-Cd types are substantially over- and sub-represented with respect to the reference values, respectively. This suggests a clustering mechanism specifically concerned with the minor Se-cluster species of the 2-Cd type, in which part of the latter clusters "decompose" into Se-clusters of 3/4-Cd types by driving off Zn atoms which eventually substitute for Cd within part of the remaining Se-clusters of the like 2-Cd type, thereby "recomposed" into Se-clusters of 1-Cd type. Altogether, the net effect is to favor Se-clusters of 1-Cd type to the detriment of Se-clusters of the 2-Cd type – as emphasized by antagonist arrows in Fig. S1b, while leaving unaffected the fraction of Se-clusters of the ultimate 0-Cd type.

The above clustering mechanism selectively impacting the minor (detectable) Se-cluster species suffices to explain all disconcerting aspects of the Se-cluster distribution revealed by the current Zn$_{0.83}$Cd$_{0.17}$Se CPMG NMR spectra. It is further globally consistent with the existing insight into the nature of the Cd↔Zn substitution in Zn$_{1-x}$Cd$_x$Se earlier gained via the Raman spectra treated within the percolation scheme using the $\kappa$-formalism[18].

By focusing on the minor Se-cluster species of 2-Cd type as the key of the clustering mechanism, a tentative (though not strictly relevant – as discussed above) $\kappa$ value of ~0.12 can be inferred for Zn$_{0.83}$Cd$_{0.17}$Se from the CPMG NMR spectrum (Fig. S1b). This is significantly less than the Raman estimate of ~0.5 mostly derived at moderate/large Cd content[18], reflecting a massive clustering. Such discrepancy occurs because the Raman and NMR techniques operate at different length scales, as briefly discussed below.

As the percolation scheme relies on a phenomenological (1D) description of a zincblende-type mixed crystal at the mesoscopic scale in terms of a composite of its two parent-like sub-regions, the Raman-based $\kappa$-insight is likewise relevant at the mesoscopic scale only. In this case $\kappa$ is determined from the intensity ratio between the two like Raman modes due to the same bond species stemming from the two regions of the composite: one experimental/observable data (the above-mentioned Raman intensity ratio) vs. one adjustable parameter ($\kappa$) leads to an exact solution, meaning that the $\kappa$-formalism[8] is naturally applicable to the Raman spectra of a mixed crystal discussed within the percolation scheme. In contrast, by addressing separately the five possible first-neighbor (3D) cluster units (centered on the invariant species, with substituents at the vertices) forming a (real) zincblende-type mixed crystal, the NMR measurements provide an insight into the nature of the atom substitution at the ultimate microscopic scale. In this case it is rather unlikely that a single physical parameter ($\kappa$) suffices to govern the relative fractions of five different cluster units measured by NMR at a given composition, meaning that, in general, the $\kappa$-formalism[8] will fall short of explaining the NMR spectrum of a mixed crystal. In fact, NMR is likely to reveal how intricate a given deviation with respect to the ideal random substitution in a mixed crystal can be at the very local scale, beyond the $\kappa$-formalism.

2. Zn$_{0.83}$Cd$_{0.17}$Se: statistical insight into the Cd-Se bond percolation

The $\kappa$ parameter of local clustering mentioned above was originally introduced in Ref. 8 for a A$_{1-x}$B$_x$C zincblende-type lattice via the probability $P_{XX}$ for the neighboring site besides a site occupied by X standing for A or B to be also occupied by X, leading to $P_{BB} = x + \kappa \cdot (1 - x)$, with a similar $P_{AA}$ definition. $\kappa = 0$ refers to the random A↔B substitution, whereas a trend to clustering is reflected by $\kappa > 0$, limited by $\kappa = 1$ corresponding to full clustering, i.e., the local phase separation. With local clustering the bond percolation threshold is presumably not the same as in the random case. As an ultimate case of clustering, consider that all B substituents dispose onto the (A,B) sublattice so as to form a continuous chain perpendicular to the [111] direction along which the (A,B)-substituent and C-invariant planes alternate. In this case, a very small fraction of B substituents suffices to create the wall-to-wall B-C bond percolation : For a $n^3$-cubic supercell with (111)-oriented basal planes, the exact amount corresponding to B-C percolation from bottom to top scales as $1/n^2$ (=$n/n^3$), tending to zero when $n$ approaches infinity.

Based on the NMR and Raman measurements (Sec. IA), the studied $Zn_{0.83}Cd_{0.17}Se$ crystal exhibits a trend towards clustering. One may well wonder whether the clustering may either hasten or delay the Cd-Se bond percolation with respect to the random case. Such information is required to get a rough idea of the mesostructure of the used $Zn_{0.83}Cd_{0.17}Se$ crystal.

Technically we proceed as follows, using MATLAB as a programming language. In the zincblende $Zn_{1-x}Cd_xSe$ lattice the (111)-oriented dense-packing $Zn_{1-x}Cd_x$-substituent planes alternate with corresponding invariant Se-type planes with similar packing, so that each invariant (Se) is tetrahedrally bonded with four substituents (Cd,Zn), and *vice versa*. Accordingly, to estimate the Cd-Se bond percolation on the $Zn_{1-x}Cd_xSe$ zincblende lattice in its $(x,\kappa)$-dependence we suppress the invariant Se-sublattice and estimate the Cd-site percolation threshold on the remaining face-centered cubic (fcc) $Zn_{1-x}Cd_x$-lattice. In fact, extensive Monte Carlo simulations[S6] done on a virtually infinite (containing $2048^3$ sites) fcc $A_{1-x}B_x$ lattice provide a refined estimate of the site percolation threshold across the high-density (111)-packing planes in case of a random A↔B substitution, *i.e.*, $x_p \sim 0.19$.

Our ambition is not to provide an exact estimate of the bond percolation threshold for a set $\kappa$ value, but rather to compare the bond percolation thresholds at different $\kappa$ values. Accordingly, finite-size fcc $Zn_{1-x}Cd_x$-supercells containing a moderate ($10^3$) number of sites, that save computer time, are sufficient for our use. The (111)-fcc stacking is of the … − 1 − 2 − 3 − 1 − 2 − 3 − … type (Fig. S2), so that each site of plane number N hangs up on top of the middle of a trio of sites from the underlying N-1 plane, forming altogether a tetrahedral arrangement. The complete zincblende-type $Zn_{1-x}Cd_xSe$ lattice is restored by (virtually) inserting one Se atom in the middle of each (Cd,Zn)-tetrahedron. In practice we build up a series of $10^3$-sites fcc $Zn_{1-x}Cd_x$-supercells with a (111)-oriented basal plane, and examine the wall-to-wall site percolation along the [111] direction from bottom to top of the supercell.

A reference curve obtained by plotting the probability $P_{(x,\kappa \sim 0)}$ for site percolation on the random fcc $Zn_{1-x}Cd_x$-lattice for $x$ spanning the entire composition domain with a constant step increase $\Delta x$=0.01, is shown in Fig. S2 (hollow symbols). Each point results from a statistical average over one hundred supercells. The supercells are generated so as to roughly match the set $\kappa$=0 value at each composition. This is achieved through enumeration of the five possible Se-centered tetrahedral units (with Zn and/or Cd at the vertices) and subsequent comparison with the predicted fractions using the binomial Bernoulli distribution at $\kappa=0$[8]. In case of significant disagreement (the accuracy is given below), a basic simulated annealing procedure is applied, consisting of successive Zn↔Cd exchanges generated at random throughout the supercell (in practice one Cd atom taken at random is replaced by a Zn atom, and, at the same time, one Zn atom, also taken at random, is replaced by a Cd atom, thus preserving the same stoichiometry), until the fractions of clusters reasonably match (within ±5%) the nominal Bernouilli's estimates at the set $\kappa$=0 value. In the Se-cluster-enumeration process special attention is awarded to the minor cluster species at the studied composition – whose fraction is dramatically $\kappa$–dependent, but all cluster fractions are eventually checked to be consistent with the Bernouilli's prediction.

The above substitution procedure is adapted to design a reference series of supercells representing the partially clustered ($\kappa$>0) $Zn_{0.83}Cd_{0.17}Se$ mixed crystal. The targeted clustering rate is $\kappa \sim 0.5$, corresponding to the upper estimate for this crystal given the Raman and RMN insights (Sec. I1) – the desired $\kappa$ value further appears to be the highest achievable $\kappa$ value for the used supercell size. In the first place, the attention is focused on the minor Se-centered cluster (with four Cd atoms at the vertices, at this composition) constituting a sensitive marker of clustering. In fact, at x=0.17 by increasing $\kappa$ from 0 to 0.5 the nominal Bernoulli's fractions[8] of Se-clusters dramatically change from (~47.5 %, ~39.0 %, ~12.0 %, ~1.7 %, ~0.3 %) to (~64.7 %, ~21.5 %, ~3.0 %, ~3.0 %, ~8.0 %), with percentages ranked in order of increasing Cd atoms at the vertices of the Se-clusters. Starting from a primary supercell generated at random, hence generally corresponding to a quasi negligible fraction of the minor Se-cluster species, the $\kappa$ value is artificially increased by leaving the few existing minor clusters unchanged and realizing a random Zn↔Cd exchange (as described above) specifically concerned with the alternative Se-cluster species. Such exchange is retained provided it generates a net increase of the $\kappa$ value. The procedure is repeated until the targeted $\kappa$ value is eventually achieved. The as-obtained "final" supercell (at $\kappa\sim 0.5$) is eventually retained only if the fractions of the four remaining Se-cluster species are further consistent (within $\pm 5\%$) with the Bernouilli's estimates at the set $\kappa\sim 0.5$ value (as specified above).

In principle, one would expect $P_{(x=0.19,\kappa\sim 0)}=1$. However, this is not so in practice due to the finite size of the used supercells (see Ref. 2, p. 17). In fact, $P_{(x=0.19,\kappa\sim 0)}$ hardly reaches $\sim 0.21$ meaning that out of the one-hundred series of selected $Zn_{0.81}Cd_{0.19}$-fcc ($Zn_{0.81}Cd_{0.19}Se$-zincblende) supercells *per* test, only $\sim 21\%$ exhibit the Cd-site (Cd-Se bond) percolation. The latter $P_{(x=0.19,\kappa\sim 0)}$ value is not interesting *per se*. It is only useful for reference purpose, notably in view of comparison with the $P_{(x=0.17,\kappa\sim 0.5)}$ value (Fig. S2). Remarkably, the two probabilities exactly match within statistical uncertainty, meaning that the Cd-Se bonds exhibit the same ability to percolate throughout the $Zn_{0.83}Cd_{0.17}Se$ -clustered and reference $Zn_{0.81}Cd_{0.19}Se$-random supercells. As the Cd-Se bond percolation is certain for an infinite ($x=0.19,\kappa\sim 0.5$) supercell, we infer that the same applies to the virtual ($x=0.17,\kappa\sim 0.5$) crystal.

We have checked that the same $\kappa$-trend persists with the wurtzite structure (full symbols in Fig. S2) that exhibits nearly the same bond percolation threshold as the zincblende structure[S6], even though the (111) planes now alternate in a … – 1 – 2 – 1 – 2 – … sequence. The $\kappa$-trend in question is thus not structure-dependent for a given (A,B) vs. C tetrahedral environment.

B.  $Zn_{1-x}Cd_xSe$: a high-pressure X-ray diffraction study

High-pressure X-ray diffraction measurements are performed on $Zn_{1-x}Cd_xSe$ with various compositions, crystallizing in the zincblende ($x<0.3$), mixed zincblende/wurtzite ($0.3\leq x\leq 0.7$) and wurtzite ($x>0.7$) structures[18] at ambient pressure at the PSICHE beamline of the SOLEIL synchrotron using the 0.3738 Å radiation, with a double aim. The first one is to determine the critical pressure corresponding to the zincblende/wurtzite ↔ rock-salt structural transition as a function of x, and the second one is to derive the x-dependence of the bulk modulus at ambient pressure for $Zn_{1-x}Cd_xSe$ taken dominantly in its zincblende structure – abbreviated $B_0(x)$ hereafter.

At each composition, a piece of $Zn_{1-x}Cd_xSe$ monocrystal is ground into a fine powder and inserted inside a 200 μm thick stainless-steel gasket preindented to 35 $\mu m$ and drilled by spark-erosion to ~150 $\mu m$ placed into the same Chervin type diamond anvil cell[39] (DAC) as that used for the high-pressure Raman measurements, with 300 μm in diameter diamond culet. Neon was preferred to methanol-ethanol-water as a pressure transmitting medium[40] because the high-pressure X-ray diffraction measurements were pushed up to pressures exceeding by far the critical pressure corresponding to the hydrostatic limit of the latter medium, *i.e.*, ~10 GPa. The measurements were performed both in the upstroke and downstroke regimes. The pressure was measured via the reference X-ray diffraction lines originating both from Au markers[S7] added besides the samples inside the cavity of the DAC or/and from the neon transmitting medium itself[S8] (depending on the pressure domain). An external setup was used to monitor the pressure inside the DAC from the computer room, *i.e.*, without entering the experimental hutch. This helped to achieve a maximum accuracy in the pressure estimate ($\pm 0.3$ GPa), while saving considerable recording time. The 0.3738 Å synchrotron radiation was focused onto a 40 $\mu m$ diameter spot at the sample position, and the diffraction pattern was recorded using a plane detector disposed at about 30 cm of the sample (measured with $LaB_6$ diffraction). The resulting 2D image plate data were then turned into intensity versus $2\theta$ plots using the software FIT2D[S9]. The peaks fitting and unit cell fitting was carried out using the software DIOPTAS[S10].

A representative series of $Zn_{0.83}Cd_{0.17}Se$ diffractograms is displayed in Fig. S3 where the diffraction lines are indexed by using the Miller indices. In the upstroke regime the first occurrence of the rock-salt phase is detected at $P_{min.}=11.3$ GPa, *i.e.*, slightly below the corresponding critical pressure for pure ZnSe (~13 GPa), as expected (see main text). The full disappearance of the native zincblende phase occurs at $P_{max.}=13.6$ GPa, corresponding to a zincblende–rock-salt coexistence domain as large as ~2.3 GPa at ambient temperature. In the downstroke regime the crystal transits to the zincblende structure, starting at ~9.5 GPa via the transient cinnabar, a common feature at small-moderate Cd content – including pure ZnSe[7,S11].

The pressure dependence of the unique (a: cubic symmetry) or double (a, c: hexagonal symmetry) lattice constants (specified in brackets) for all studied Zn$_{1-x}$Cd$_x$Se mixed crystals on the way forth to the cubic/rock-salt phase, and, from there, on the way back to the native cubic/zincblende or hexagonal/wurtzite phases at ambient pressure, via in certain case the hexagonal/cinnabar phase, is recapitulated in Fig. S4. This offers a direct insight into the existence domains of each phase in both the upstroke and downstroke regimes, and helps to determine the composition dependence of the critical pressure transition to rock-salt, denoted $P_T$, reported in Fig. S5 and in Fig. 1 as the average between the pressures corresponding to the onset of the rock-salt phase and the subsequent disappearance of the zincblende/wurtzite phase of all studied systems in the upstroke regime. An error bar is accordingly assigned to each $P_T$ value. The basic trend is that $P_T$ decreases when the Cd content increases, as expected – referring to (v) / main text.

The data displayed in Fig. S4 are further exploited to derive the $B_0(x)$ dependence for the studied mixed crystals throughout the zincblende domain, including Zn$_{0.63}$Cd$_{0.37}$Se – on account that this mixed crystal is dominantly of the zincblende type (only traces of the wurtzite structure are visible in its X-ray diffractogram taken at a near-ambient pressure – see Fig. 1 of Ref. 18) – besides Zn$_{0.925}$Cd$_{0.075}$Se and Zn$_{0.83}$Cd$_{0.17}$Se that both exhibit a pure-zincblende structure. $B_0(x)$ was inferred by fitting the variation of the unit-cell volume ($a^3$) to a Murnaghan equation of state maintaining for the pressure derivative of the bulk modulus $B'=4$[S12]. The corresponding $B_0(x=0)$ value of the pure ZnSe compound with zincblende structure[S13] is added for reference purpose.

The symbols in Fig. S5 refer to the $B_0(x)$ values resulting from averaging over various estimates obtained at all studied pressures at a given composition. The upper and lower limits of the error bars assigned to the average $B_0(x)$ values refer to the maximum and minimum estimates throughout all studied pressures at the considered composition, respectively. If we omit the error bars and focus on the average $B_0$ values, it seems that $B_0$ goes through a maximum around the Cd-Se bond percolation threshold, in echo to the predicted singularity in the $L(x)$ dependence by *ab initio* calculations[11]. Such anomaly cannot be accounted for by the smooth VCA-like Vegard's law predicted based on *ab initio* calculations done with periodically-repeated ordered (8-atom) Zn$_{1-x}$Cd$_x$Se supercells (x=0.25, 0.50, 0.75 together with the end compounds)[S14]. Additional measurements are needed to decide whether the currently observed experimental singularity in $B_0(x)$ is intrinsic to the statistics behind the substitutional disorder in Zn$_{1-x}$Cd$_x$Se, *i.e.*, the result of the Cd-Se bond percolation throughout the ZnSe-like host matrix, or merely fortuitous, *i.e.*, reflecting a weak experimental insight that may not deserve attention given the too small number of studied compositions so far.

At this stage, we mention that reaching a maximum in a mechanical property of Zn$_{1-x}$Cd$_x$Se by crossing the Cd-Se bond percolation threshold is not so intuitive. Indeed the Cd-Se bond is more ionic than the Zn-Se one (by ~12%, see Ref. 23) and thus less mechanically resistant to the inherent stresses in shear and compression/tension resulting from the contrast in bond length/stiffness of the two species. This gets reflected both at the macroscopic scale through a smaller value of the bulk modulus for CdSe than for ZnSe (by ~19% if we refer to the comparative *ab initio* insight of both cited systems taken in the zincblende structure given in Ref. S14), and also, at the microscopic scale, based on our estimate of the effective bond-stretching force constant (by ~7%) as the reduced mass of the bond multiplied by the square TO Raman frequency (using the values given in Ref. 18). On the above basis, one would rather expect a softening of the Zn$_{1-x}$Cd$_x$Se lattice on crossing the Cd-Se bond percolation threshold, and not a hardening.

C. Zn$_{0.83}$Cd$_{0.17}$Se: a high-pressure Raman scattering study

By inserting a tiny piece of a zincblende-type single crystal with parallel (110)-oriented crystal faces (obtained by cleavage) inside a diamond anvil cell and combining the backward and near-forward scattering geometries, one is in a position to address the pressure dependence of the long-wavelength (q~0) optic modes in their full variety, as detailed below.

The classical backscattering (reflection-like) setup, corresponding to the maximum value of the wavevector transferred to the crystal, of the order of ~1% of the Brillouin zone size, probes the transverse optic modes in their asymptotic purely-mechanical regime (abbreviated TO) away from the center Γ (q=0) of the Brillouin zone. By adopting the alternative near-forward scattering (transmission-like) geometry, the wavevector transferred to the crystal is reduced by about two orders of magnitude and reaches minimum, offering an access to the same transverse optic modes but now taken in their asymptotic phonon-polariton (PP) regime close to Γ. Obviously, the near-forward scattering geometry can be implemented only if the used crystal is transparent to the incoming laser beam. In practice, for PP detection the scattering angle $\theta$ between the wavevectors of the incident laser beam and of the detected scattered light inside the crystal must not exceed a few degrees. Otherwise, one falls short of penetrating the actual phonon-polariton regime and remains stuck inside the asymptotic backscattering-like regime of the purely-mechanical TO modes[10,31,32].

Though the LO modes are theoretically forbidden at (nearly) normal incidence/detection onto the (110) crystal faces of a zincblende crystal[30], they show up clearly in both the backward and forward scattering geometries, due to multi-reflection of the laser beam between parallel crystal faces[31]. The multi-reflection is likewise responsible for the co-emergence of the TO and PP features in a high-pressure Raman spectrum. Basically, on its way forth to the top crystal face (detector side) the laser beam generates the PP (forward-like) Raman signal, which superimposes onto the TO (backscattering-like) Raman signal produced by the laser beam on its way back to the rear crystal face after reflection off the top surface.

Summarizing, provided the scattering angle reaches minimum with a relevant laser line, the TO, LO and PP modes may well come together in a high-pressure Raman spectrum, offering an overview of all Γ-like optic modes in a single shot. A representative series of such "complete" near-forward Raman spectra taken at various pressures with a (110)-oriented $Zn_{0.83}Cd_{0.17}Se$ single crystal by using a green laser excitation (514.5 nm) – from which are selected those shown in Fig. 2 – is displayed in Fig. S6a. A corresponding series of PP-deprived high-pressure Raman spectra taken in the backscattering geometry by using another green laser line (532.0 nm) and with the same crystal now ground as a powder is shown in Fig. S6b, for reference purpose. Note that in the latter powder-based experiment both the TO and LO modes are allowed owing to the lack of crystal orientation.

1. Contour modeling of the high-pressure backward/near-forward $Zn_{0.83}Cd_{0.17}Se$ Raman spectra

Contour modeling of the three-mode {Cd-Se, (Zn-Se)$^{Zn}$, (Zn-Se)$^{Cd}$} $Zn_{0.83}Cd_{0.17}Se$ high-pressure Raman lineshapes in the main text, covering the purely-mechanical (TO) ones and their polar variants in the transverse (PP) and longitudinal (LO) symmetries, was achieved by using our generic expression of the multi-mode Raman cross section given in Ref. 18 and established in Ref. S15. Only, the resonance term, i.e., $Im\{-[\varepsilon_r(\omega,x) - q^2c^2/\omega^2]^{-1}\}$ in which $\omega = \omega_i - \omega_s$ is the transferred frequency in a Raman experiment (defined in the far-infrared/phonon spectral range), differs in each case, depending on the magnitude of the transferred wavevector $\vec{q}$ in a Raman experiment, with $q \to \infty$ for the TO modes, $q$ finite for the PP modes and $q = 0$ for the LO ones (see detail for a pure compound, e.g., in Ref. 14). In principle the pressure dependence of the resonance term can occur on the one hand, through $\varepsilon_r(\omega,x)$ that captures the whole phonon behavior of $Zn_{1-x}Cd_xSe$ and through $q$ that refers to the used scattering geometry. For a given external scattering geometry the $q$ value is further dependent on the dispersion of the refractive index $n(\omega)$ around the used laser excitation. In fact the pressure dependence of $n(\omega)$ matters only for the PP modes characterized by finite $q$ values, whereas the pressure dependence of $\varepsilon_r(\omega,x)$ is crucial for all modes.

Additional input parameters involved in the pre-factor of the resonance term of the Raman cross section are the Faust-Henry coefficients $C_{F-H}$ of the Cd-Se and Zn-Se bonds, that represent the relative Raman efficiencies of the non-polar TO modes to the polar PP and LO ones. Such coefficients scale as the fractions of corresponding oscillators[S16] in the considered three-mode {Cd-Se, (Zn-Se)$^{Zn}$, (Zn-Se)$^{Cd}$} system, i.e., as $\{x, (1-x)^3 + 2x(1-x)^2, x^2(1-x)\}$, correspondingly, reflecting a sensitivity of the Zn-Se vibration to its local 1D-environment at the second-neighbor scale[18]. The parent $C_{F-H}$ values at ambient pressure are given in Ref. 18. As the $C_{F-H}$ coefficients do not count for the resonance term itself, and thus play a minor role in our calculations, we consider that they are not pressure dependent in the following, in a crude approximation.

1-a. Pressure dependence of $\varepsilon_r(\omega, x)$ in the resonance term of the Raman cross section

The main ingredient in the resonance term of the Raman cross section is the relative dielectric function $\varepsilon_r(\omega, x)$ of $Zn_{1-x}Cd_xSe$. As the latter system appears to be of the "opening" type under pressure (see main text), meaning that no pressure-induced crossing occurs between any of its three TO oscillators, these can be considered as independent (uncoupled) in a first approximation. In this case, $\varepsilon_r(\omega, x)$ takes the classical form

$$\varepsilon_r(\omega, x) = \varepsilon_\infty(x) + \sum_{p=1}^{3} x_p \cdot S_{0,p} \cdot \frac{\omega_{T,p}^2}{\omega_{T,p}^2(x) - \omega^2 - j\gamma_p \omega} \tag{1}$$

which includes an electronic background $\varepsilon_\infty(x)$ that scales linearly with $x$ between the parent $\varepsilon_{\infty,p}$ values – representing the asymptotic limit of $\varepsilon_r(\omega, x)$ at frequencies well-beyond the phonon resonances – together with three Lorentzian functions standing for the various TO oscillators. In Eq. (1), $x_p$, $\gamma_p$, $\omega_{T,p}(x)$ and $S_{0,p}$ are the fraction of oscillator $p$ in $Zn_{1-x}Cd_xSe$ (given above), the phonon damping (coming via a friction force in the force assessment per bond) – corresponding in practice to the full width at half maximum of the $p$-like TO Raman peak, the frequency of the observed TO mode due to oscillator $p$ in the Raman spectra of the mixed crystal, and the oscillator strength awarded to the related pure compound, namely CdSe ($p$=1) or ZnSe ($p$=2,3), respectively. In fact, $S_{0,p}$ is expressed as $\varepsilon_{\infty,p} \cdot \Omega_p^2 / \omega_{T,p}^2$ where $\Omega_p^2 = \omega_{L,p}^2 - \omega_{T,p}^2$ refers to the TO-LO splitting of the $p$-type compound[8]. In all reported simulations $\gamma_p$ is taken minimal (1 cm$^{-1}$), for a clear resolution of neighboring features.

Regarding the pressure dependence of $\varepsilon_\infty(x)$, we assume that the slight bowing ($b$=-1.55) evidenced at ambient pressure using *ab initio* calculations persists at any pressure[S17]. Only, our reference $\varepsilon_\infty$ value for ZnSe is rescaled to that measured at ambient pressure using spectroscopic ellipsometry[S18]. The latter value does not exhibit any significant pressure dependence throughout the studied pressure domain (0 – 10 GPa) in the *ab initio* calculations reported in Ref. S19 (no dependence) and in Ref. 41 (the variation is less than 2%). We assume the same for CdSe taken in the zincblende structure – for which the corresponding data are lacking in the literature.

The pressure dependence of the phonon oscillator strengths $S_{0,p}$ awarded to the zincblende ZnSe compound is well documented. The TO and LO Raman frequencies of pure ZnSe were studied in detail by between ambient pressure and 10 GPa, and found to adopt the generic form $\omega(P) = \omega_0 + \alpha P - \beta P^2$, with $(\alpha, \beta)$=(5.50, 0.0497) and (4.79, 0.137), respectively[46]. The resulting $\varepsilon_\infty$ (see above) and $\omega(P)$ trends for ZnSe generate a quasi linear collapse of the ZnSe oscillator strength versus pressure, with a final oscillator strength at 10 GPa scaled down the reference value at ambient pressure by ~36%.

Equivalent $\omega(P)$ data are not available for CdSe because this compound does not crystallize in bulk in the zincblende structure at ambient pressure. Only thin films can be grown, which are not so convenient for optical measurements under pressure. Now, a linear pressure dependence of the LO Raman frequency was measured up to 5 GPa using CdSe clusters (35-55 Å in diameter) with zincblende structure[S20] testified by X-ray diffraction. At ambient pressure the X-ray diffractograms of such clusters indicate a lattice constant (6.05±0.6 Å) matching the bulk value (6.052 Å)[S21]. Therefore, along with the cited authors, we consider that the linear dependence is presumably valid for the bulk zincblende CdSe crystal. By assuming further that the linearity is preserved up to 10 GPa, the predicted upward shift of the LO frequency between ambient pressure and 10 GPa reaches 43 cm$^{-1}$. A similar variation for the TO frequency is still lacking in the literature. However, *ab initio* calculations reveal that the TO-LO splitting of CdSe near $\Gamma$ is virtually identical in the zincblende and wurtzite structures[S22]. Only, in the wurtzite case the TO-LO splitting is somewhat blurred by the increased diversity of optical phonon branches. This is due to the lowering of the crystal symmetry when shifting from the zincblende (cubic – isotropic) structure to the wurtzite (hexagonal – uniaxial) one, which duplicates the TO and LO modes into their so-called $A_1$ and $E_1$ variants – corresponding to ion vibrations along and perpendicular to the singular crystal axis, respectively. Despite the blurring, one may well think of resorting to the wurtzite structure for a crude insight into the pressure dependence of the TO-LO splitting. However, such approach is not applicable since the wurtzite CdSe crystal is not documented with this respect in the literature. At this point we can hardly proceed, except by analogy with another crystal. CdS is a natural candidate owing to its proximity to CdSe, regarding not only the lattice dynamical properties (the phonon dispersions of CdSe and CdS resemble very much, in both the zincblende and wurtzite structures) but also the elastic and electronic ones[S23-S25]. Moreover CdSe and CdS take the same remarkable path via the transient zincblende structure (cubic, coordination number 4) to transite from wurtzite (hexagonal, coordination number 4) to rock-salt (cubic, coordination number 6) under pressure[S26]. Further, the pressure domain over which the latter two-step structural transition develops is roughly the same for both systems[S27] (1.5–3 GPa). To recollect with the raised issue, recent calculations performed with the wurtzite-type CdS crystal using a shell model-based interatomic potential indicate a quasi linear reduction of the TO-LO splitting by increasing pressure up to 5 GPa[S28], *i.e.*, by ~37.5%. We assume that the TO-LO bands of the wurtzite- and zincblende-type CdSe crystals shrink at the same rate versus pressure up to 10 GPa, in a crude approximation. In this case the TO-LO splitting of the zincblende-type CdSe crystal, estimated to ~20 cm$^{-1}$ at ambient pressure[S22,S23], falls to ~8 cm$^{-1}$ at 10 GPa. With this, the remaining amount of oscillator strength for the zincblende-type CdSe crystal at 10 GPa hardly represents ~30% of the reference value at ambient pressure. We further assume a linear dependence on pressure.

1-b. Pressure dependence of $n(\omega)$ in the resonance term of the Raman cross section

The dispersion $n(\omega)$ of the refractive index of Zn$_{0.83}$Cd$_{0.17}$Se measured at ambient pressure throughout the visible spectral range by applying spectroscopic ellipsometry to a large single crystal taken from the bulk ingot is shown in Fig. S7 (symbols; a truncated version of the current data set – sufficient for our use – has earlier been reported in Ref. 18). A maximum is observed close to the optical band gap, estimated at $E_g$~2.488 eV at ambient pressure from a direct (model-free) numerical inversion of the measured raw ellipsometry angles. The full data set is adjusted via a polynomial fit for future analytical use (plain line).

The reported $n(\omega)$ dispersion is a major ingredient into the multi-PP Raman cross section of Zn$_{0.83}$Cd$_{0.17}$Se, coming via the expression of the dimensionless parameter $y = qc/\omega_1$ that conveniently substitutes for the magnitude $q$ of the wavevector transferred to the crystal in a Raman experiment. In this expression $\omega_1$ arbitrarily refers to the TO Raman frequency of pure ZnSe (~207 cm$^{-1}$)[18] and $c$ is the speed of light in vacuum. The conservation of impulsion that governs the Raman scattering, *i.e.*, $\vec{q} = \vec{k}_i - \vec{k}_s$, where $\vec{k}_i$ and $\vec{k}_s$ refer to the wavevectors of the incident and scattered lights forming the scattering angle $\theta$ inside the crystal, leads to $q = \left(k_i^2 - k_s^2 - 2k_i k_s cos\theta\right)^{1/2}$ with $k_{i,s} = c^{-1}n(\omega_{i,s})\omega_{i,s}$ and $\omega_{i,s}$ referring to the frequencies of the incident (laser) and scattered lights – with obvious subscripts. The resulting $q(\omega_i - \omega_s) = q(\omega)$ dispersion achieved experimentally with the used laser line ($\omega_i$) for a certain scattering angle $\theta$ provides the so-called ($\omega_i, \theta$)-Raman scan line (used in the main text). By injecting the experimental $q(\omega)$ dispersion inside the resonance term of the Raman cross section, one obtains the Zn$_{1-x}$Cd$_x$Se multi-PP Raman cross section in its ($x, y \equiv q, \omega$) dependence at ambient pressure.

For a given visible laser excitation $\omega_i$ the minimal $q$ value is achieved in the perfect forward scattering geometry ($\theta = 0°$), corresponding to $q_{min} = |n(\omega_i) \cdot \omega_i - n(\omega_s) \cdot \omega_s|$. As the $n(\omega)$ dispersion of Zn$_{0.83}$Cd$_{0.17}$Se is positive ($n$ increases with $\omega$) throughout the visible (where operates the Raman scattering), $|n(\omega_i) - n(\omega_s)|$ works along $|\omega_i - \omega_s|$ in the above expression, so that $q_{min} = 0$ can never be achieved experimentally. As the $dn(\omega)/d\omega$ derivative increases with $\omega$ below the optical band gap of Zn$_{1-x}$Cd$_x$Se, a smaller $q_{min}$ value is achieved by adopting the Stokes scattering geometry ($\omega_i > \omega_s$, our approach) than the anti-Stokes one ($\omega_i < \omega_s$). Indeed, in this case the $|n(\omega_i) - n(\omega_s)|$ difference is minimal for a given $|\omega_i - \omega_s|$ frequency gap (whichever laser excitation – in reference to $\omega_i$, is used), with concomitant impact on the $q_{min}$ value, being also minimal.

We have discussed elsewhere in extensive detail[18] that out of our available near-infrared (785.0 nm), red (632.8 nm), green (514.5 nm) and blue (488.0 nm) laser excitations, optimal conditions to probe the sensitive bottleneck region of the PP dispersion of $Zn_{0.0925}Cd_{0.075}Se$ ($E_g$~2.615 eV) are achieved by using the green excitation at nearly normal incidence/detection onto/from (110)-oriented crystal faces. This remains basically valid for the current $Zn_{0.83}Cd_{0.17}Se$ system, with a close composition.

In Fig. 2 the relevant $\theta$ value behind the experimentally detected PP Raman modes at a given pressure is estimated theoretically, *i.e.*, by adjusting the Raman scan line (via $\theta$) until it intercepts the PP dispersion of the crystal right at the observed PP frequencies. Care must be taken that both the PP dispersion and the Raman scan line are pressure dependent, *i.e.*, via both the TO frequencies and the refractive index of the crystal. In our approach a rough $\theta$ estimate, sufficient for our use, is obtained by neglecting any distortion of the light path due to the diamonds framing the studied $Zn_{0.83}Cd_{0.17}Se$ crystal in the DAC. Only the refractive index of $Zn_{0.83}Cd_{0.17}Se$ is taken into account. In fact, the difference in refractive index is small between diamond and $Zn_{0.83}Cd_{0.17}Se$ (less than 8% at 700 nm at ambient pressure), and the change in refractive index versus pressure for diamond is negligible compared to that of $Zn_{0.83}Cd_{0.17}Se$ (by a factor of ~15)[S29], meaning that most of the pressure dependence of the refractive index of the studied diamond/ $Zn_{0.83}Cd_{0.17}Se$ /diamond system in the DAC is due to $Zn_{0.83}Cd_{0.17}Se$. The approximation is further justified in that the provided $\theta$ values in this work are indicative only; these are not discussed as significant physical parameters *per se*.

An experimental measurement of the pressure dependence of the $n(\omega)$ dispersion of $Zn_{0.83}Cd_{0.17}Se$ is a difficult task. Indeed, effective spectroscopic ellipsometry measurements usually require a large sample area, typically of the order of several $mm^2$, and therefore cannot be made in a DAC. The difficulty can be circumvented by resorting to theory. *Ab initio* calculations done on pure ZnSe[39] reveal that its optical band gap widens under pressure, dragging with it the $n(\omega)$ dispersion in an overall translation towards low wavelength (high frequency/energy) at a rate of ~5 nm/GPa. We assume the same for the currently studied $Zn_{0.83}Cd_{0.17}Se$ crystal with a large ZnSe content, with the optical band gap of this mixed crystal at ambient pressure taken as a set point. The $n(\omega)$ dispersions of $Zn_{0.83}Cd_{0.17}Se$ at 5 GPa (dotted line) and 9 GPa (dashed line) resulting from such overall translations are displayed in Fig. S7 besides the experimentally measured $n(\omega)$ curve at ambient pressure (symbols), used as the reference/starting curve.

2. $Zn_{0.83}Cd_{0.17}Se$: experimental results and discussion

By applying pressure, the PP dispersion is right-shifted in Fig. 2 (*i.e.*, horizontally) due to a high-frequency shift of the asymptotic TO (away from Γ) and LO (close to Γ) modes resulting from a strengthening of the related effective bond force constants. In contrast the Raman scan line is upward-shifted towards Γ (*i.e.*, vertically) due to the reduction of the $n(\omega)$ dispersion around the used laser excitation (Fig. S7). The Raman scan lines obtained in the perfect forward scattering geometry ($\theta = 0°$) depending on pressure are shown in Fig. 2, for reference purpose. For any detected PP signal the experimentally achieved $\theta$ values lie in the range 0.5 – 0.8°. This falls close to the minimal achievable value $\bar{\theta}_{min}$~0.15° taking into account an experimental limitation that the scattered light is not strictly detected perpendicularly to the sample surface but fits into a pencil cone due to the finite numerical aperture of the used microscope objective for the detection (see Methods).

At ambient pressure (0 GPa) no PP is detected. Most probably, the reason is that as soon as a PP mode emerges, it interferes destructively via a Fano-type coupling with the spurious two-phonon continua of transverse acoustic modes – abbreviated 2TA – that shows up nearby. In fact, the native TOs behind the searched PPs are already corrupted by such Fano interference, testified by their far-from-perfect TO Raman selection rules (see Fig. 7b of Ref. 18). The TO-distortion via the Fano-coupling involving the 2TA continuum is a general feature of ZnSe-based systems. In the case of $Zn_{0.83}Cd_{0.17}Se$ the coupling is merely seen through a slight subsidence of the baseline on the low-frequency tail of the TO mode, as observed in the pure TO symmetry at ambient pressure (marked by a star in Fig. 2). In some cases the subsidence can develop into a pronounced antiresonance separating the TO mode from the incriminated 2×TA band, then showing up strongly in the Raman spectra, as independently observed with $Zn_{1-x}Be_xSe$[20] – an example is given below, and with $Zn_{1-x}Cd_xSe$ as well[S30].

Under pressure the zone-edge TA modes soften – the trend is aggravated for the 2TA band – in contrast with the zone-center optic (TO, LO) modes and the related (PP) features that harden[33], as already mentioned. This leads to Fano-decoupling, offering a chance for PP detection (Fig. S6a). In fact, various PPs are detected at intermediate (~5 GPa) and high pressure (~9 GPa), if none at ambient pressure (~0 GPa).

At intermediate pressure (~5 GPa), the PP dispersion is probed on approach to the sensitive bottleneck region, corresponding to $\theta$~0.7°. The native $TO_{Cd-Se}$ and $TO_{Zn-Se}^{Zn}$ modes behind $PP^-$ and $PP^{int}$ remain close so that the $PP^{int} \rightarrow PP^-$ transfer of oscillator strength mediated by the macroscopic PP-like transverse electric field $\vec{E}_T(q)$ is massive. This results in the clear emergence of $PP^-$, showing up strong and sharp, at the cost of $PP^{int}$, absent.

At maximum pressure (~9 GPa), the Raman scan line now corresponding to $\theta$~0.5° probes the PP dispersion closer to the bottleneck where $PP^-$, still distinct and sharp, starts to vanish away from its native $TO_{Cd-Se}$ mode (off-shifted by ~20 cm$^{-1}$). Remarkably $PP^{int}$ now shows up as a weak but still pronounced shoulder significantly beneath its native $TO_{Zn-Se}^{Zn}$ mode (by ~15 cm$^{-1}$). This can be explained only if the native $TO_{Cd-Se}$ mode behind $PP^-$ breaks away from $TO_{Zn-Se}^{Zn}$ under pressure. In this case the $\vec{E}_T(q)$-coupling between $PP^-$ and $PP^{int}$ is partially relaxed, with a consequence that $PP^{int}$ retains sufficient oscillator strength to emerge as a distinct Raman feature.

We have checked that, the detected $PP^-$ mode at ~5 GPa vanishes to full disappearance as soon as the scattering angle $\theta$ increases, as achieved by departing the laser beam from normal incidence at the rear crystal face (Fig. S6a). Also, at ~9 GPa, both $PP^-$ and $PP^{int}$ slightly "retreat" towards their native TOs by changing the laser line from green (514.5 nm) to blue (488.0 nm), while keeping the same external incidence of the laser beam at the rear of the crystal (Fig. S6a). Altogether, such trends ascertain the PP nature of the discussed features. In fact, owing to their finite $q$ value (see above), the PP modes are strongly laser- and $\theta$-sensitive, regarding both their Raman frequency and their Raman intensity (in contrast the TO and LO modes consist of "robust" features that emerge with similar characteristics whichever Raman scattering setup is implemented, at least out of resonance conditions).

Last, we address the LO modes with special attention to the minor intermediate $LO^{int}$ one (Fig. 2). Our ambition is to infer the pressure-dependence of the native percolation-type ($TO_{Zn-Se}^{Zn}$, $TO_{Zn-Se}^{Cd}$) doublet behind the LOs from the experimentally observed strengthening/upward-shift-towards-$LO^+$ of $LO^{int}$ under pressure (Fig. 2). It is a matter to decide whether the spacing between the two ZnSe-like TOs reduces (scenario1, closure case), is preserved as such (scenario 2, invariant case), or enlarges (scenario 3, opening case) when the pressure increases. A comparative study of $LO^{int}$ is carried out between ambient pressure and 9 GPa using the LO-variant of the Raman cross section given in Ref. 18. The input parameters are the $TO_{Cd-Se}$ and $TO_{Zn-Se}^{Zn}$ frequencies taken from Fig. 2, and also the $TO_{Zn-Se}^{Cd}$ one with some flexibility depending on the used scenario (1: spacing reduced by 10 cm$^{-1}$, 2: spacing maintained at 20 cm$^{-1}$, spacing enlarged by 10 cm$^{-1}$, with a proper rescaling of the ZnSe-like available oscillator between ambient pressure and 9 GPa as specified in Sec. ID1. As apparent in Fig. S8, the combined strengthening/upward-shift of $LO^{int}$ are simultaneously reproduced only under scenario 3, while neither of the trends comes out under scenarios 1 or 2. We conclude to the widening of the ($TO_{Zn-Se}^{Zn}$, $TO_{Zn-Se}^{Cd}$) percolation doublet under pressure.

Last, we discuss briefly the PP-deprived high-pressure backscattering Raman spectra taken with powders (Fig. S6b), for the sake of completeness. A careful examination focusing on selected pressures corresponding to a reasonable resolution of the minor $TO_{Cd-Se}$ feature supports via a direct insight into the underlying (broad, poorly-defined) TO modes behind the PPs the conclusion drawn from observation of the latter (sharp, well-resolved) modes (Figs. 2 and S6a) that $TO_{Cd-Se}$ breaks away from $TO_{Zn-Se}^{Zn}$ at increasing pressure (Δ enlarges, as emphasized by open-arrows). As for the upper $TO_{Zn-Se}^{Cd}$ mode, a direct insight remains forbidden due to a screening by the (now allowed) $LO^{int}$ signal, as by near-forward Raman scattering. Now, $LO^{int}$ offers a convenient substitute for $TO_{Zn-Se}^{Cd}$ due to their quasi degeneracy, as already mentioned[18]. As independently observed by near-forward Raman scattering (Fig. S6a) $LO^{int}$ gets closer and closer to the dominant $LO^+$ feature at increasing pressure ($\delta$ enlarges, as emphasized by plain-arrows – the trend is visible at least up to 8.1 GPa) and by doing so $LO^{int}$ reinforces, as emphasized by the large upward arrow at 8.1 GPa. The LO nature of the pointed feature is verified as it collapses together with the upper $LO^+$ mode at maximum pressure, as emphasized by large downward arrows. In brief, the (TO,LO,PP) near-forward and (TO,LO) backward Raman spectra are consistent on the main trends for what regards the pressure-dependence of the underlying compact ($TO_{Cd-Se}$, $TO_{Zn-Se}^{Zn}$, $TO_{Zn-Se}^{Cd}$) triplet.

## II.    $Zn_{1-x}Be_xSe$

A reference insight into the pressure-induced closing of the percolation-type Raman doublet in a mixed crystal can be achieved by focusing on the Be-Se doublet of $Zn_{0.5}Be_{0.5}Se$, as a case study. Generally, $Zn_{1-x}Be_xSe$ exhibits an unusually large contrast in its bond physical properties (length, reduced mass, stiffness), so that the percolation doublet due to its short/light/stiff Be-Se bond is, at the same time[20], well resolved – with a splitting as large as ~45 cm$^{-1}$ – and well-separated from the Raman signal due to the long/heavy/soft Zn-Se bond – vibrating at a lower frequency by as much as ~200 cm$^{-1}$. Moreover, at intermediate composition (~50 at. %Be) the fractions of the two Be-Se oscillators are identical in the crystal – given by $x^2$ and $x \cdot (1-x)$ in order of ascending frequency (reflecting a sensitivity of the Be-Se vibrations up to first-neighbors in $Zn_{1-x}Be_xSe$), with concomitant impact on the TO Raman intensities[S31]. This is ideal to achieve a reliable insight into the pressure dependence of both individual submodes forming the Be-Se Raman doublet. Additional interest for a focus at ~50 at. %Be arises from the theoretical point of view, as detailed in the course of the presentation of the model at a later stage (see Sec. IIA).

Fig. S9a displays a representative series of high-pressure Raman spectra taken in the upstroke regime with a $Zn_{0.48}Be_{0.52}Se$ powder from ambient pressure up to ~25 GPa, whichs remains below the critical pressure corresponding to the zincblende→ rock-salt structural transition, *i.e.*, $P_T$~35 GPa[S32]. A zoom into the BeSe-like TO and LO Raman frequencies in their pressure dependence is provided in Fig. S9b (squares), for clarity. The assignment of various TO and LO modes in the Zn-Se and Be-Se spectral range is the same as in Ref. 20. Additional high-pressure Raman data currently taken both in the upstroke (hollow symbols) and downstroke (filled symbols) regimes with the same sample now prepared as a single crystal (circles) are added, for the sake of completeness. Both series of high-pressure Raman measurements were performed by using methanol/ethanol/water (16:3:1) as a pressure transmitting medium. This remains hydrostatic up to ~10.5 GPa (Methods) and can be treated as quasihydrostatic until ~30 GPa.

The detailed study of the Zn-Se Raman signal initiated in Ref. 20, not central in the current study, was pursued in Ref. S33. In fact, part of the $Zn_{0.48}Be_{0.52}Se$ Raman spectra shown in Fig. S9a were already published at this occasion (up to 7 GPa). The interest at the time was to clarify the rather confusing multi-mode Raman pattern in the Zn-Se spectral range, using pressure as a convenient tool, notably in search for possible ZnSe-like PP modes. In the current study the focus has shifted to the Be-Se Raman doublet of $Zn_{0.48}Be_{0.52}Se$, with the aim to understand its pressure dependence, in the same spirit as in our original contribution on $Zn_{0.76}Be_{0.24}Se$[20]. A noticeable improvement with respect to the latter pioneering contribution, that was centered on the TOs and qualitative only, is that the TOs and LOs are now treated on equal footing and on a quantitative basis.

Unambiguous experimental trends emerge from the reported data in Figs. S9a and S9b, listed as follows. Recollecting with the used notation in the main text, these enunciate as (i) the gradual convergence of the lower mode onto the upper one under pressure, (ii) accompanied by its progressive collapse until extinction at the crossing/resonance ($\omega_c$), (iii) that occurs around the critical pressure $P_c$~15 GPa, (iv) corresponding to an apparent freezing of the lower oscillator independently testified by *ab initio* calculations done with $Zn_{1-x}Be_xSe$[20] and $ZnSe_{1-x}S_x$[21] (using relevant impurity motifs in each case). Apparently, (iv) only the upper mode remains visible above $P_c$; the lower one has disappeared. However, it is not clear whether both modes survive as degenerated features above $P_c$ or whether only one has survived and the other has been 'killed' by pressure. Preliminary results obtained by examining the pressure dependence of the PP modes of ZnSe$_{1-x}$S$_x$ tend to support the latter option[21]. Indeed, fair contour modeling of the PP Raman signal behind the Zn-S percolation-doublet depending on pressure could be achieved only by considering a progressive loss of the oscillator strength awarded to the lower mode until full extinction at the resonance. However, the PPs are not so reliable markers with respect to the addressed issue in that both the PP frequencies and intensities are so sensitive to the scattering angle and the used laser line, in contrast with the TOs and the LOs.

More generally, besides the persisting question of the "survival/death" of the lower mode *post $P_c$*, neither the origin of its progressive collapse *ante $P_c$*, nor the mechanism behind its crossing with the upper mode at $P_c$, could be explained so far. The latter pending issues are addressed hereafter by focusing on the TO and LO modes constituting more robust markers than the PPs, using Zn$_{0.5}$Be$_{0.5}$Se as a model system.

A. The Be-Se doublet of Zn$_{\sim 0.5}$Be$_{\sim 0.5}$Se viewed as a damped system of coupled harmonic 1D-oscillators

Generally, the above experimental/*ab initio* trends (i–iv) may look confusing at first. The progressive collapse of the lower TO mode when it gets closer to the upper mode under pressure reveals that the two (purely-mechanical) TOs are mechanically coupled. In this case one would expect a strong repulsion of both oscillators at the resonance corresponding to perfect tuning of the bare/uncoupled oscillators. In fact, the two modes do cross at $P_c$, suggesting that the two TOs do not "see" each other and hence are uncoupled. The picture which emerges is that the presumed mechanical coupling between the two TOs is totally screened by an overdamping of the lower mode at the resonance (see main text). In this case the TOs are virtually decoupled by overdamping at the resonance, so that an actual crossing can occur. Altogether, this guides towards a coupled model of two damped 1D-harmonic (spring, mass) oscillators, as recently proposed[25].

The cited authors developed an exhaustive analytical study of a viscoelastic system consisting of two damped purely-mechanical 1D-harmonic oscillators (with damping coefficients noted $\gamma_n$; $n$=1,2) corresponding to distinct masses coupled mechanically via a damped spring (with stiffness $k'$ and damping coefficient $\gamma'$) – thereby introducing some anharmonicity – and to the lab. frame as well (with spring stiffness $k_n$, $n$=1 or 2, being not dependent on the mass displacement – in reference to the harmonic character of the bare oscillators).

Such description can be transposed almost literally to the treatment of the purely-mechanical TO modes forming the BeSe-like Raman doublet of Zn$_{0.5}$Be$_{0.5}$Se. In this case, the lab. frame refers to the surrounding Zn-Se bonds that constitute an obstacle to the Be-Se vibrations, simply because they naturally vibrate at a far-off frequency[20]. Only, in this case the two oscillators have identical mass ($\mu$) – corresponding to the reduced mass of a Be-Se bond – because they both refer to the same bond species. Each Be-Se oscillator is attached to the lab frame via an effective spring-like bond force constant ($k$) that varies depending on whether the Be-Se bond vibrates in the hetero (Zn-type: upper oscillator, numbered 1) or homo (Be-type: lower oscillator, numbered 2) environments – using the terminology of the percolation scheme. The vibration frequency of each bare, *i.e.*, uncoupled, TO oscillator taken alone takes the classical form $\omega_{T,n} = \sqrt{\mu^{-1}k_n}$, where $k_n$ assimilates with the effective Be-Se bond stretching force. For purposes of the model, a damping term ($\gamma_n$) is introduced by adding into each scalar (1D) equation of motion per oscillator a friction force (anti-proportional to the velocity $\dot{u}_n$, with proportionality factor $\alpha_n = \mu\gamma_n$; $n$=1,2) besides the spring-like restoring force to the lab frame $-k_n u_n$, where $u_n$ refers to the relative displacement of oscillator $n$ with respect to its position at rest representing in fact the effective Be-Se bond-stretching behind oscillator $n$. In practice, $\gamma_n$ monitors the full width at half maximum of the TO Raman peak related to the $n$-like Be-Se oscillator. In contrast, the damping term ($\gamma'$) added to the coupling-spring ($k'$) between the two (BeSe-like) oscillators has no physical meaning *per se* in that it does not relate directly to any oscillator. It is considered only as a convenient way to introduce the concept of exceptional point, which marks a separation between two regimes depending on whether the coupling (in reference to $k'$) dominates the damping (in reference to $\gamma_1$ and $\gamma_2$), or vice versa (a schematic view of the coupled/damped oscillator system is provided in Fig. 1) – see below.

1. Damping-dependent (TO,LO) coupled mode frequencies

The non-polar TO-like description worked out in Ref. 25 and outlined above can be generalized to their polar (electrical-mechanical) LO counterparts by adding to the force assessment per oscillator a Coulomb force created by the LO-like long range electric field ($E$) due to the ionic character of the chemical bond in a zincblende crystal[S34]. The Coulomb force involves the Born/dynamic charge of the Be-Se bond, noted $Z$, which monitors the magnitude of the TO-LO splitting related to the unique phonon of the BeSe end compound of $Zn_{1-x}Be_xSe$[9]. In fact, the LO-like Coulombian interaction reinforces the purely-mechanical TO-like restoring bond force constant, so that the LO mode vibrates at a higher frequency (noted $\omega_L$ for the pure compound) than the TO one[S34] (noted $\omega_T$). The as-expanded generic set of scalar (1D-like) equations of motion covering both the TO and LO modes of the considered two oscillator BeSe-like system takes the following form,

$\mu \ddot{u}_1 = -k_1 u_1 - k'(u_1 - u_2) - \alpha_1 \dot{u}_1 + ZE$ (2a)
$\mu \ddot{u}_2 = -k_2 u_2 - k'(u_2 - u_1) - \alpha_2 \dot{u}_2 + ZE$. (2b)

where $Z$ is taken independent of composition[S16], in a crude approximation. The justification for using a scalar version of the equations of motion per oscillator when discussing the long wavelength optical modes as detected by Raman scattering is given in the main text. We specify further that at such length scale the like Be-Se bonds of a given species all vibrate in phase throughout the crystal, so that a unique equation *per* oscillator (a Be-Se bond taken in a given environment, *i.e.*, Be- or Zn-like) suffices to capture the whole dynamics of the like oscillators throughout the crystal[8].

An additional equation is needed to govern the transverse (TO) or longitudinal (LO) character of the macroscopic electric field $E$. This is derived from the Maxwell's equations via the relative dielectric function $\varepsilon_r(\omega)$ of the crystal, given by,

$\varepsilon_0 \varepsilon_r(\omega) E = \varepsilon_0 E + \varepsilon_0 \chi_\infty E + NZ(x_1 u_1 + x_2 u_2)$. (3)

In this equation $\varepsilon_0$ is the permittivity of vacuum; $\omega = \omega_i - \omega_s$ is the transferred energy to the crystal in a Raman experiment; $N$ is the number of Be-Se bonds per volume crystal unit, and $x_n$ the fraction of oscillator $n$ (=1,2) in the crystal ($x_1 + x_2 = 1$). The last term of Eq. (3) represents the (ionic, phonon-like) contribution from both oscillators to the polarization of the crystal (considering each Be-Se bond as a permanent dipolar momentum). The intermediate term involves the susceptibility of a Be-Se bond at $\omega \gg \omega_{T,1}, \omega_{T,2}$, concerned with the electronic part of the crystal polarization.

We emphasize that the involved electric field in the actual equations of motion per oscillator is not the local electric field ($E_l$) that actually "feel" the individual Be-Se bonds at the microscopic scale but the macroscopic one ($E$, averaged over several unit cells of the crystal), as governed by Maxwell's equations from which is directly derived Eq. (3). As such, Eqs. (2) result from a local field correction (using the Lorentz's approach), with the consequence that all physical parameters *per* oscillator therein, including the spring stiffness ($k_n$; n=1,2) and the Born effective dynamic charge ($Z$), are only effective ($E$-related) with no straightforward meaning at the microscopic ($E_l$-related) scale[10].

By operating Eqs. (1) and (2) on a uniform ($q \sim 0$, *i.e.*, long-wavelength) plane wave ($u_n, E \sim e^{i\omega t}$), representing a Raman-active optic mode, one arrives at

$+ (K_1 - \omega^2) u_1 - K u_2 = Z\mu^{-1} E$ (4a)
$- K u_1 + (K_2 - \omega^2) u_2 = Z\mu^{-1} E$ (4b)
$\varepsilon_r(\omega) = \varepsilon_\infty + NZ\varepsilon_0^{-1}(x_1 u_1 E^{-1} + x_2 u_2 E^{-1})$ (5)

where $\varepsilon_\infty = 1 + \chi_\infty$ is the relative dielectric constant of the crystal far beyond the phonon resonance, and $K_n$ and $K$ stand for $\omega_{T,n}^2 + \omega'^2 + i\gamma_n \omega$ and $\omega'^2 = k'\mu^{-1}$, respectively. The TO (purely-mechanical, $E = 0$) and LO ($E$ finite) characters are expressed via $\varepsilon_r(\omega) \to \infty$ and $\varepsilon_r(\omega) = 0$, respectively (detail is given, *e.g.*, in Ref. 14, notably in the supplementary section 1.1 therein), eventually leading to the following set of two generic (TO,LO) coupled equations,

$+ (E_1 - \omega^2)u_1 - V_{12}u_2 = 0$ (6a)

$- V_{21}u_1 + (E_2 - \omega^2)u_2 = 0$ (6b)

with $E_n = K_n + x_n\Omega^2$, $V_{nm} = K - x_m\Omega^2$, considering that $\Omega^2$ is zero for a TO mode and takes the finite value $Z^2(V\varepsilon_0\varepsilon_\infty\mu)^{-1} = \omega_L^2 - \omega_T^2$ for a LO mode, where *V* is the volume of the crystal unit cell[S35]. The TO system is always symmetrical, but the LO one is symmetrical only at intermediate composition ($x_1 = x_2 = 0.5$), the reason for the current focus on $Zn_{0.5}Be_{0.5}Se$. At this limit the two (BeSe-like) oscillators coexist with identical fractions in $Zn_{1-x}Be_xSe$[S33], justifying their analytical treatment on equal footing (implicit in the TO-approach of Ref. 25), in both symmetries (TO and LO). In this case, $V_{12} = V_{21} = K - \Omega^2/2$, from now on simplified as $V$. In fact, $V$ represents a competition between, on the one hand, the mechanical (spring-like) coupling mediated by $k'$ (via $K$, related to $\omega'$) between the two (BeSe-like) oscillators, and, on the other hand, the so-called ionic plasmon coupling[S16] behind the effective Born/dynamic charge $Z$ of a (Be-Se) bond (via $\Omega^2$).

The above system has non-trivial solution ($u_1, u_2 \neq 0$) only if its determinant (abbreviated Det) vanishes. In its present form, due to the linear dependence on $\omega$ brought by the friction force appearing in the $K_n$-terms, Det=0 consists of a general quartic equation. A more convenient biquadratic form likely to provide solutions coming in pairs – as ideally expected for a system of two coupled oscillators – was achieved[25] by removing the friction forces and substituting imaginary restoring forces (not in phase with the mass displacements, *i.e.*, with the Be-Se bond stretching in our case) for the real ones in the force assessment per oscillator – referring to Eqs. (2). This lead to introduce the concept of anelastic damping, as formalized by operating a series of transformations on the $K_n$-terms, schematically summarized as follows: $\omega_{T,n}^2 + \omega'^2 + i\gamma_n\omega$ (original expression) $\rightarrow \omega_{T,n}^2 + \omega'^2$ (by omitting the friction force) $\rightarrow \omega_{T,n}^2(1 + i\phi_n) + \omega'^2(1 + i\phi)$ (each genuine spring constant is equipped with an imaginary part) $\rightarrow \omega_{T,n}^2 + i\gamma_n\omega_{T,n}$ (the disruptive linear dependence on $\omega$ is suppressed).

At the end of the above $K_n$-transformations, $\omega_{T,n}^2$ re-appears but with a different meaning as in its original $K_n$-form. The final $\omega_{T,n}^2$ term is dictated not only by the intrinsic spring constant of the considered oscillator (*i.e.*, $k_n$), as in the starting $K_n$-expression, but also by the coupling spring constant between the two oscillators (*i.e.*, $k'$). In our case it is important to preserve the starting $\omega_{T,n}$ terms because these carry the frequency information on the bare TO oscillators *ante* coupling (see above). Therefore, we generally proceed as in Ref. 25 but take care to retain the original meaning for each individual term inside $K_n$. This leads to consider $K_n^* = \omega_{T,n}^2 + \omega'^2(1 + i\phi) + i\gamma_n\omega_{T,n}$ as the final form for $K_n$, with concomitant impact on $E_n$ now expressed as $E_n^* = K_n^* + \Omega^2/2$. $V$ is likewise re-written as $V^* = K(1 + i\phi) - \Omega^2/2$. A star added as a superscript reflects the imaginary character of each term, for a clear distinction from its original (real) form.

Adopting $K_n^*$ comes to consider a constant friction force at any frequency $\omega$, in fact that suffered by the bare-uncoupled (TO) oscillator. Such approximation is valid as long as the coupling spring constant ($k'$) remains small with respect to the intrinsic spring constants of both oscillators ($k_n$), meaning that the frequencies of the coupled TO modes do not substantially differ from those of the bare-uncoupled ones. In fact, this is actually so with $Zn_{0.5}Be_{0.5}Se$, as discussed below. In this case $\gamma_i$ keeps the meaning of a friction-like damping parameter corresponding to a physical observable in the Raman data, namely the full width at half height of a Raman peak, as already mentioned. As for the anelastic damping of the coupling spring constant apparent in the central term of $K_n^*$, its reason for being is that it becomes useful at a later stage in view to introduce the concept of "exceptional point".

Introducing $E_0^* = (K_1^* + K_2^*)/2$ and $\delta^* = (K_1^* + K_2^*)/2$ eventually leads to the compact system

$+[(E_0^* + \delta^*) - \omega^2]u_1 - V^*u_2 = 0$ (7a)

$-V^*u_1 + [(E_0^* - \delta^*) - \omega^2]u_2 = 0$ (7b)

whose biquadratic equation Det=0 has solutions given by,

$\omega_\pm^2 = E_0^* \pm (\delta^* + V^{*2})^{\frac{1}{2}}$, (8a)

expanding into

$$\omega_\pm^{*2} = \tfrac{1}{2}\left\{\omega_{T,1}^2+\omega_{T,2}^2+2\omega'^2+\Omega^2+i(\gamma_1\omega_{T,1}+\gamma_2\omega_{T,2})\pm\left[\left(\omega_{T,1}^2-\omega_{T,2}^2+i(\gamma_1\omega_{T,1}-\gamma_2\omega_{T,2})\right)^2+(2\omega'^2-\Omega^2)^2\right]^{\tfrac{1}{2}}\right\}, \quad (8b)$$

if we omit the $(1+i\phi)$-dependence of $V^*$, for simplicity. The Raman frequencies of the coupled TO (noted $\omega_{T,-}$ and $\omega_{T,+}$) and LO (noted $\omega_{L,-}$ and $\omega_{L,+}$) modes with $(u_1,u_2)$-mixed character readily follow by taking the real parts of $\omega^*$ given by Eq. (8b), noted $Re(\omega^*)$, taking $\Omega^2=0$ and $\Omega^2=\omega_L^2-\omega_T^2$, respectively.

2. The exceptional point: a pivot between the underdamped and overdamped regimes

The exceptional point of the TO-coupled system ($\Omega^2=0$, specified via subscript $T$), corresponding to $Re(\omega_{T,+}^*)=Re(\omega_{T,-}^*)$, is achieved when the square-root-term in Eq. (8b) is at zero – meaning that both its real and imaginary parts are at zero – taking care to replace $\omega'^2$ by $\omega'^2(1+i\phi)$. The condition on the imaginary part leads to $\gamma_1\omega_{T,1}-\gamma_2\omega_{T,2}=-4\omega'^2\phi(\omega_{T,1}^2-\omega_{T,2}^2)^{-1}$. When injected in the equality to zero applying to the real part, the latter equality leads to a biquadratic equation on $X=(\omega_{T,1}^2-\omega_{T,2}^2)^2$ whose only admissible solution obeys $\omega_{T,1}^2-\omega_{T,2}^2=\pm 2\omega'^2\phi$. The above two equations relate via $\phi$ (justifying a posteriori its introduction as an anelastic damping term related to $k'$ – see above), leading to $\gamma_1\omega_{T,1}-\gamma_2\omega_{T,2}=\pm 2\beta_T\omega_c^2$ near the resonance ($\omega_c$), where

$$\beta_T=\omega'^2/\omega_c^2. \quad (9a)$$

In fact, $\beta$ measures the strength of the mechanical-coupling between the two oscillators (via $k'$) compared to the intrinsic strength of each mechanical oscillator at the resonance (in reference to $k_1$ and $k_2$, being equal at this limit). Strictly at the resonance, i.e., $\omega_{T,1}=\omega_{T,2}=\omega_c$, the critical $\beta$ value

$$\beta_{cr}=|\gamma_1-\gamma_2|/\omega_c \quad (9b)$$

is achieved. This gives a measure of the damping compared to the intrinsic strength of each mechanical oscillator at the resonance. Different regimes are achieved depending on whether $\beta_T>\beta_{cr}$ (coupling dominates over damping) or $\beta_T<\beta_{cr}$ (vice versa), separated by the so-called exceptional point characterized by $\beta_T=\beta_{cr}$ at the resonance[25].

From now on the interest for $\phi$ disappears, for our use at least. Hence $\phi$ is taken equal to zero in the next sections, with an immediate consequence that $V$ becomes real.

Incidentally, the exceptional point for the LO mode can be derived in the same way by adopting the relevant $\Omega^2$ value (i.e., $\omega_L^2-\omega_T^2$) and substituting $(\omega'^2-\tfrac{\Omega^2}{2})(1+i\phi)$ for $(\omega'^2-\tfrac{\Omega^2}{2})$ in Eq. (8) then. This leads to introduce a characteristic $\beta$-like parameter for the LO modes, noted $\beta_L=\left|\omega'^2-\tfrac{\Omega^2}{2}\right|/\omega_c^2$. The exceptional point for the LO modes, corresponding to an actual crossing of the LO frequencies, is then achieved for $\beta_L=\beta_{cr}$. Note, that owing to the $\Omega^2$ term, the exceptional point cannot be achieved simultaneously for the TO ($\Omega^2=0$) and LO ($\Omega^2\neq 0$) modes.

2-a. $\beta\gg\beta_{cr}$: the reference insight at zero damping ($\gamma_1=\gamma_2=0$; $\beta_{cr}=0$)

The absence of damping ($\gamma_1=\gamma_2=0$), corresponding to $\beta\gg\beta_{cr}$, is treated in the first place prior to examining the $\beta>\beta_{cr}$ and $\beta<\beta_{cr}$ cases in the subsequent Secs., for reference purpose. The general ($\omega_1\neq\omega_2$) and resonant cases ($\omega_1=\omega_2$) are successively considered, for the sake of completeness and for future use. Generally, the used notation without the "$T$" or "$L$" subscripts indicates that the developed treatment is valid for both the TO and LO modes depending on the used $\Omega^2$ value.

At $\gamma_1=\gamma_2=0$, all parameters become real, so that the star added as a superscript to mark the imaginary character in Eq. (8) can be removed. The frequencies of the coupled modes, identifying with the eigenvalues of the relevant dynamical matrices, are $\omega_\pm^2=E_0\pm(\delta+V^2)^{\tfrac{1}{2}}$, and the corresponding unit wave vectors take the general form

$$|u^{\pm}\rangle = \begin{pmatrix} u_1^{\pm} \\ u_2^{\pm} \end{pmatrix} = \left\{V^2 + \left[-\delta \pm \sqrt{\delta^2 + V^2}\right]^2\right\}^{-\frac{1}{2}} \begin{pmatrix} -V \\ -\delta \pm \sqrt{\delta^2 + V^2} \end{pmatrix}. \tag{10}$$

As orthogonal unit vectors, $|u^{\pm}\rangle$ can be written as, e.g., $|u^+\rangle = \begin{pmatrix} \cos\sigma \\ \sin\sigma \end{pmatrix}$ and $|u^-\rangle = \begin{pmatrix} \sin\sigma \\ -\cos\sigma \end{pmatrix}$. In this case, $(\cos\sigma)^2$ and $(\sin\sigma)^2$ conveniently represent the relative contributions of oscillators 1 and 2 to $|u^+\rangle$ in each symmetry (TO or LO), respectively. The same applies also to $|u^-\rangle$. Eventually $(\tan\sigma)^2$ represents the degree of mixing of the two oscillators in the coupled modes[S16].

Besides, the unit vector defined as

$$|2\sigma\rangle = \begin{pmatrix} \sin 2\sigma \\ \cos 2\sigma \end{pmatrix} = \frac{1}{\sqrt{\delta^2 + V^2}} \begin{pmatrix} -V \\ \delta \end{pmatrix} \tag{11}$$

is interesting as well since the ratio of its components, i.e., $|\tan 2\sigma| = V/\delta$ provides a direct insight into the "strength of the coupling" (represented by $V$) depending on the proximity to the resonance between the uncoupled (TO or LO) oscillators[S16] (governed by $\delta$). Basically the "strength of the coupling" is all the greater that $V$ (dictated by $k'$) is large and that $\delta$ is small (meaning that the resonance is close).

Strictly at the resonance ($\omega_1 = \omega_2 = \omega_c$, being clear that $\omega_c$ is not the same for the TO and LO modes), where $\delta = 0$, the real parts of the eigenvalues, corresponding to physical observable, express as[25]

$$\omega_+ \sim \omega_c + \beta \tag{12a}$$

$$\omega_- \sim \omega_c + \frac{1}{2} \cdot \frac{\Omega^2}{\omega_c}, \tag{12b}$$

based on a first order Taylor expansion assuming $\omega' \ll \omega_c$ (verified experimentally with Zn$_{~0.5}$Be$_{~0.5}$Se – see below), with the usual condition on $\Omega^2$ to distinguish between the TO or LO modes. As for the related wave vectors, they simplify to

$$|u^{\pm}\rangle = \frac{1}{\sqrt{2}} \begin{pmatrix} 1 \\ \pm 1 \end{pmatrix}. \tag{13}$$

The mass displacements are the same in magnitude for oscillators 1 and 2, being either in the same or in opposite direction(s), i.e., in-phase or out-of-phase, corresponding to the so-called symmetric (SYM) and antisymmetric (ASYM) normal modes of vibrations of the coupled system, vibrating at frequencies $\omega_+$ and $\omega_-$, respectively.

Note that in the ASYM mode the mechanical coupling is not active; the corresponding spring $k'$ behaves in fact like a rigid bar connecting the two masses ($\mu$), the reason why $\omega'$ does not contribute to $\omega_-$. In contrast $k'$ is "active" in the SYM mode, being at the origin of a finite frequency gap between the two coupled (TO or LO) modes. If we (virtually) neglect the change in $\omega_c$ when shifting from the TO to the LO modes, the upper LO frequency is shifted upward the lower TO one by an amount fixed by the ionic plasmon coupling (in reference to $\Omega^2$), whereas the lower LO frequency matches the upper TO one (within the current approximation on $\omega_c$).

2-b. Resonant conditions at finite damping ($\gamma_1, \gamma_2 \neq 0$; $\omega_1 = \omega_2 = \omega_c$)

At the resonance ($\omega_1 = \omega_2 = \omega_c$), $\delta = 0$, so that, noting $\begin{pmatrix} u_1 \\ u_2 \end{pmatrix} = |u\rangle$, the system (6) reduces to

$$\begin{pmatrix} \omega_c^2 + \omega'^2 + \frac{1}{2}\Omega^2 + i\gamma_1\omega_c & -(\omega'^2 - \frac{1}{2}\Omega^2) \\ -(\omega'^2 - \frac{1}{2}\Omega^2) & \omega_c^2 + \omega'^2 + \frac{1}{2}\Omega^2 + i\gamma_2\omega_c \end{pmatrix} |u\rangle = \omega^2 |u\rangle, \tag{14}$$

that can be developed into

$$\omega_c^2 \left\{ \left[1 + \beta + \frac{\Omega^2}{\omega_c^2} + i\left(\frac{\gamma_1 + \gamma_2}{2\omega_c}\right)\right] \tilde{I} + \tilde{M} \right\} |u\rangle = \omega^2 |u\rangle, \tag{15}$$

where $\tilde{I}$ is the identity matrix, $\tilde{M} = \begin{pmatrix} -i\beta_{cr} & -\beta \\ -\beta & +i\beta_{cr} \end{pmatrix}$, $\beta = \frac{\omega'^2 - \frac{1}{2}\Omega^2}{\omega_c^2}$ identifying with $\beta_T$ or $\beta_L$ depending on the used $\Omega^2$ value, and with $\beta_{cr}$ defined as $\frac{\gamma_2 - \gamma_1}{2\omega_c}$ on account that the lower oscillator (specified via subscript "2") is an overdamped one in the case of Zn$_{~0.5}$Be$_{~0.5}$Se ($\gamma_2 > \gamma_1$, see below).

The diagonalization of the relevant TO ($\Omega^2 = 0$) and LO ($\Omega^2 \neq 0$) dynamical matrices behind Eq. (15) leads to the complex set of (starred) eigenvalues[25],

$$\omega_{\pm}^{*2} = \omega_0^2 \left[1 + \beta + \frac{\Omega^2}{\omega_c^2} + i\left(\frac{\gamma_1 + \gamma_2}{2\omega_c}\right) \pm \sqrt{\beta^2 - \beta_{cr}^2}\right], \tag{16}$$

with two variants, examined below, depending on the relative importance of the damping and coupling effects, reflected by the $\frac{\beta_{cr}}{\beta}$ ratio. The two variants in question develop on each side of the exceptional point ($\frac{\beta_{cr}}{\beta}$=1), where $\Delta= |Re(\omega_+^*) - Re(\omega_-^*)| = 0$ (see above).

- $\beta > \beta_{cr}$: by assuming that all terms within the above bracket are small with respect to unity – which is valid for Zn$_{\sim0.5}$Be$_{\sim0.5}$Se (see below) – and by subsequently performing a first-order Taylor expansion (that is sufficient to keep the information on the $\gamma_i$-damping terms), one eventually arrives at

$$\Delta= \omega_c\beta\sqrt{1-\left(\frac{\beta_{cr}}{\beta}\right)^2}, \tag{17a}$$

providing information on the $\frac{\beta_{cr}}{\beta}$ –dependence of the frequency gap between the coupled modes (with TO and LO types depending on the considered $\Omega^2$ value).

- $\beta < \beta_{cr}$: In this case the square root in Eq. (15) becomes purely imaginary. By repeating the above procedure and extending the Taylor expansion up to the second order (because the first order falls short of generating any dependence of the eigenvalues on the $\gamma_i$-damping terms), $\Delta$ takes the form

$$\Delta= \frac{(\gamma_2^2-\gamma_1^2)}{8\omega_c}\sqrt{1-\left(\frac{\beta_{cr}}{\beta}\right)^{-2}}. \tag{17b}$$

It is further instructive to determine the eigenvectors of the dynamical matrix, that coincide in fact with those of $\widetilde{M}$, given by,

$$|u_\pm^*\rangle = \left\{\beta^2 + \left[-i\beta_{cr} \pm \sqrt{\beta^2 - \beta_{cr}^2}\right]^2\right\}^{\frac{-1}{2}} \begin{pmatrix} \beta \\ -i\beta_{cr}\pm\sqrt{\beta^2-\beta_{cr}^2} \end{pmatrix}, \tag{18}$$

*i.e.*, complex ones, hence justifying the addition of a star as a superscript. At $\beta_{cr} \to 0$, one recovers the SYM and ASYM normal modes. An actual displacement, *i.e.*, a real one, is preserved for oscillator 2 as long as the coupling dominates over the damping ($\beta > \beta_{cr}$). From the exceptional point onwards ($\beta \leq \beta_{cr}$), the displacement of oscillator 2 becomes purely imaginary, meaning that the latter overdamped oscillator is, in reality, frozen; only oscillator 1 vibrates, as sketched out in Fig. 1. The selective freezing of oscillator 2 at the exceptional point, applying as well in the TO and LO symmetries (using the relevant $\beta$ value in each case – see Sec. IIA1), is rather intuitive. Indeed, at this limit the coupling is exactly screened by the damping so that the original SYM and ASYM normal modes of the undamped system ($\gamma_1 = \gamma_2 = 0$) representing the in-phase and out-of-phase motions of the two masses, respectively, merge into an unique mode with mixed SYM/ASYM character. The resulting degenerated normal mode has to achieve a balanced compromise between its two native normal modes, corresponding to one mass in motion and the other inert.

B. The Be-Se doublet of Zn$_{\sim0.5}$Be$_{\sim0.5}$Se : a high-pressure Raman study

1. Ambient pressure – combined TO/LO insight into the mechanical coupling ($k'$)

The pressure dependence of the BeSe-like TO and LO frequencies of Zn$_{0.48}$Be$_{0.52}$Se displayed in Fig. S9b is discussed hereafter within the above damped model of coupled harmonic 1D-oscillators. The coupling can be either purely-mechanical (in reference to $k'$) or both electrical-mechanical (including $E$ on top of $k'$) whether considering the TO or LO modes, respectively. The LO-like $E$-coupling is not an issue. It is determined by $\Omega^2 = \omega_L^2 - \omega_T^2$, whose pressure dependence for BeSe can be determined from existing data in the literature (detail is given below). This leaves $k'$ as the only unknown parameter governing the entire set of TO and LO frequencies in their pressure dependence.

A crude $k'$-estimate can be achieved by assuming that $k'$ is not pressure dependent, and by placing the discussion at ambient pressure, *i.e.*, far from the resonance ($\omega_c, P_c$). The justification is that at ambient pressure the two BeSe-like TO Raman peaks of Zn$_{\sim0.5}$Be$_{\sim0.5}$Se exhibit similar linewidths at full maximum ($\gamma_1 = \gamma_2$, Fig. S9a). Hence $\beta_{cr} \ll \beta_T$, meaning that the $k'$ −coupling is not yet challenged by overdamping of either mode. This is ideal for a direct $k'$ −insight. We anticipate that the mechanical coupling (in reference to $k'$) is weak, i.e., that $\omega' \ll \omega_c$, since the two BeSe-like TO modes of Zn$_{\sim 0.5}$Be$_{\sim 0.5}$Se exhibit comparable Raman intensities at ambient pressure, as expected in absence of coupling in the ideal case of a random Zn↔Be substitution[S31].

In fact, a combined (TO, LO) study is required to achieve a tentative $k'$-estimate. Concerning the LO insight we refer more specifically to the lower/minor $LO^-_{Be-Se}$ mode as explained below. One problem is that the latter feature is not visible in Fig. S9a, being screened by the dominant TO modes that emerge on each side of it. For a pure LO insight we resort to earlier Raman measurements done at normal incidence/detection onto the (100)-face of a ~1µm-thick Zn$_{\sim 0.5}$Be$_{\sim 0.5}$Se/GaAs epitaxial layer, corresponding to a LO-allowed/TO-forbidden scattering geometry[S36]. Generally, we have shown elsewhere that the TO and LO Raman modes detected with Zn$_{1-x}$Be$_x$Se epitaxial layers and single crystals of similar compositions are remarkably similar (see Fig. 1 of Ref. S30), and can thus be treated on equal footing.

A two-step self-consistent (TO,LO) procedure is used to estimate $k'$.

The first step is concerned with the TOs. The $\omega'$-dependence of the frequencies of the bare/uncoupled ($\omega_{T,2}$ and $\omega_{T,1}$, ranked in order of increasing frequency) TO modes (not observable experimentally) behind the experimentally observed ($\omega_{T,-}, \omega_{T,+}$) Raman frequencies of the coupled TO modes detected with our Zn$_{\sim 0.5}$Be$_{\sim 0.5}$Se epilayer, used as setpoints, is derived for $\omega'$ varying continuously between 0 and 150 cm$^{-1}$ (note that, in this case, the conditions $\omega' \ll \omega_c$ is fulfilled) using the TO-version of Eq. (8) implemented in absence of damping ($\gamma_1 = \gamma_2 = 0$). For each $\omega'$ value one relevant ($\omega_{T,2}, \omega_{T,1}$) pair of frequencies is generated corresponding to theoretical ($\omega_{T,-}, \omega_{T,+}$) frequencies matching the experimental values (marked by plain arrows) within $\pm 0.5$ cm$^{-1}$ (the arbitrary set accuracy). We have checked that the same ($\omega_{T,2}, \omega_{T,1}$) pair is generated (within less than ~1.5 cm$^{-1}$) for any given $\omega'$ value whether implementing the research procedure on the targeted ($\omega_{T,-}, \omega_{T,+}$) values by incrementing (upward pointing open triangles) or by decrementing (downward pointing) the ($\omega_{T,2}, \omega_{T,1}$)-test values from arbitrary sets of starting test frequencies. This provides confidence in the reported curves in Fig. S10. Remarkably the ($\omega_{T,2}, \omega_{T,1}$) vs. $\omega'$ curves (plain symbols) meet at $\omega'$~140 cm$^{-1}$, beyond which critical $\omega'$ value no ($\omega_{T,2}, \omega_{T,1}$)-solution can be found.

The next step is concerned with the LOs. The theoretical ($\omega_{L,-}, \omega_{L,+}$) LO frequencies generated via the LO-version of Eq. (8) in absence of damping throughout the spanned $\omega'$-domain by using the above ($\omega_{T,2}, \omega_{T,1}$) predetermined pairs as input parameters are shown in Fig. S10. Note that the as-obtained theoretical curves in absence of mechanical coupling ($k'$=0) underestimate by far the experimental $LO^-_{Be-Se}$ and $LO^+_{Be-Se}$ Raman frequencies of the used Zn$_{\sim 0.5}$Be$_{\sim 0.5}$Se epilayer (pointed by external arrows). This overall discrepancy between experiment and theory concerning the LO frequencies has originally been attributed to a discrete fine structuring of the native TO modes behind the LOs, the result of inherent fluctuations in the local composition in a disordered system such as a mixed crystal[S36]. In this case the transfer of available Be-Se oscillator strength mediated by the LO-like macroscopic electric field from the lower to the upper Be-Se sub-mode of the series, is both of intra- and inter-mode types, and thus emphasized in comparison with a mere inter-mode transfer taking place in absence of fine structuring, with concomitant impact on the overall $TO - LO^+$ splitting, being emphasized. In fact, the $LO^+_{Be-Se}$ shift cannot be explained by $k'$, because increasing $\omega'$ softens $LO^+$ (Fig. S10). In contrast, $k'$ and fine-structuring work along for $LO^-$, contributing both to its hardening. Therefore, an upper $\omega'$ estimate, useful to fix ideas, can be achieved by focusing on the $LO^-_{Be-Se}$ shift and omitting the fine structure effect. In this approximation the perfect matching between the theoretical and experimental $LO^-_{Be-Se}$ frequency is achieved for $\omega'$=65 $\pm$ 20 cm$^{-1}$. The $LO^-_{Be-Se}$ mode is overdamped in Zn$_{1-x}$Be$_x$Se at any x value so that the $LO^-_{Be-Se}$ frequency is marred by a large error[S36] (refer to the dashed area in Fig. S10), with concomitant impact on the accuracy of the $\omega'$ value.

In a crude approximation we assume that $\omega'$ is not pressure dependent and transpose the study near the resonance ($\omega_c$~590 cm$^{-1}$, $P_c$~15 GPa), corresponding to the actual crossing of the two BeSe-like coupled TO modes, visible in Fig. S9b – within experimental uncertainty. As anticipated, the mechanical coupling is weak ($\omega' \ll \omega_c$). The corresponding dependence of $\Delta_T = |Re(\omega^*_{T,+}) - Re(\omega^*_{T,-})|$ on $\frac{\beta_{cr}}{\beta_T}$ at the resonance for the considered damped system of coupled BeSe-like TO modes,

given by Eq. (17), is displayed in Fig. S11. An important feature dictated by $\omega'$ is the starting $\Delta_T$ value at zero damping ($\beta_{cr}$=0), currently falling into the range 3 – 11 cm$^{-1}$, that fixes the overall shape of the admissible $\Delta_T$ –domain (framed by dotted curves). The main feature is the exceptional point ($\beta_T = \beta_{cr}$), corresponding to $\Delta_T$=0, as observed in the Raman spectra.

Based on Fig. S11, ae are now in a position to discuss the pending issues (i-iv) on a quantitative basis. The source of overdamping ($\beta_{cr} \neq 0$) in our coupled system of two BeSe-like oscillators is the lower mode revealed by the gradual collapse of this mode when forced into proximity of the upper mode by pressure (Fig. S9a) – in reference to (ii), being clear that at ambient pressure such overdamping does not exist ($\beta_{cr}$=0, since $\gamma_1 \sim \gamma_2$). The trend persists until a complete screening of the mechanical coupling ($\beta = \beta_{cr}$) is eventually achieved at the resonance ($\omega_c$, $P_c$) – see Fig. S9b, hence characterized by an actual crossing of the two modes – in reference to (i). From this critical pressure onwards the coupled system of BeSe-like oscillators resonantly locks into its unique overdamped normal mode vibrating at the $\sim\omega_c$ frequency [see Eq. (11), in which $\omega' \ll \omega_c$] which keeps varying with pressure. The exceptional mode achieves a compromise between the distinct SYM and ASYM normal modes of the same-but-undamped system at the resonance (Sec. IIA2b). As such, it is characterized by the overdamped oscillator, *i.e.*, the lower Be-Se one in our case, being inert/frozen, which nicely recollects with the *ab initio* insight – in reference to (iv). As for the particular value of the critical pressure ($P_c$) corresponding to the resonance ($\omega_c$) – in reference to (iii), it does not appear to be so remarkable in fact given the explanation provided in the main text. The only pending issue relates to the survival/death of the lower mode beyond $P_c$.

2. Pressure dependence of the BeSe-like TO and LO Raman frequencies

What remains unclear from the experimental data at this stage is whether only the upper oscillator "survives" and the lower one is "killed" at the resonance – scenario 1, as predicted[25] (see Sec. IIA), or whether both modes survive as degenerated features into the same mode from $P_c$ onwards – scenario 2.

A decisive marker is the amount of Be-Se oscillator strength awarded to the exceptional mode ($P \geq P_c$). In Zn$_{0.5}$Be$_{0.5}$Se the available amount of BeSe-like oscillator strength represents half the parent value at any pressure (feature 1)[8], dictated by the Be content. This amount is equally shared between the two submodes forming (feature 2) the Be-Se doublet as long as these do not couple each other[S36]. This is apparently the case at ambient pressure, reflected by the nearly equal Raman intensities of the two BeSe-like TOs (Fig. S9a). Scenario 1 or 2 applies depending on whether the exceptional mode carries the totality of the available Be-Se oscillator strength or only half of it, respectively.

By approaching $P_c$ the two BeSe-like TOs get closer and couple. The coupling presumably modifies the sharing of the available BeSe-like oscillator strength between the two modes meaning that the Raman intensities of the coupled TOs do not reflect any more the oscillator strengths of the underlying bare/uncoupled TOs.

In this case one may alternatively resort to the LOs, changing the focus from the Raman intensity to the Raman frequency then. In fact, in a pure zincblende compound such as BeSe, the oscillator strength monitors the magnitude of the TO-LO splitting of the unique phonon mode. The situation is not as simple with Zn$_{0.5}$Be$_{0.5}$Se since the Be-Se Raman signal is bi-modal. Additional complexity arises in that the polar (electrical-mechanical) LO modes couple very easily via their long-range electric field $\vec{E}_L$ – in contrast with the non-polar (purely-mechanical) TO modes that need to be very close to develop a mechanical coupling (see above). As already mentioned (Sec. IIB1), the effect of the $\vec{E}_L$-coupling is to channel most of the available Be-Se oscillator strength towards a giant $LO^+$ feature, hence shifted at a high frequency far off its native TO doublet[S36] – abbreviated $TO^{(2)}$ hereafter. In other words, the magnitude of the $TO^{(2)} - LO^+$ splitting is directly monitored by the available amount of Be-Se oscillator strength at a given pressure, offering a valuable test for scenarios 1 and 2. Basically the $TO^{(2)} - LO^+$ splitting will be large if $LO^+$ attracts the full amount of available BeSe-like oscillator strength at a given pressure – along scenario 2. In contrast, if one Be-Se oscillator freezes on mixing at the resonance – as predicted under scenario 1, half of the available Be-Se oscillator strength will be lost for the transfer to $LO^+$, with concomitant impact on the magnitude of the $TO^{(2)} - LO^+$ splitting, dropped by half with respect to scenario 2.

2-a. Scenario 1 vs. scenario 2: reference BeSe-like (TO,LO) data sets

Useful sets of theoretical Zn$_{0.5}$Be$_{0.5}$Se TO and LO curves to test scenarios 1 and 2, displayed in Fig. S12, to compare with experimental Zn$_{0.48}$Be$_{0.52}$Se TO (hollow symbols) and LO (full symbols) Raman frequencies taken from Fig. S9a (sample 1), comprises the following: (a) polynomial adjustments of the experimentally observed coupled TO frequencies; (b) corresponding bare/uncoupled TO frequencies ($\omega_{T,1}$, $\omega_{T,2}$) at $P \leq P_c$; (c) bare/uncoupled TO frequency of the exceptional mode at $P > P_c$; (d) frequencies of the individual uncoupled LO modes at $P \leq P_c$; (e) Coupled LO frequencies ($\omega_{-,L}, \omega_{+,L}$) at $P \leq P_c$; (f) Frequency of the upper coupled mode LO mode ($\omega_{+,L}$) at $P \leq P_c$ in its dependence on the phonon damping; (g)/(h) LO frequency of the exceptional mode considering that this attracts all/half the available BeSe-like oscillator strength.

Technically, the theoretical data sets (b–h) are generated as follows. The separate ($\omega_{T,1}, \omega_{T,2}$) frequencies of the bare-uncoupled TO oscillators at $P \leq P_c$ (b) are derived from the experimentally observed/coupled ($\omega_{T,+}, \omega_{T,-}$) ones by using the undamped ($\gamma_1=\gamma_2=0$) TO-version ($\Omega^2=0$) of Eq. (8) in case of a weak mechanical coupling ($\omega'$=65 cm$^{-1}$ ≪ $\omega_c$=590 cm$^{-1}$, see Figs. S10 and S12 for the $\omega'$ and $\omega_c$ estimates, respectively), as explained in Sec. IIB1. The $\omega_{T,1} = \omega_{T,2}$ frequency of the uncoupled TO mode behind the exceptional TO mode at $P > P_c$ (c) is derived on the same basis in absence of damping, by cancelling the square root term in Eq. (8). The individual frequencies of the uncoupled LO modes at $P \leq P_c$, in reference to set (d), are obtained via the $Im\{-\varepsilon_r^{-1}(\omega)\}$-like form of the generic Raman cross section given in Ref. 18 using a truncated version of the relative dielectric function $\varepsilon_r(\omega)$ of Zn$_{0.83}$Cd$_{0.17}$Se, being reduced to the sole considered Be-Se oscillator as characterized by its coupled TO frequency, i.e., $\omega_{-,T}$ (any $P$) or $\omega_{+,T}$. The nominal pressure-dependent (see above) amount of oscillator strength is preserved for each mode throughout the pressure domain. As for the individual phonon dampings ($\gamma_1, \gamma_2$) they have strictly no impact on the LO frequency (they only modify the broadening of the Raman peak – of no interest for our concern). The frequencies of the coupled LO modes displayed as set (e) are calculated along the same approach but by using the full expression of $\varepsilon_r(\omega)$ taking into account the two BeSe-like oscillators, as characterized by their coupled TO frequencies ($\omega_{-,T}, \omega_{+,T}$). This suffices to materialize the $\vec{E}_L$-coupling between neighboring individual LO modes. The individual phonon dampings play no role regarding set (e), as for set (d). The situation changes for the LO frequency of the upper coupled mode at $P \leq P_c$, i.e., $\omega_{+,L}$ displayed as set (f). This is generated by injecting the uncoupled ($\omega_{T,1}, \omega_{T,2}$) to frequencies reported as set (b) into Eq. (8) taken in its LO version with the nominal pressure-dependent amount of available strength ($\Omega^2 \neq 0$) and parametrized with a weak mechanical coupling ($\omega'$=65 cm$^{-1}$), further considering a finite pressure-dependent damping ($\gamma_1, \gamma_2 \neq 0$). In fact $\gamma_1$ is taken constant throughout the entire pressure domain (~20 cm$^{-1}$), consistently with experimental observations (Fig. S9a) but $\gamma_2$ is pressure dependent. At ambient pressure $\gamma_2$ is minimal and matches $\gamma_1$, testified by identical width at half height of the two BeSe-like TO Raman peaks (Fig. S9a). It roughly doubles (~35 cm$^{-1}$) at the resonance ($\omega_c, P_c$) corresponding to the exact screening of the mechanical coupling by overdamping testified by the emergence of the "exceptional mode". The exact $\gamma_2$ value at this limit is derived from the $\beta' = \beta_{cr}$ correspondence for the current weak mechanical coupling ($\omega'$=65 cm$^{-1}$). A linear dependence of $\gamma_2$ on pressure is further assumed, in a crude approximation. At $P > P_c$, the LO frequency of the exceptional mode, corresponding to set (g), is also calculated via Eq. (8) from which the square root term is omitted using the same relevant damping set at any pressure ($\gamma_1$=20 cm$^{-1}$, $\gamma_2$=35 cm$^{-1}$), considering that the exceptional mode is awarded the full amount of the pressure-dependent BeSe-like oscillator strength from $P_c$ onwards. The remaining set (h) refers to the same parameter obtained along the same line but on the basis of a dead loss of oscillator strength for the lower mode from $P_c$ onwards, meaning that the exceptional mode retains only half of the available BeSe-like oscillator strength.

In building up data sets (b-h) care is taken that most parameters coming into Eqs. (8) and (19) are pressure dependent, *i.e.*, not only the frequencies $\omega_{T,p}(x)$ of the BeSe-like TO modes of Zn$_{1-x}$Be$_x$Se, whether coupled or uncoupled depending on the used model (see above), but also several input parameters related to the end compounds, including $\omega_{T,BeSe}$ and $\omega_{L,BeSe}$ together with $\varepsilon_{\infty,ZnSe}$ and $\varepsilon_{\infty,BeSe}$.

The pressure dependence of $\varepsilon_{\infty,ZnSe}$, abbreviated $\varepsilon_{\infty,ZnSe}(P)$, is already known (Sec. IC), as derived via *ab initio* calculations[41]. As for BeSe, due to the lack of experimental data in the literature, we resort to the careful *ab initio* study of the lattice dynamics of BeSe taken in its zincblende structure depending on pressure up to 60 GPa[S37]. This notably includes a detailed description of $\varepsilon_{\infty,BeSe}(P)$ – as defined well beyond the phonon resonance (*i.e.*, at $\omega \gg \omega_{T,BeSe}$) – that is remarkably consistent with the original *ab initio* prediction[41,] and thus presumably valid. A direct calculation of $\omega_{T,BeSe}(P)$ is further provided in Ref. S37. The missing insight into $\omega_{L,BeSe}(P)$ is eventually gained from the prediction in the same work of the pressure dependence of the Born/dynamic effective charge of BeSe, given[S35] by $Z(P)=\mu\varepsilon_0\varepsilon_\infty(P)V(P)\Omega^2(P)$, which monitors the magnitude of the TO-LO splitting via $\Omega^2_{BeSe}(P) = \omega^2_{L,BeSe}(P) - \omega^2_{T,BeSe}(P)$, with $Z$, $\mu$, $V$ and $\Omega$ introduced in Sec. IIA1.

The $V(P)$ dependence for zincblende BeSe has been determined based on a careful high-pressure X-ray diffraction study throughout the zincblende domain, *i.e.*, from ambient pressure up to ~60 GPa, corresponding to adoption of the NiAs structure by BeSe[S38]. Besides, the basic trend for BeSe is that $Z$ reduces when $P$ increases (*i.e.*, by as much as ~45% from ambient pressure to 50 GPa) – as observed with most semiconductor compounds, whereas $\varepsilon_\infty(P)$ slightly enlarges (*i.e.*, by ~7.5%, correspondingly). By referring to ambient pressure, where $\omega_{L,BeSe}$ and $\omega_{L,BeSe}$ are estimated at ~510 and ~578 cm$^{-1}$ in Ref. S37– consistently with experiment[S36] (within less than 10 cm$^{-1}$), the available $V(P)$, $\varepsilon_\infty(P)$, $\omega_T(P)$ and $Z(P)$ dependencies eventually offer an access to $\omega_L(P)$ throughout the studied pressure domain. The pressure dependence of the BeSe oscillator strength $S_0(P) = \varepsilon_\infty \omega_T^{-2} \Omega^2$ comes off as a by-product, corresponding to a quasi linear loss of ~1.25% per GPa under pressure throughout the studied domain (0 – 30 GPa).

Considering Zn$_{0.5}$Be$_{0.5}$Se as a system of three-independent {1x(Zn-Se),2x(Be-Se)} TO oscillators, $\varepsilon_r(\omega,x)$ takes the classical form (in absence of damping) given by Eq. (1), in which $x_p$, $S_{0,p} = \varepsilon_{\infty,p} \cdot \Omega_p^2/\omega_{T,p}^2$ and $\omega_{T,p}(x)$ represent the fraction of oscillator $p$ in Zn$_{1-x}$Be$_x$Se, the oscillator strength awarded to the related pure compound[8], *i.e.*, ZnSe ($p$=1) or BeSe ($p$=2,3), and the frequency of the observed (uncoupled) TO mode due to oscillator $p$ in the Raman spectra, respectively. For the current Be content of ~50 at.%, $x_1$~0.5; $x_2 = x_3$~0.25 (reflecting the sensitivity of the Be-Se vibration to its local environment at the first-neighbor scale, in case of a random Zn↔Be substitution[S31]) and $\varepsilon_\infty(x)$ is the average between $\varepsilon_{\infty,ZnSe}$ and $\varepsilon_{\infty,BeSe}$. Additional input parameters coming into the used generic $\varepsilon_r(\omega,x)$-dependent expression of the Raman cross section given in Ref. 18 are the parent ZnSe and BeSe $C_{F-H}$ values, given in Ref. S36. The Zn-Se and Be-Se spectral ranges are well separated, *i.e.*, by as much as 200 cm$^{-1}$, so that for a specific Be-Se insight, one may well omit the Zn-Se oscillator in $\varepsilon_r(\omega,x)$, in a crude approximation.

We emphasize that no spurious LO-like transfer of oscillator strength between the Zn-Se and Be-Se spectral ranges, *i.e.*, likely to challenge the above description, can occur, due to the prohibitive frequency gap. This guarantees that the Be-Se and Zn-Se oscillators can be treated as independent, as implicitly considered above. To fix ideas, the difference in the theoretical frequency of the upper (GaP-like) $LO^+$ mode of GaAs$_{1-x}$P$_x$ whether considering or not the LO-like coupling between the Ga-As and Ga-P spectral ranges, separated by ~70 cm$^{-1}$, is hardly discernible[9].

2-b. Discussion – experiment vs. theory

In Fig. S12 the various predicted pressure dependencies of the $LO^+_{Be-Se}$ frequency – referring to the data sets (d, e, f) – are superimposed onto the corresponding experimental data (full symbols), for comparison. Our guideline is that at ambient pressure, the experimental $LO^+_{Be-Se}$ frequency stays well above any theoretical estimate – by as much as $\alpha$~15 cm$^{-1}$, the result of natural fluctuations in the local

Zn$_{0.5}$Be$_{0.5}$Se composition giving rise to a fine structuring of the two Be-Se TO sub-modes – see Sec. IIB1. The $LO^+_{Be-Se}$-shift is intrinsic to alloy disorder, and thus presumably not pressure dependent. Hence, it should persist as such throughout the entire pressure domain.

In fact, the $LO^+_{Be-Se}$-shift $\alpha$~15 cm$^{-1}$ between experiment and theory can be observed at high pressure only if the exceptional mode carries half of the available BeSe-like oscillator strength, meaning that only one sub-mode "survived" the resonance, as ideally expected under scenario 1. Somewhat ideally, the $LO^+_{Be-Se}$-shift stabilizes around ~15 cm$^{-1}$ from $P_c$ onwards, meaning that the alternative submode is "killed" right at the resonance, *i.e.*, on emergence of the "exceptional mode". For a tentative assignment of the active ("surviving") and passive ("killed") submodes involved in the exceptional mode beyond $P_c$ we resort to the TOs. The pressure dependence of the exceptional TO mode at $P \geq P_c$ is in continuation of that related to the upper mode at $P < P_c$. Besides, the progressive collapse of the lower sub-mode with increasing pressure at $P < P_c$ is a precursor sign of its full extinction at $P_c$. On this basis, the active and passive TO oscillators behind the exceptional mode *post* resonance can be traced back to the upper and lower ones *ante* resonance, respectively.

However, the actual "freezing" of part of the available oscillator strength above $P_c$ makes no physical sense. The "freezing" must be apparent only. In fact, we have checked that the TO sub-mode given up for "dead" beyond $P_c$ is "revived", *i.e.*, regains its original oscillator strength, when the pressure drops below $P_c$ in a downstroke Raman experiment (pressure decrease). Our present view is that when the lower mode is forced in the proximity of the upper mode by pressure, it "evades" an effective coupling by becoming overdamped right at the resonance. By doing so, the lower mode retains its original oscillator strength – though in an overdamped form, hence turned away from the inherent transfer of oscillator strength accompanying an actual coupling with the upper mode.

It is interesting to examine in detail how the "2-mode→1-exceptional-mode" transition develops on approach to $P_c$. In absence of damping, the $TO^{(2)} - LO^+_{Be-Se}$ splitting is expected to drop suddenly by half when the two distinct Be-Se sub-modes ($P < P_c$) resonantly lock into the exceptional mode ($P > P_c$) – compare the data sets (e) and (h). Such sharp transition opposes to experimental findings, indicating a smooth transition (full symbols). Now, the absence of damping is unrealistic since the overdamping is the *sine qua non* condition for the emergence of the exceptional mode (Sec. IIB). In fact, by incorporating damping in the two-mode regime ($P < P_c$) – in reference to the data set (f), the $LO^+_{Be-Se}$ mode tends to soften already close to ambient pressure, and the softening progressively increases with pressure until coincidence with the $LO^+_{Be-Se}$ frequency of the exceptional mode exactly at the resonance ($P_c$) – at the origin of the data set (h), as ideally expected.

Generally, the current validation of scenario 1 based on a combined understanding of the TO and LO Raman data in the Be-Se spectral range of Zn$_{1-x}$Be$_x$Se reveals that the so-called "exceptional point" of a coupled system of damped harmonic (1D-)oscillators is not a purely theoretical feature but can be observed experimentally.

Last, it is worth to mention that, given the $\omega'$ (~65 cm$^{-1}$), $\omega_{L,BeSe}$ (~501 cm$^{-1}$) and $\omega_{T,BeSe}$ (~450 cm$^{-1}$) values, the coupling between the polar LO modes mediated by their common macroscopic electric field remains quasi unaffected by overdamping ($\beta_L \gg \beta_{cr}$, Sec. IIA1), though this suffices to bring the system of non polar (purely mechanical) TO modes at its exceptional point ($\beta = \beta_{cr}$).

## Supplementary Information – only references

S1. Vodopyanov L. K. *et al*. Optical phonons in $Zn_{1-x}Cd_xSe$ alloys. *Phys. Stat. Sol. C* **1**, 3162–3165 (2004)

S2. Massiot, D. *et al*. Modelling one- and two-dimensional solid-state NMR spectra[+]. *Magn. Reson. Chem.* **40**, 70–76 (2002).

S3. Cadars, S. *et al*. Atomic positional versus electronic order in semiconducting ZnSe nanoparticles. *Phys. Rev. Lett.* **103**, 136802-1–136802-4 (2009).

S4. Berettini, M. G., Braun, G., Hu, J. G. & Strouse, G. F. NMR analysis of surfaces and interfaces in 2-nm CdSe. *J. Am. Chem. Soc.* **126**, 7063–7070 (2004).

S5. Larsen, F. H., Jakobsen, H. J., Ellis, P. D. & Nielsen, N. C. Sensitivity-enhanced quadrupolar-echo NMR of half-integer quadrupolar nuclei. Magnitudes and Relative orientation of chemical shielding and quadrupolar coupling tensors. *J. Phys. Chem. A* **101**, 8597–8606 (1997).

S6. Lorenz, C. D., May, R. & Ziff, R. M. Similarity of percolation thresholds on the HCP and FCC lattices. *J. Statist. Phys*. **98**, 961–970 (2000).

S7. Takemura, K. & Dewaele, A. Isothermal equation of state for gold with a He-pressure medium. *Phys. Rev. B* **78**, 104119-1–104119-13 (2008).

S8. Dewaele, A., Datchi, F., Loubeyre, P. & Mezouar, M. High pressure-high temperature equations of state of neon and diamond. *Phys. Rev. B* **77**, 094106-1–094106-9 (2008).

S9. Hammersley, A. P., Svensson, S. O., Hanfland, M., Fitch, A. N. & Hausermann, D. Two-dimensional detector software: from real detector to idealised image or two-theta scan. *High Press. Res.* **14**, 235–248 (1996).

S10. Prescher, C. & Prakapenka, V. B. *DIOPTAS*: a program for reduction of two-dimensional X-ray diffraction data and data exploration. *High. Press. Res*. **35**, 223–230 (2015).

S11. Pellicer-Porres, J. *et al*. Observation of the cinnabar phase in ZnSe at high pressure. *High. Press. Res*. **22**, 355–359 (2002).

S12. Pellicer-Porres, J., Martinez-Garcia, D., Ferrer-Roca, Ch., Segura, A. & Muńoz-Sanjosé, V. High-pressure phase diagram of $ZnSe_xTe_{1-x}$ alloys. *Phys. Rev. B* **71**, 035210-1–035210-7 (2005).

S13. Karzel, H. *et al.* Lattice dynamics and hyperfine interactions in ZnO and ZnSe at high external pressures. *Phys. Rev. B* **53**, 11425–11438 (1996).

S14. Ameri, M. *et al*. Physical properties of the $Zn_xCd_{1-x}Se$ alloys: *Ab-initio* method. *Mat. Sci. & Appl.* **3**, 768–778 (2012).

S15. Hajj Hussein, R., Pagès, O. Firszt, F. Paszkowicz, W. & Maillard, A. Near-forward Raman scattering by bulk and surface phonon-polaritons in the model percolation-type ZnBeSe alloy. *Appl. Phys. Lett*. **103**, 071912-1–071912-5 (2013).

S16. Groenen, J. *et al*. Optical-phonon behavior in $Ga_{1-x}In_xAs$ : the role of microscopic strains and ionic plasmon coupling. *Phys. Rev. B* **58**, 10452–10462 (1998).

S17. Bouamama, K., Djemia, P., Lebga, N. & Kassali, K. *Ab initio* calculations of the elastic properties and the lattice dynamics of the $Zn_xCd_{1-x}Se$ alloy. *Semicond. Sci. Technol*. **24**, 045005-1–045005-5 (2009).

S18. Suzuki, K.-I. & Adachi, S. Optical constants of $Cd_xZn_{1-x}Se$ ternary alloys. *J. Appl. Phys*. **83**, 1018–1022 (1998).

S19. Asad, A. & Afaq, A. Structural and optical properties of ZnSe under pressure. *Res. & Rev.: J. Phys.* **5**, 6–10 (2016).

S20. Alivisatos, A. P., Harris, T. D., Brus, L. E. & Jayaraman, A. Resonance Raman scattering and optical absorption studies of CdSe microclusters at high pressure. *J. Chem. Phys.* **89**, 5979–5982 (1988).

S21. Steigerwald, M. L. *et al*. Surface derivatization and isolation of semiconductor cluster molecules. *J. Am. Chem. Soc.* **110**, 3046–3050 (1988).

S22. Cuscó, R., Consonni, V., Bellet-Amalric, E., André, R. & Artůs, L. Phase discrimination in CdSe structures by means of Raman scattering. *Phys. Stat. Solidi RRL* **11**, 1700006-1–1700006-5 (2017).

S23. Deligoz, E., Colakoglu, K. & Ciftci, Y. Elastic, electronic, and lattice dynamical properties of CdS, CdSe, and CdTe. *Physica B* **373**, 124–130 (2006).

S24. Fan, Z. *et al*. A transferable force field for CdS-CdSe-PbS-PbSe solid systems. *J. Chem. Phys.* **141**, 244503-1–244503-14 (2014).

S25. Feng, S.-Q., Li, J.-Y., & Cheng, X.-L. The structural, dielectric, lattice dynamical and thermodynamic properties of zincblende Cd*X* (*X*=S, Se, Te) from first-principles analysis. *Chin. Phys. Lett.* **32**, 036301-1–036301-5 (2014).

S26. Yu, W. C. & Gielisse, P. J. High pressure polymorphism in CdS, CdSe and CdTe. *Mat. Res. Bull.* **6**, 621–638 (1971).

S27. Benkhetou, N., Rached, D. & Rabah, M. *Ab initio* calculations of stability and structural properties of cadmium chalcogenides CdS, CdSe and CdTe under high pressure. *Czech. J. Phys.* **56**, 409–418 (2006).

S28. Güler, M. & Güler, E. Elastic, mechanical and phonon behavior of wurtzite Cadmium Sulfide under pressure. *Crystals* **7**, 164-1–164-11 (2017).

S29. Balzaretti, N. M. & da Jornada, J. A. H. Pressure dependence of the refractive index of diamond, cubic silicon, carbide and cubic boron nitride. *Solid State Commun*. **99**, 943–948 (1996).

S30. Valakh, M. Y., Lisitsa, M. P., Sidorenko, V. I. & Polissky, G. N. Antiresonance in the phonon spectrum of mixed crystals $Zn_{1-x}Cd_xSe$. *Phys. Lett.* **78A**, 115–116 (1980).

S31. Pagès, O. *et al*. Non-random Be-to-Zn substitution in ZnBeSe alloys: Raman scattering and *ab initio* calculations. *Eur. Phys. J. B* **73**, 461–469 (2010).

S32. Bhalerao, G. M. *et al*. High-pressure x-ray diffraction and extended x-ray absorption fine structure studies on ternary $Zn_{1-x}Be_xSe$. *J. Appl. Phys*. **108**, 083533-1–083533-7 (2010).

S33. Dicko, H. *et al.* Near-forward/high-pressure-backward Raman study of $Zn_{1-x}Be_xSe$ (x~0.5) – evidence for percolation behavior of the long (Zn-Se) bond. *J. Raman Spectrosc*. **47**, 357–367 (2016).

S34. Geurts, J. Analysis of band bending at III-V semiconductor interfaces by Raman spectroscopy. *Surf. Sci. Rep*. **18**, 1–89 (1993).

S35. Trommer, R., Müller, H., Cardona, M. & Vogl., P. Dependence of the phonon spectrum of InP on hydrostatic pressure. *Phys. Rev. B* **21**, 4869–4878 (1980).

S36. Pagès, O. *et al*. Long-wave phonons in ZnSe-BeSe mixed crystals : Raman scattering and percolation model. *Phys. Rev. B* **70**, 155319-1– 155319-10 (2004).

S37. Dahbi, S., Mankad, V. & Jha, P. K. A first principles study of phase stability, bonding, electronic and lattice dynamical properties of beryllium chalcogenides at high pressure. *J. Alloys and Compounds* **617**, 905–914 (2014).

S38. Luo, H., Ghandehari, K., Greene, R. G. & Ruoff, A. L. Phase transformations of BeSe and BeTe to the NiAs structure at high pressure. *Phys. Rev. B* **52**, 7058–7064 (1995).

# Supplementary-figure captions

**Figure S1 | Zn$_{0.83}$Cd$_{0.17}$Se $^{77}$Se NMR study. a** One-dimensional $^{77}$Se solid-state NMR spectra of ZnSe in direct acquisition (bottom spectrum), of Zn$_{0.83}$Cd$_{0.17}$Se in direct acquisition using short pulses (15°) and ~0.1 T$_1$ recycle delay, recorded in 90 hours (central spectrum), and of Zn$_{0.83}$Cd$_{0.17}$Se in a CPMG experiment with ~3 T$_1$ recycle delay, recorded in 90 hours (top spectrum). In the latter case the deconvoluted spectrum (red curve) is superimposed onto the raw data (black curve), for a direct comparison. **b** Binomial distribution of the five possible Se-centered tetrahedron clusters – as sketched out – forming the zincblende Zn$_{0.83}$Cd$_{0.17}$Se crystal in its $\kappa$-dependence on a scale going from slight anticlustering (a-clust., $\kappa<0$) to full clustering (clust., $\kappa=1$) on each side of the random case ($\kappa=0$).

**Figure S2 | Probability for B-C percolation on the zincblende and wurtite A$_{1-x}$B$_x$C lattices depending on clustering.** Probability for wall-to-wall B-C percolation in large (10×10×10) A$_{1-x}$B$_x$C supercells with zincblende (hollow symbols) and wurtzite (full symbols) structures – corresponding to different 1-2-3-1… and 1-2-1… sequences of high-density (111)-type packing planes (in the zincblende case) from top to bottom – in the ideal case of a random A↔B substitution ($\kappa\sim0$, at any x) and in case of a pronounced trend towards local clustering ($\kappa\sim0.5$, at x=0.17). Both types of atom arrangement are optimized by simulated annealing (see Methods). Side and front views of a B-C percolative path – emphasized (grey scale) – in the zincblende lattice, with a distinction between like substituents (B) from the bottom (-) and top (+) planes, for clarity.

**Figure S3 | High-pressure X-ray Zn$_{0.83}$Cd$_{0.17}$Se diffractograms.** Selection of high-pressure Zn$_{0.83}$Cd$_{0.17}$Se powder X-ray diffraction diffractograms taken in the upstroke (↑) regime with the 0.378 Å radiation showing the phase transition sequence from zincblende (ZB) to rock-salt (RS). The reported diffractograms cover successively the pure zincblende (ZB) phase, the progressive emergence of the rock-salt (RS) phase, the disappearance of the zincblende phase and the pure rock-salt phase, from bottom to top. The final diffractogram obtained at nearly ambient pressure at the term of the downstroke (↓) regime is added (bottom curve), for reference purpose. Circles and stars mark diffraction lines due to Au used for pressure calibration and Neon used as the pressure transmitting medium, respectively.

**Figure S4 | Pressure dependence of Zn$_{1-x}$Cd$_x$Se lattice parameters. a** Zn$_{0.925}$Cd$_{0.075}$Se. **b** Zn$_{0.83}$Cd$_{0.17}$Se. **c** Zn$_{0.63}$Cd$_{0.37}$Se. **d** Zn$_{0.34}$Cd$_{0.66}$Se. **e** Zn$_{0.12}$Cd$_{0.88}$Se. Various cubic (zincblende-ZB, rock-salt-RS) and hexagonal (Wurtzite-WU, Cinnabar-CI) crystal structures, corresponding to one (a) and two (a, c) lattice parameters, respectively, are revealed in the upstroke (↑, hollow symbols) or downstroke (↓, full symbols) regimes.

**Figure S5 | Composition dependence of Zn$_{1-x}$Cd$_x$Se structural/mechanical parameters.** Bulk modulus at ambient pressure of Zn$_{1-x}$Cd$_x$Se taken in the (dominant) zincblende phase (B$_0$, left axis) and Zn$_{1-x}$Cd$_x$Se zincblende→rock-salt pressure transition (right axis) in the upstroke regime. The average B$_0$ values (circles) derived from the current high-pressure X-ray data (full symbols, in reference to Fig. S4) and by applying the same procedure to the ZnSe data taken from Ref. S13 (hollow symbol) are marred by error bars extending up/down to extreme B$_0$ values, as indicated. The transition pressure is identified as the mean value between the critical pressures corresponding to first emergence of the rock-salt phase and subsequent disappearance of the zincblende/wurtzite phase in the upstroke regime. The related error bar is defined accordingly.

**Figure S6 | High-pressure Zn$_{0.83}$Cd$_{0.17}$Se near-forward Raman spectra. a** Selection of high-pressure Zn$_{0.83}$Cd$_{0.17}$Se near-forward and backward Raman spectra taken at (nearly) the same sample spot in the

upstroke regime by using a 514.5 nm laser beam at (nearly) at normal incidence/detection through/onto the parallel (110)-oriented faces of a single crystal obtained by cleavage. Various spectra at ~5 GPa are taken by departing progressively from the (nearly) perfect forward scattering geometry up to maximum scattering angle (backward scattering geometry, thin curve), as sketched out. At high pressure, the near-forward Raman spectra taken at the same sample spot with the 488.0 nm and 514.5 nm are juxtaposed, for comparison. The Raman spectrum recorded at the term of the downstroke regime (bottom curve) is added for reference purpose. **b** Selection of high-pressure $Zn_{0.83}Cd_{0.17}Se$ backward Raman spectra taken with a powder by using the 532.0 nm laser line. The spacings between the two submodes forming the Zn-Se TO doublet ($\delta$) and between the (lower) Cd-Se singlet and the (upper) Zn-Se doublet ($\Delta$) are emphasized (horizontal paired arrows), as well as remarkable intensity interplays (vertical arrows).

**Figure S7 │ $Zn_{0.83}Cd_{0.17}Se$ refractive index depending on pressure.** $Zn_{0.83}Cd_{0.17}Se$ refractive index measured at ambient pressure throughout the visible range by spectroscopic ellipsometry (symbols) adjusted by a polynomial fit (solid curve). Corresponding replicas translated at 5 GPa and 10 GPa are guided by corresponding shifts of the fundamental optical band gap of pure ZnSe as predicted in Ref. 41. The fundamental optical band gap of $Zn_{0.83}Cd_{0.17}Se$ at 0 GPa corresponding to the onset of absorption in the ellipsometry data is indicated, for reference purpose. By exciting the Raman spectra with the 488.0 and 514.5 nm laser lines (the hatched areas cover ~200 cm$^{-1}$, of the order of the detected PP frequencies – Fig. 2) in the used Stokes geometry (corresponding to $\omega_i < \omega_s$, see arrows) the addressed spectral ranges fall close to the gap-related singularity in the dispersion of the refractive index throughout the studied pressure domain.

**Figure S8 │ Theoretical (TO,LO) $Zn_{0.83}Cd_{0.17}Se$ Raman spectra depending on pressure. a.** Reference theoretical insight into the $Zn_{0.83}Cd_{0.17}Se$ (TO,LO) Raman spectra at 0 GPa. **b** Same spectra at 10 GPa depending on whether the $TO_{Zn-Se}^{Zn} - TO_{Zn-Se}^{Cd}$ frequency gap ($\delta$) closes (scenario 1), remains stable (scenario 2) or opens (scenario 3) under pressure. The spectra are calculated *via* the generic form of the Raman cross section given in Ref. 18, taking an arbitrary decrement (scenario 1) / decrement (scenario 3) in the $\delta$ value of 10 cm$^{-1}$ at 10 GPa with respect to that at 0 GPa, using the same $TO_{Cd-Se} - TO_{Zn-Se}^{Zn}$ frequency gap ($\Delta$) at 10 GPa in all scenarios (taken from Fig. 2).

**Figure S9 │ High-pressure $Zn_{0.48}Be_{0.52}Se$ (TO,LO) Raman spectra/frequencies. a** Selection of high-pressure $Zn_{0.48}Be_{0.52}Se$ Raman spectra taken in the backscattering geometry with a powder in the upstroke regime. **b** Corresponding TO and LO Raman frequencies (hollow squares) and independent data set obtained by using a $Zn_{0.48}Be_{0.52}Se$ single crystal (circles) both in the upstroke (hollow symbols) and downstroke (full symbols) regimes. The second data set is marred by error bars (shaded areas) reflecting an uncertainty in the modeling of the bimodal TO signal using Lorentzians. The resonance ($\omega_c$,$P_c$) corresponding approximately to the actual crossing – within experimental error – of the two TOs is indicated.

**Figure S10 │ Mechanical coupling between the two BeSe-like TO oscillators of $Zn_{0.5}Be_{0.5}Se$.** Predicted bare/uncoupled $Zn_{0.48}Be_{0.52}Se$ ($\omega_{T,1}, \omega_{T,2}$) TO frequencies behind the experimentally observed ($\omega_{T,+}, \omega_{T,-}$) ones depending on the magnitude of the mechanical coupling (determined by $\omega'$) and corresponding variation of the coupled ($\omega_{L,+}, \omega_{L,-}$) LO frequencies. The experimental TO and LO values are taken from Ref. S36 (specifically Fig. 1 therein). The experimental $\omega_{L,-}$ frequency, though marred by a rather large error bar (shaded area), helps to estimate $\omega'$ (as schematically shown).

**Figure S11 │ Exceptional phonon point of $Zn_{0.5}Be_{0.5}Se$.** Frequency gap between the TO-like SYM. and ASYM. normal modes at the resonance depending on competition between mechanical coupling ($\beta_T$: fixed by $k'$) and overdamping ($\beta_{cr}$: depending on $\gamma_1$ vs. $\gamma_2$), as emphasized (shaded area). At exact

compensation between gain (coupling) and loss (overdamping) the phonon exceptional point is achieved corresponding to degeneracy of the coupled TOs (dark spot).

**Figure S12 │ Pressure-dependence of the Zn$_{\sim0.5}$Be$_{\sim0.5}$Se Raman frequencies – Theory vs. experiment.** Experimental TO (hollow symbols) and LO (full symbols) Zn$_{0.48}$Be$_{0.52}$Se Raman frequencies depending on pressure (Fig. S7). **a** Polynomial adjustments of the experimental TO frequencies. **b** Bare/uncoupled TO frequencies ($P \leq P_c$). **c** Bare/uncoupled TO frequency of the exceptional mode ($P > P_c$). **d** Uncoupled LO frequencies ($P \leq P_c$). **e** Coupled LO frequencies ($P \leq P_c$). **f** Coupled $\omega_{+,L}$ frequency obtained by considering a progressive overdamping of the lower oscillator until the exceptional point is achieved at the resonance ($\omega_c, P_c$). **g** LO frequency of the exceptional mode being awarded the full amount of Be-Se oscillator strength. **h** Corresponding LO frequency in case of a dead loss of oscillator strength for the lower mode from the resonance onwards.

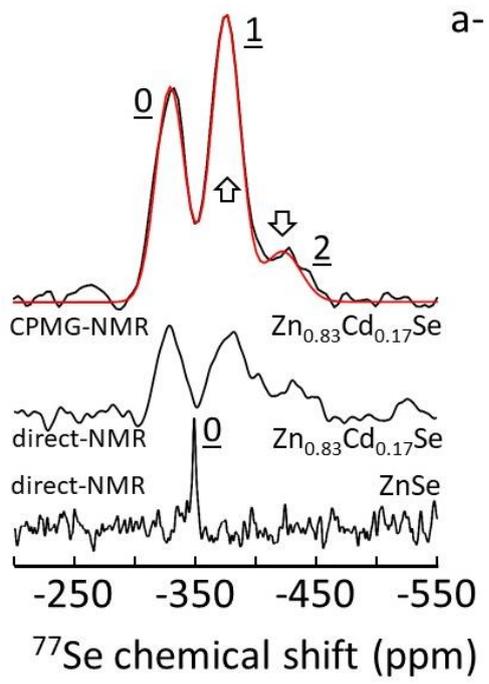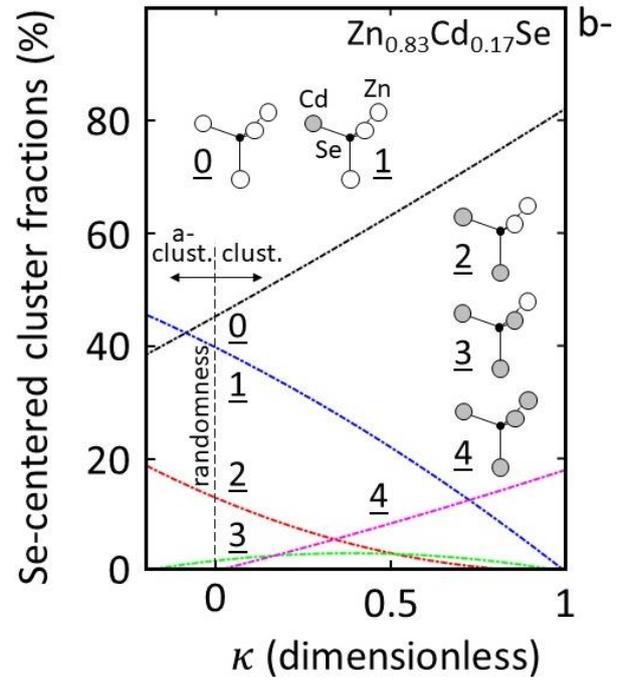

**Figure S1**

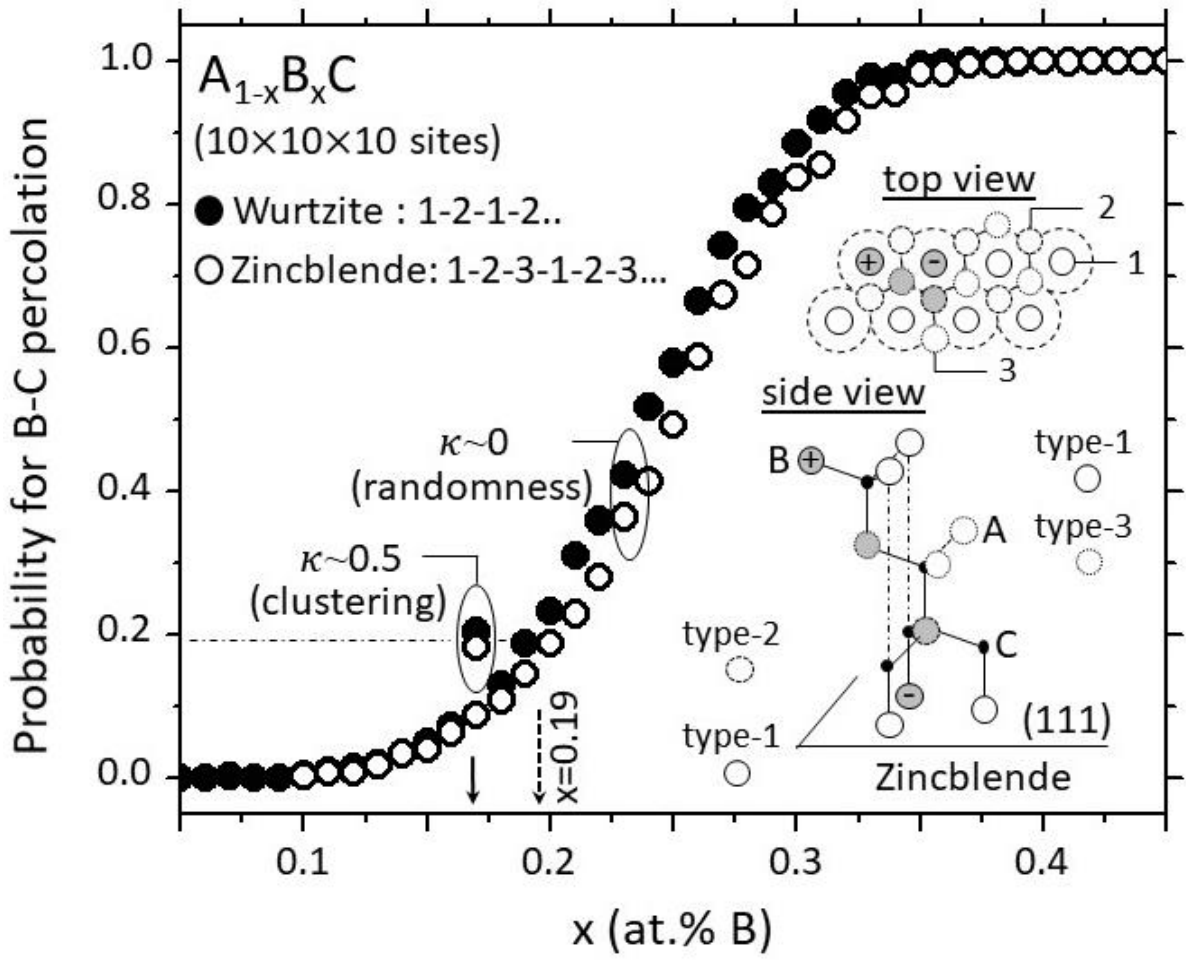

**Figure S2**

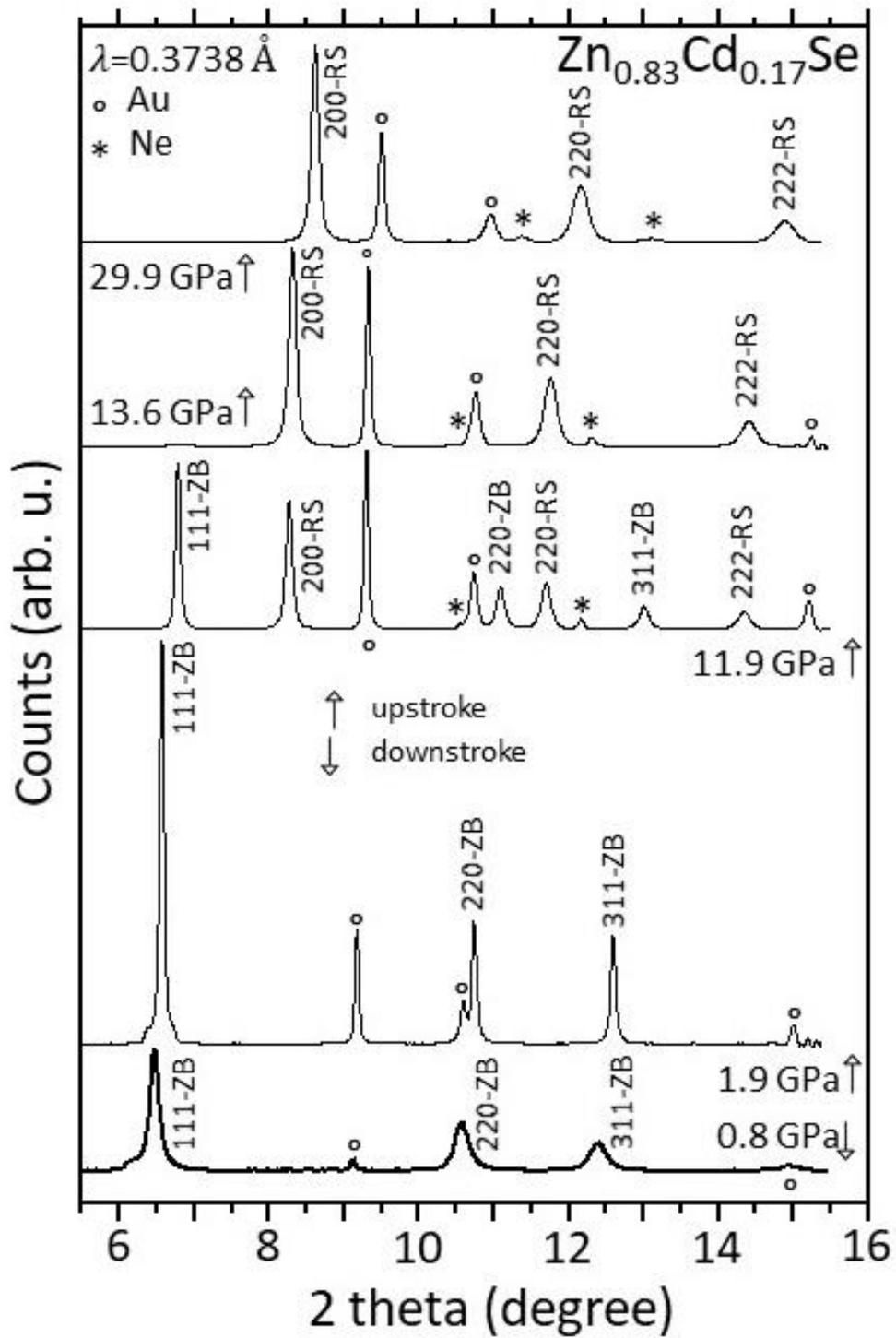

**Figure S3**

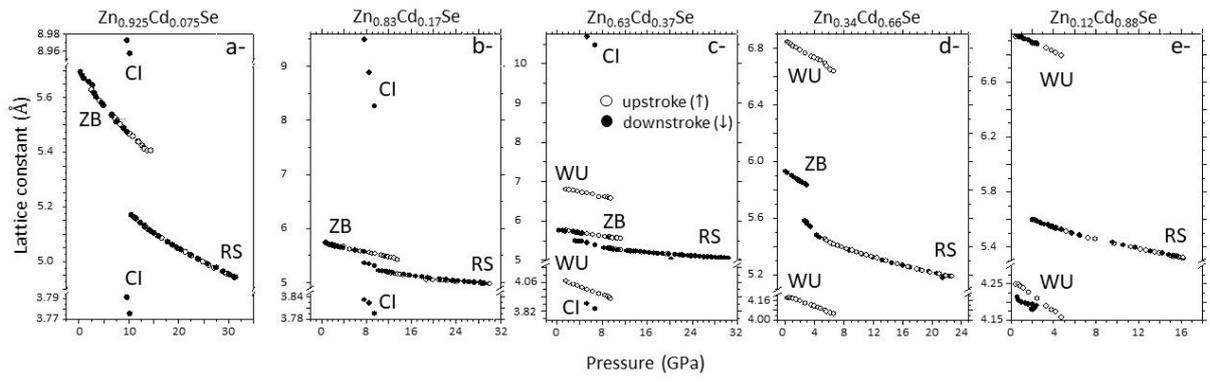

**Figure S4**

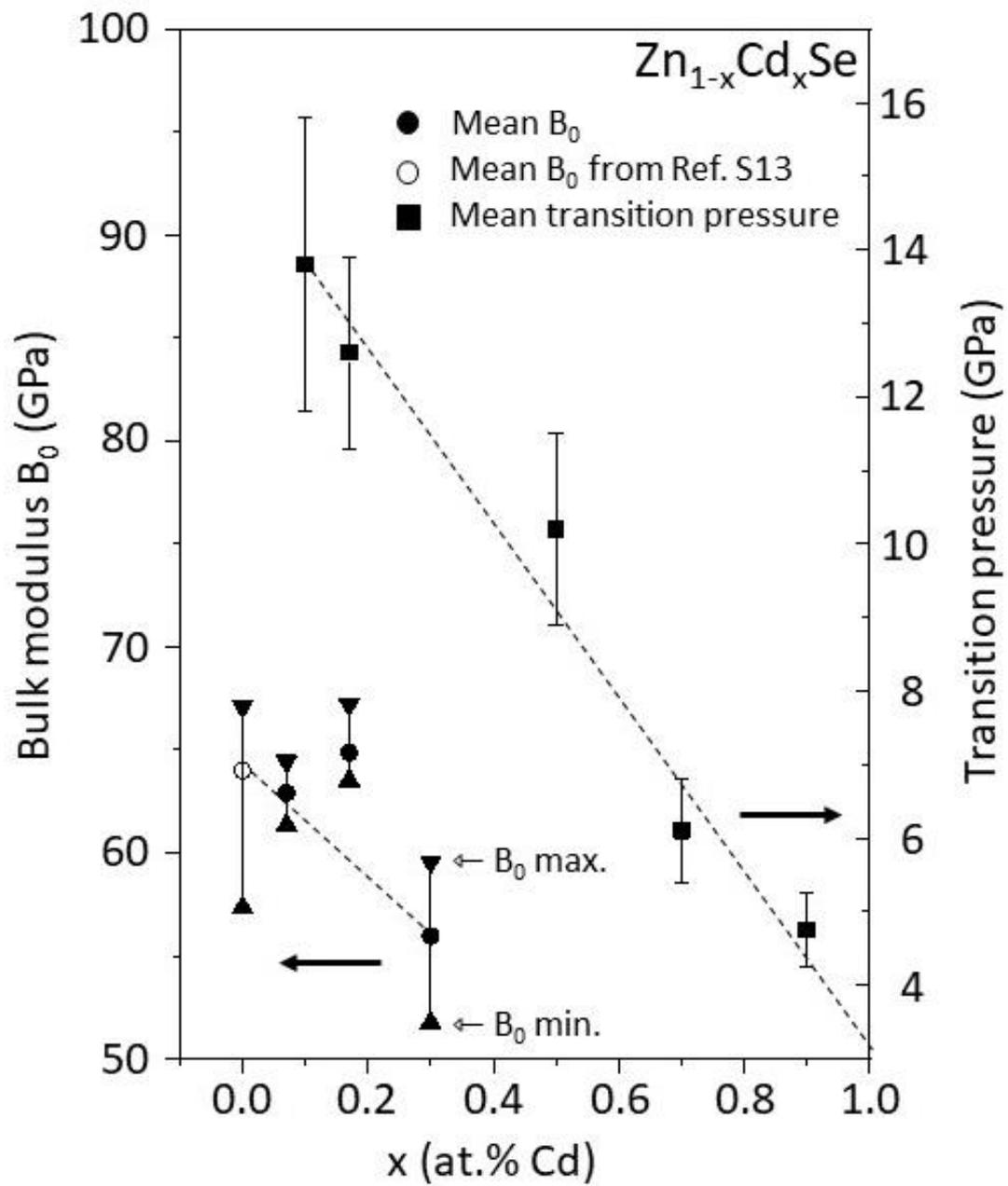

**Figure S5**

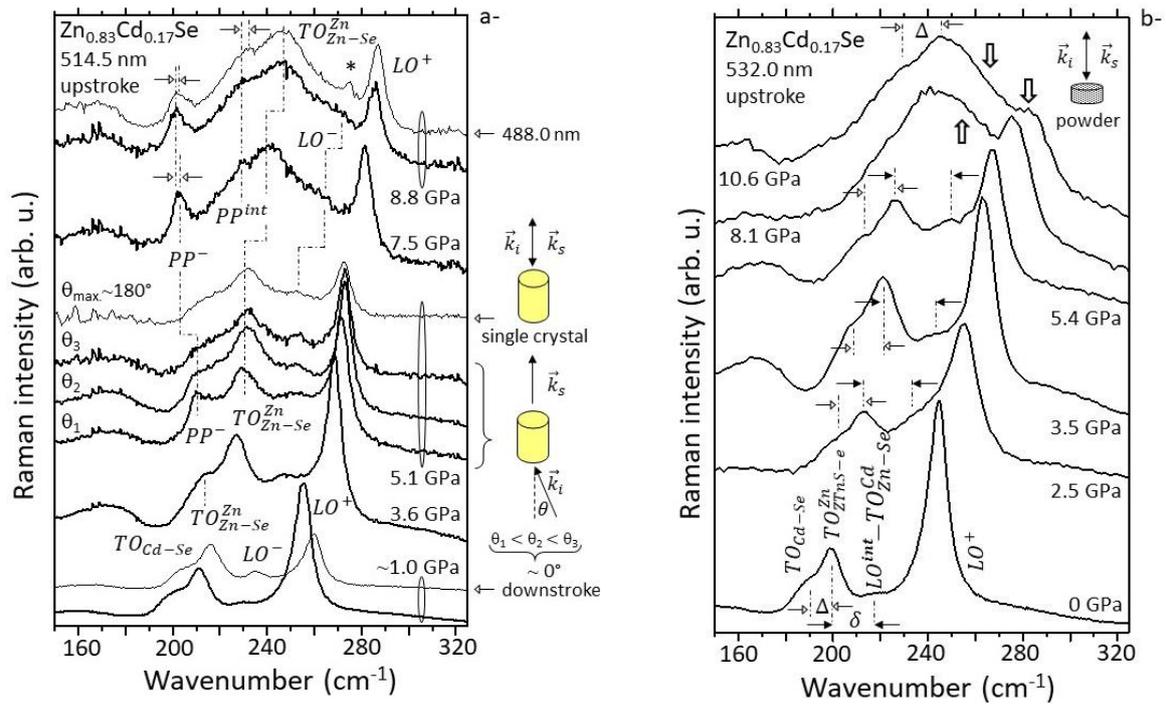

**Figure S6**

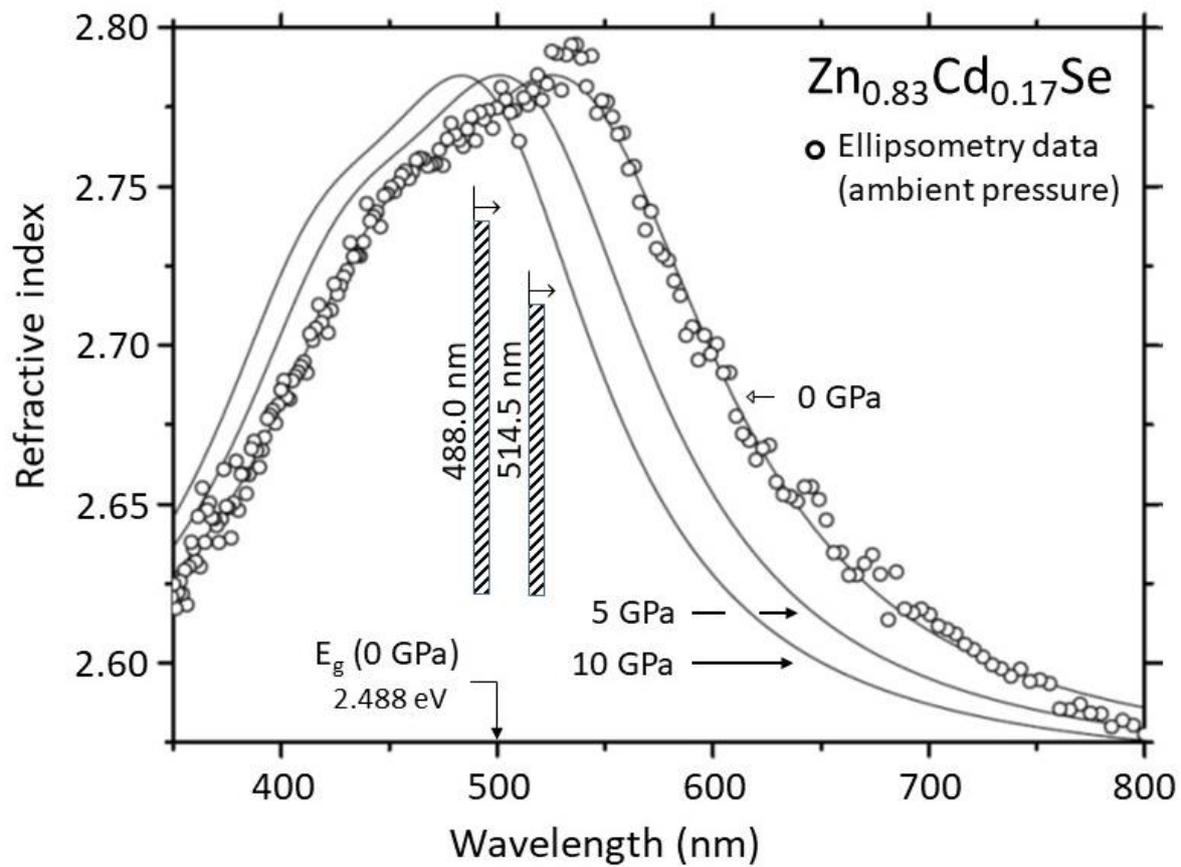

**Figure S7**

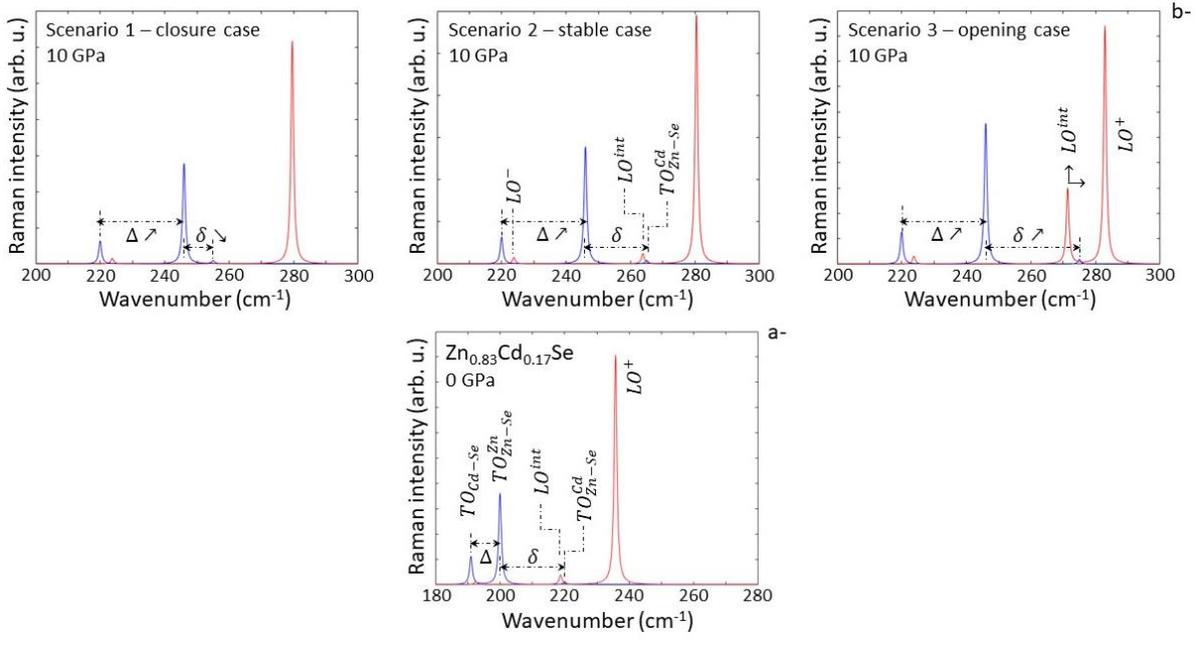

**Figure S8**

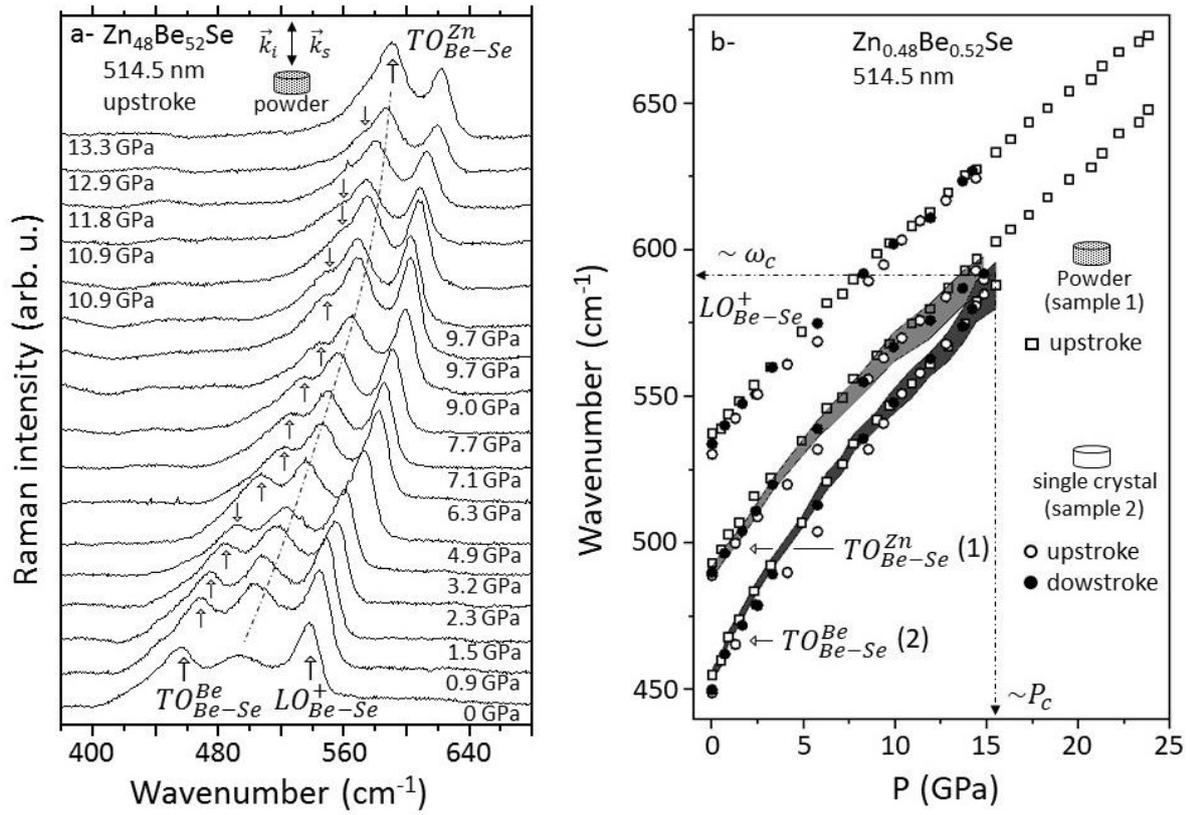

**Figure S9**

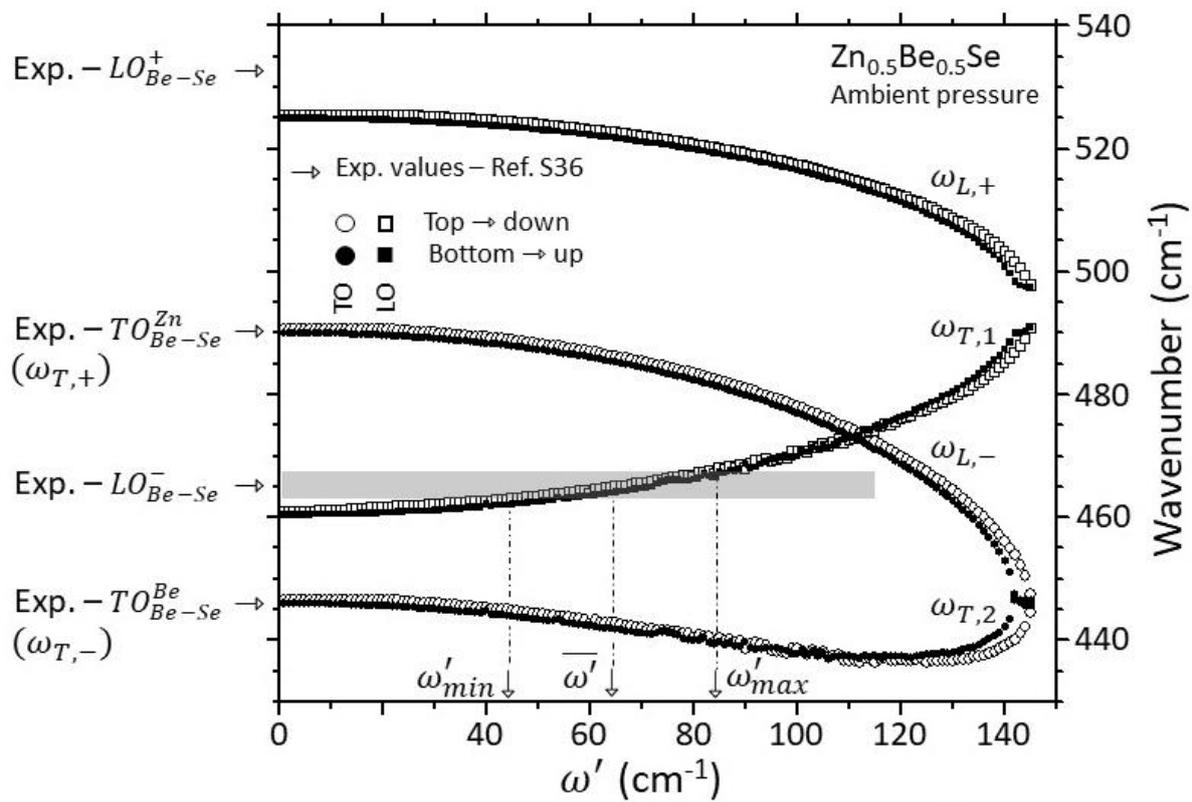

**Figure S10**

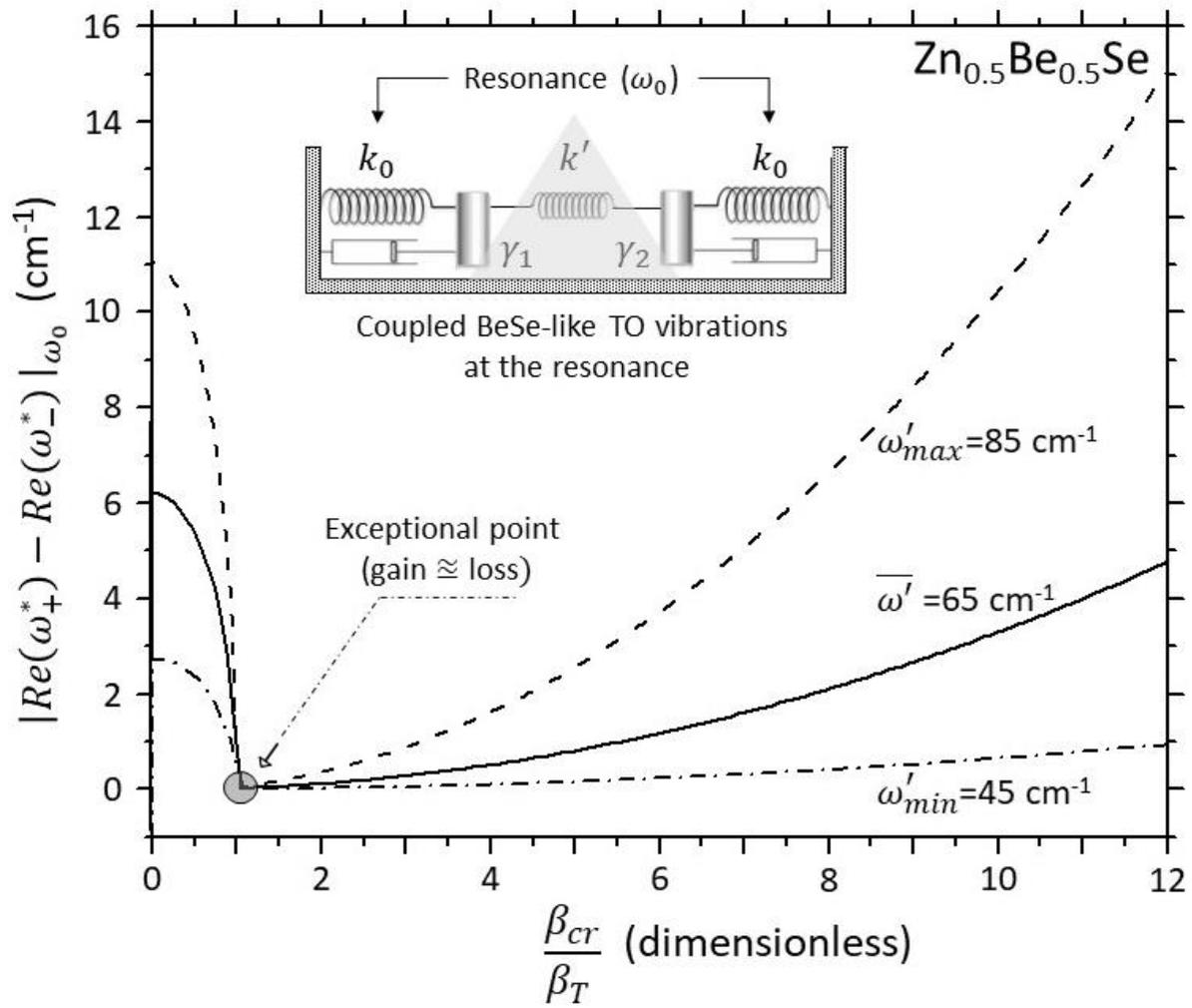

**Figure S11**

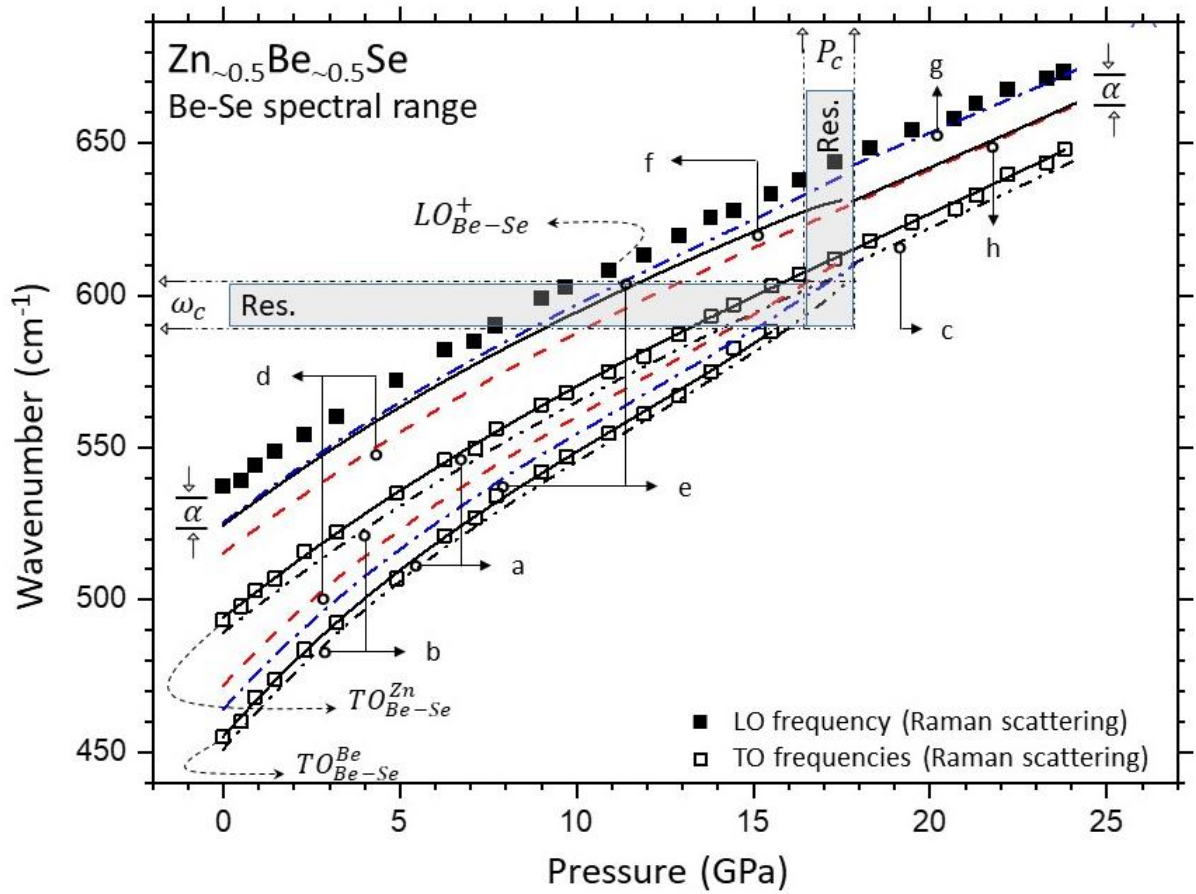

**Figure S12**